\documentclass[twocolumn,tighten]{aastex701}
\usepackage{CJK}

\DeclareRobustCommand{\GalThreeD}{\texttt{Gal3D}}
\graphicspath{{./}{figure/}}
\usepackage{amsmath}

\begin{document}
\begin{CJK*}{UTF8}{gbsn}

\title{\texttt{Gal3D}: Superellipsoid Modeling of Radial 3D Galaxy Structure in IllustrisTNG and EAGLE Simulations}

\author[orcid=0009-0006-5658-414X,gname=Shuai,sname=Lu]{Shuai Lu (卢帅)}
\email{lushuai@stu.xmu.edu.cn}
\affiliation{Department of Astronomy, Xiamen University, Xiamen, Fujian 361005, P.R. China}

\author[orcid=0000-0001-9953-0359,gname=Min,sname=Du]{Min Du (杜敏)}
\affiliation{Department of Astronomy, Xiamen University, Xiamen, Fujian 361005, P.R. China}
\email[show]{dumin@xmu.edu.cn}
\correspondingauthor{Min Du}

\begin{abstract}

Galaxy morphology and structure are key tracers of galaxy formation and evolution, making accurate measurements of intrinsic three-dimensional (3D) shape essential for linking morphology to galaxy assembly and for comparing numerical simulations. We present \texttt{Gal3D}, a framework that reconstructs smoothed density fields from particle data and quantifies the radial 3D structure of simulated galaxies by fitting superellipsoids to iso-density surfaces. The method recovers axis ratios, orientations, center offsets, and superellipsoid indices ($S_a$, $S_b$, $S_c$), enabling a flexible characterization of diverse galactic structures such as disks, classical bulges, box/peanut bulges, and triaxial components.
Applying \texttt{Gal3D} to galaxies in the IllustrisTNG and EAGLE simulations, we find that the radial extent of flattened disk regions increases with stellar mass up to $M_{*,30}\sim10^{11}\,M_\odot$ and then declines sharply, with EAGLE galaxies showing a saturation at $M_{*,30}\sim10^{10.5}\,M_\odot$. The bar-related $ \varepsilon_{ab}\equiv 1-b/a$ strengthens above $M_{*,30}\sim10^{10.5}\,M_\odot$ in both simulations, but remains systematically weaker in EAGLE. In TNG, outer bar regions are commonly associated with elevated $S_a$ and $S_c$, indicating enhanced boxiness and more prominent box/peanut-shaped bulges, whereas such higher-order signatures are weak or absent in EAGLE. At the highest stellar masses, flattened disks become less prominent, while inner prolate or triaxial structures remain common and massive EAGLE galaxies have more prolate or triaxial outer stellar bodies than their TNG counterparts. These results demonstrate that \texttt{Gal3D} provides a practical framework for quantifying intrinsic radial 3D structure and comparing morphology across cosmological simulations.

\end{abstract}

%\keywords{\uat{Galaxies}{573} --- \uat{Cosmology}{343} --- \uat{High Energy astrophysics}{739} --- \uat{Interstellar medium}{847} --- \uat{Stellar astronomy}{1583} --- \uat{Solar physics}{1476}}


\section{INTRODUCTION}

The intrinsic three-dimensional (3D) shape of a galaxy is a fundamental tracer of its assembly history and other physical properties \citep[e.g.,][]{rydenIntrinsicShapeSpiral2006,weijmansATLAS3DProject2014,vandesandeRelationCharacteristicStellar2018}. 
Galaxies span a wide range of morphologies, from irregular dwarfs and thin disks to triaxial early-type systems \citep{hubbleExtragalacticNebulae1926,freemanDisksSpiralS01970,sandageCarnegieAtlasGalaxies1994,kormendySecularEvolutionDisk2013}, and their shape often changes strongly with radius as bulges, disks, bars, and halos dominate different regions \citep{kormendySecularEvolutionFormation2004,kruitGalaxyDisks2011,kormendySecularEvolutionDisk2013,sellwoodSecularEvolutionDisk2014}. Measuring how intrinsic 3D shape varies with radius is therefore essential for understanding the formation and evolution of galactic structure.

In observations, however, intrinsic 3D structure cannot be measured directly and must be inferred from projected two-dimensional light distributions. Common approaches, such as isophotal ellipse fitting \citep{jedrzejewskiCCDSurfacePhotometry1987,buskoErrorEstimationElliptical1996,ciamburEllipsesAccuratelyModelling2015} and parametric light decomposition \citep{simardGIM2DIRAFPackage1998,pengDetailedStructuralDecomposition2002,desouzaBUDDANewTwodimensional2004,pignatelliGASPHOTToolGalaxy2006,pengDetailedDecompositionGalaxy2010,erwinIMFITFASTFLEXIBLE2015}, provide valuable information on apparent morphology. However, projection couples intrinsic geometry to viewing angle, so recovering the underlying 3D structure generally requires assumptions about symmetry, orientation, and structural simplicity, often supplemented by statistical inversion over large samples \citep[e.g.,][]{sandageIntrinsicFlatteningSpiral1970,binneyApparentTrueEllipticities1981,bakIntrinsicShapeDistribution2000}. 
As a result, existing methods usually constrain the intrinsic shape distributions of specific galaxy populations \citep[e.g.,][]{lambasTrueShapesGalaxies1992,padillaShapesGalaxiesSloan2008,chakrabortyIntrinsicShapesVery2011,rodriguezIntrinsicShapeGalaxies2013,roychowdhuryIntrinsicShapesDwarf2013,weijmansATLAS3DProject2014} or of specific galactic structures \citep[e.g.,][]{costantinIntrinsicShapeBulges2018,mendez-abreuIntrinsicThreedimensionalShape2018,favaroIntrinsicFlatteningGalaxy2024}, rather than detailed radial 3D shape profiles for individual multicomponent galaxies.

This limitation is especially important for structures that are intrinsically 3D and often overlap in projection. Warps and tilted stellar components remain difficult to interpret observationally \citep[e.g.,][]{binneyWarps1992,dehnenTwistedPrecessingCepheid2023,hanTiltedDarkHalos2023,sanchez-saavedraFrequencyWarpedSpiral1990,garcia-ruizNeutralHydrogenOptical2002,annWarpedDisksSpiral2006}, while bars are linked to orbital families and related features such as double bars, box/peanut-shaped bulges, disky bulges, and nuclear disks \citep{erwinDoublebarredGalaxiesCatalog2004,athanassoulaNatureBulgesGeneral2005,shenOurMilkyWay2010,erwinPeanutsAngle2013,duFORMINGDOUBLEBARREDGALAXIES2015,athanassoulaBoxyPeanutBulges2016,erwinDependenceBarFrequency2018,schultheisNuclearStellarDiscs2025}. Even elliptical galaxies can show isophotal twists and boxy or disky deviations whose physical origins remain debated \citep[e.g.,][]{benacchioTriaxialityEllipticalGalaxies1980,zeeuwStructureDynamicsElliptical1991,naabFormationBoxyDisky1999,naabPropertiesEarlyTypeDry2006,mitsudaIsophoteShapesEarlytype2017,monteiro-oliveiraNoEvidenceDichotomy2025}.
More generally, projected data alone make it difficult to disentangle overlapping components, measure their radial 3D shapes robustly, and connect morphology to the underlying physical processes.

Numerical simulations provide a powerful complement because the intrinsic 3D structure is directly accessible. Early collisionless N-body simulations established the theoretical foundation for understanding galaxy dynamics and structure formation \citep[e.g.,][]{binneyGalacticDynamics1987,zeeuwStructureDynamicsElliptical1991,sellwoodDynamicsBarredGalaxies1993,binneyGalacticDynamicsSecond2008,sellwoodSecularEvolutionDisk2014,sellwoodGALAXYPackageNbody2014}.
Modern cosmological hydrodynamical simulations \citep[e.g.,][]{vogelsbergerCosmologicalSimulationsGalaxy2020,valentiniHydrodynamicMethodsSubresolution2025,teyssierNumericalCosmology2025,feldmannCosmologicalSimulationsGalaxies2026}, such as IllustrisTNG \citep{marinacciFirstResultsIllustrisTNG2018,pillepichFirstResultsIllustrisTNG2018,nelsonFirstResultsIllustrisTNG2018,naimanFirstResultsIllustrisTNG2018,springelFirstResultsIllustrisTNG2018} and EAGLE \citep{schayeEAGLEProjectSimulating2015,crainEAGLESimulationsGalaxy2015}, produce large and diverse galaxy populations and broadly reproduce many observed galaxy properties. 
They naturally contain disks \citep[e.g.,][]{sotillo-ramosDiscFlaringTNG502023,semenovFormationGalacticDisks2024,chenDownbendingBreaksGalactic2026}, bars \citep[e.g.,][]{algorryBarredGalaxiesEAGLE2017,rosas-guevaraBuildupStronglyBarred2020,zhaoBarredGalaxiesIllustrisTNG2020,luIllustrisTNGInsightsFactors2025}, bulges \citep[e.g.,][]{gargiuloHighLowSersic2022,andersonInterplayAccretionGalaxy2023,zhangDiversePhysicalOrigins2025}, warps \citep[e.g.,][]{semczukTidallyInducedWarps2020,zeeWarpedDiskGalaxies2022,hanTiltedDarkHalos2023}, stellar halos \citep[e.g.,][]{folsomCosmologicalSimulationsStellar2025,heMorphologyChallengesDecomposing2025}, and other structural features in a fully cosmological context. They therefore provide an excellent laboratory for systematic comparisons of radial 3D morphology across galaxy populations and simulation models.
The challenge, however, lies in reliably characterizing such intrinsic 3D structure.
%A robust measurement of radial 3D shape in simulations is non-trivial. 

The most widely used approach is the iterative inertia-tensor method \citep{katzDissipationlessCollapseExpanding1991,dubinskiStructureColdDark1991,zempDeterminingShapeMatter2011}, which selects particles in shells and repeatedly diagonalizes the shape tensor to align each shell with the principal axes of the mass distribution. Although this method was originally developed for nearly ellipsoidal systems and has been widely applied to dark matter halos \citep[e.g.,][]{bailinInternalExternalAlignment2005,chuaShapeDarkMatter2019,giocoliAIDATNGProject3D2026}, it has also been used as a first-order estimate of stellar shapes in some studies \citep[e.g.,][]{bassettProspectsRecoveringGalaxy2019,yongGalaxy3DShape2024,kleinShapeFIREboxGalaxies2026}. However, it can be computationally expensive for particle-rich galaxies and sensitive to noise in low-density regions. More importantly, stellar systems often contain multiple dynamical components and strongly non-ellipsoidal features, such as boxiness and diskiness, that violate the assumptions underlying the method. Traditional tensor-based approaches therefore have limited ability to characterize the radial structure of complex multicomponent galaxies.

In this work, we extend the iterative shape-tensor framework to adaptively smoothed density fields and introduce a superellipsoid description of iso-density surfaces. The superellipsoid formalism recovers axis ratios, orientations, center offsets, and boxiness/diskiness indices, enabling a more flexible characterization of non-ellipsoidal galactic structures. The rest of the paper is organized as follows. Section~\ref{sec:method} presents the extended iterative method and the iso-density superellipsoid formalism. Section~\ref{sec:data} describes the simulations and galaxy samples. Section~\ref{sec:examples} applies the method to representative galaxies and presents their radial 3D shape profiles. Section~\ref{sec:statistics} reports statistical trends across different simulations. Section~\ref{sec:summary} summarizes the main results.

\section{\GalThreeD: Three-dimensional shape modeling}
\label{sec:method}

\begin{figure*}
\centering
{\includegraphics[width=0.99\textwidth]{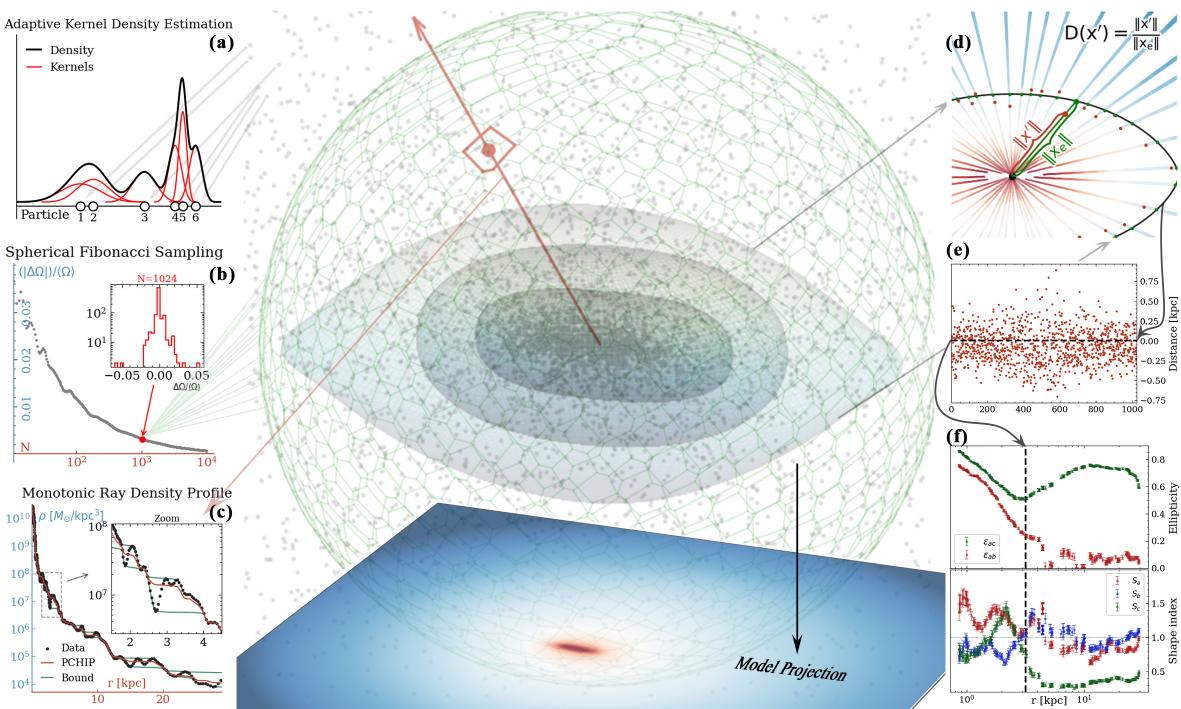}}
\caption{Overview of the \texttt{Gal3D} methodology. The central 3D rendering shows the stellar particle distribution (gray points), the spherical Voronoi cells associated with the ray directions (green mesh), and three nested fitted superellipsoid models at different density levels; the projected image beneath the rendering shows the corresponding model projection. Panels (a)--(c) show the three preprocessing steps: (a) adaptive kernel density estimation from particles, (b) spherical Fibonacci sampling, and (c) construction of a monotonic density profile along each ray. Panels (d)--(f) show the subsequent fitting outputs: (d) the surface-radius-ratio construction used in the fit, (e) the residuals for one representative iso-density surface, and (f) the resulting radial profiles of ellipticity and superellipsoid shape indices.}
	\label{fig:gal3d_method}
\end{figure*}

\begin{figure*}
\centering
{\includegraphics[width=0.95\columnwidth]{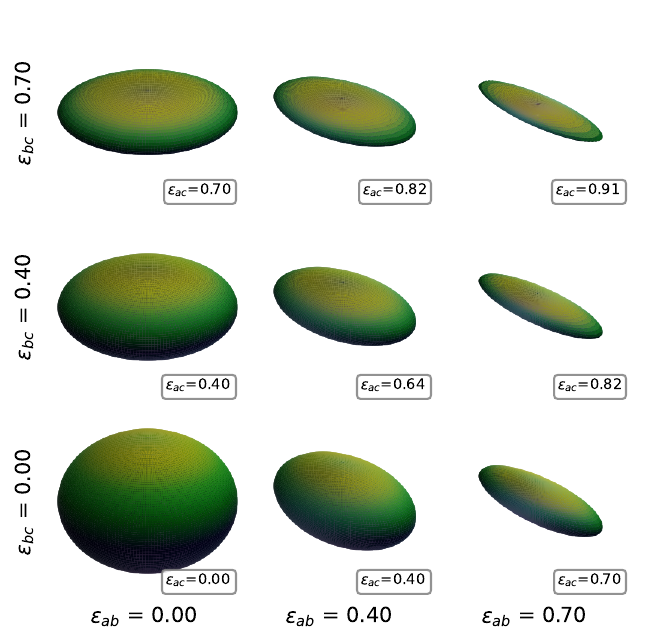}}
{\includegraphics[width=0.95\columnwidth]{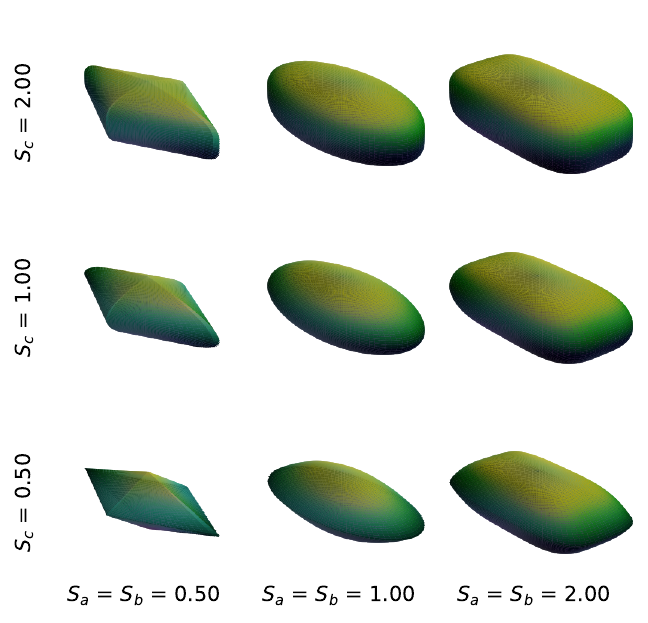}} 
    \caption{Examples of generalized ellipsoids (superellipsoids) with varying ellipticities and shape indices. Left: standard ellipsoids ($S_a=S_b=S_c=1$) with varying ellipticities $\varepsilon_{ab} \equiv 1-b/a$ and $\varepsilon_{bc} \equiv 1-c/b$, with the value of $\varepsilon_{ac}\equiv 1-c/a$ annotated in each panel. Moving upward increases flattening, while moving from left to right increases elongation. Right: models with fixed axis ratios $a:b:c=10:5:4$ and varying shape indices. The $x$-axis labels give $S_a$ and $S_b$ (with $S_a=S_b$), and the $y$-axis labels give $S_c$. This panel shows the case in which the two indices ($S_a$ and $S_b$) vary together, while Figure~\ref{fig:superellipsoid_shape_sa_sb_examples} in Appendix~\ref{sec:superellipsoid_sa_sb} shows independent variations in $S_a$ and $S_b$ at fixed $S_c=1$. Values $S<1$ produce more pointed or diamond-like shapes, whereas $S>1$ produces boxier shapes. The ellipticities therefore control flattening and elongation, while the shape indices quantify boxiness and diskiness.}
	\label{fig:superellipsoid_shape}
\end{figure*}

\begin{figure*}[ht!]
\plotone{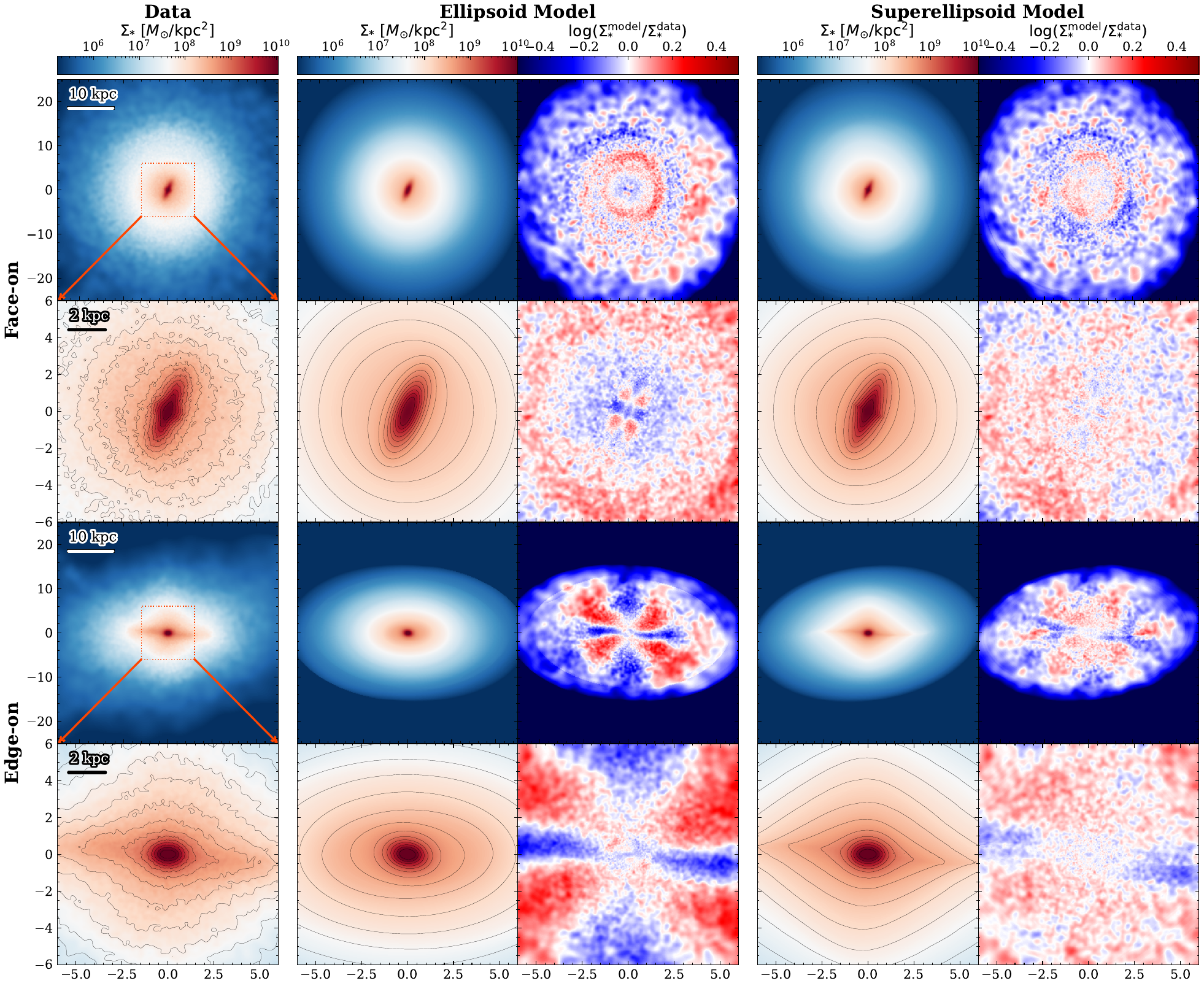}
	\caption{Comparison of the projected stellar surface-density maps
produced by the ellipsoid and superellipsoid models for a
representative simulated galaxy.  Rows are grouped into face-on
(top two rows) and edge-on (bottom two rows) projections; within
each pair, the upper row shows the wide-field view
($25\,\mathrm{kpc}$ on a side) and the lower row shows the
zoomed central region ($6\,\mathrm{kpc}$ on a side), as
indicated by the dashed rectangle in the wide-field data panel.
From left to right, the five columns show: (1) the simulation
stellar surface density $\Sigma_*$ (SPH-projected from the particle data); (2) the ellipsoid model
$\Sigma_*$; (3) the ellipsoid model logarithmic residual
$\log(\Sigma_{*}^{\mathrm{model}}/\Sigma_{*}^{\mathrm{data}})$; (4) the superellipsoid model
$\Sigma_*$; and (5) the corresponding superellipsoid
logarithmic residual. For both models, $\Sigma_*$ is obtained by integrating the fitted 3D density along the line of sight. Positive (red) and negative (blue) residuals
indicate local over- and under-prediction of the surface density,
respectively.  Contour lines in the zoomed panels are isophotes
spaced uniformly in $\log\Sigma_*$.
	\label{fig:ellipsoid_vs_superellipsoid}}
\end{figure*}

In this section, we present the \texttt{Gal3D} pipeline for measuring radial 3D galaxy structure from particle data.\footnote{\url{https://github.com/GalaxySimAnalytics/gal3d}} The method first reconstructs a continuous stellar mass density field with adaptive kernel density estimation, and then provides two complementary approaches: an iterative shape-tensor method that refines a trial ellipsoidal shell on the density field, and a ray-based superellipsoid fit that matches a superellipsoid surface to extracted isodensity contours. The shape tensor provides an ellipsoidal reference measurement of axis ratios and orientations, while the superellipsoid fit is our fiducial method for capturing non-ellipsoidal structure through additional shape parameters. The fitted 3D models can also be projected and compared directly with simulated stellar surface-density maps. Figure~\ref{fig:gal3d_method} summarizes the workflow and the resulting radial shape measurements.

\subsection{Kernel Density-Field Construction}
\label{sec:density_estimator}
We begin with the 3D positions and masses of stellar particles from the simulations and construct a continuous density field. This suppresses discreteness noise in sparsely sampled regions and allows the density to be evaluated at arbitrary spatial positions.

We estimate the density field with an adaptive kernel density estimator with the cubic-spline kernel \citep{monaghanRefinedParticleMethod1985, springelGADGETCodeCollisionless2001}:
\begin{equation}
W(r, h) \;=\; \frac{8}{\pi h^{3}} \begin{cases}
1 - 6\left(\frac{r}{h}\right)^{2} + 6\left(\frac{r}{h}\right)^{3}, & 0 \leq \frac{r}{h} \leq \frac{1}{2}, \\
2\left(1 - \frac{r}{h}\right)^{3}, & \frac{1}{2} < \frac{r}{h} \leq 1, \\
0, & \text{otherwise},
\end{cases}
\end{equation}
This kernel is widely used in smoothed particle hydrodynamics (SPH) simulations.

The local density at an evaluation point $\boldsymbol{x}_i \in \mathbb{R}^3$ is then computed by summing the contributions from neighboring particles,
\begin{equation}
\rho(\boldsymbol{x}) \;=\; \sum_{j=1}^{k} m_{j}\; W\!\bigl(\,\| \boldsymbol{x} - \boldsymbol{x}_j\|,\, h(\boldsymbol{x})\bigr),
\end{equation}
where $m_j$ and $\boldsymbol{x}_j$ are the mass and position of the $j$th neighboring particle, and $h(\boldsymbol{x})$ is the adaptive smoothing length given by the distance from $\boldsymbol{x}$ to its $k$th nearest neighbor. Panel (a) of Figure~\ref{fig:gal3d_method} illustrates this step. Neighbor searches are performed with a KD-tree \citep{maneewongvatanaAnalysisApproximateNearest1999}, as implemented in \texttt{scipy.spatial.KDTree} \citep{2020SciPy-NMeth}.

Throughout this work, we adopt $k=32$, the number of nearest neighbors in the adaptive kernel density estimation, which provides a practical compromise between spatial resolution and noise suppression. Appendix~\ref{sec:numerical_caveats} shows that the recovered shape profiles are stable for $k \gtrsim 16$, so this choice does not affect the main conclusions. All subsequent analyses use this smoothed field, which allows the density to be evaluated on trial ellipsoidal surfaces and along arbitrary radial rays.

\subsection{Angular Sampling on the Sphere}
\label{sec:angular_sampling}

Both the density-field shape tensor and the ray-based superellipsoid fit require nearly uniform angular coverage. We therefore generate a common set of directions with a spherical Fibonacci lattice.
For $i = 1, 2, \ldots, N$, the unit vector $\hat{\boldsymbol{n}}_i = (n_{i,x},\, n_{i,y},\, n_{i,z})$ for direction $i$ is defined by
\begin{equation}
\label{eq:fibonacci}
\begin{aligned}
n_{i,x} &= \sqrt{1 - z_i^2} \, \cos\left(2\pi i \phi^{-1}\right), \\
n_{i,y} &= \sqrt{1 - z_i^2} \, \sin\left(2\pi i \phi^{-1}\right), \\
n_{i,z} &= \frac{2i - 1}{N} - 1,
\end{aligned}
\end{equation}
where $\phi = (\sqrt{5}+1)/2$ is the golden ratio. Unless otherwise noted, we adopt $N=1024$ directions. Panel (b) of Figure~\ref{fig:gal3d_method} shows that these directions provide nearly uniform angular coverage: the spherical Voronoi-cell areas are concentrated within about $2\%$ of the mean. We compute these cells with \texttt{scipy.spatial.SphericalVoronoi} \citep{2020SciPy-NMeth}, following the Delaunay--Voronoi construction described by \citet{caroliRobustEfficientDelaunay2009}. Appendix~\ref{sec:numerical_caveats} shows that the inferred shape profiles are insensitive to reasonable changes in the number of angular directions.

\subsection{Iterative shape-tensor method}
\label{sec:shape_tensor}

Iterative shape tensor analysis \citep{katzDissipationlessCollapseExpanding1991, dubinskiStructureColdDark1991, zempDeterminingShapeMatter2011} measures the local 3D geometry of a mass distribution by iteratively refining a trial ellipsoidal shell. The iteration is initialized with a spherical shell of radius $a$, centered on the galaxy center determined in Section~\ref{sec:data}. The shape tensor is then computed as
\begin{equation}
S_{ij} = \frac{\int_{V} w(\boldsymbol{x})\, \rho(\boldsymbol{x})\, x_i\, x_j\, dV}
              {\int_{V} w(\boldsymbol{x})\, \rho(\boldsymbol{x})\, dV},
\end{equation}
where $\rho(\boldsymbol{x})$ is the density field, $w(\boldsymbol{x})$ is an optional weight, 
$V$ denotes the volume of the current trial ellipsoidal shell, and $x_i$, $x_j$ denote the Cartesian components of the position vector $\boldsymbol{x}$. Following \citet{zempDeterminingShapeMatter2011}, who found that the unweighted tensor gives the most reliable results, we adopt $w(\boldsymbol{x})=1$ throughout. Diagonalizing $S_{ij}$ yields the principal axes of the mass distribution. For a uniform ellipsoidal shell, the eigenvalues scale as $\lambda_i \propto a^2$, $b^2$, and $c^2$, where $a$, $b$, and $c$ are the semi-major, semi-intermediate, and semi-minor axes of the ellipsoid ($a \ge b \ge c$). The trial shell is then rotated into the principal-axis frame and updated with the new axis ratios while keeping the semi-major axis fixed. Iteration continues until the fractional changes in the axis ratios fall below $10^{-3}$ or until 30 iterations are reached. The latter is a conservative upper limit, because we find that convergence is typically reached within $<20$ iterations across our galaxy sample, consistent with previous studies \citep{katzDissipationlessCollapseExpanding1991,debattistaCausesHaloShape2008}. We estimate uncertainties from the differences between the final two iterations.

In the conventional particle-based implementation, the integrals above are replaced by sums over particles in the shell. These estimates can become noisy in narrow radial bins or low-density outer regions \citep{dubinskiStructureColdDark1991, zempDeterminingShapeMatter2011}, where few particles contribute. We therefore adopt a density-field implementation based on the smoothed field introduced in Section~\ref{sec:density_estimator}. In this form, the tensor is evaluated on the trial ellipsoidal surface,
\begin{equation}
\label{eq:angular_shape_tensor}
S_{ij} = \frac{\displaystyle\oint \rho(\boldsymbol{x})\, x_i\, x_j\, d\Omega}
              {\displaystyle\oint \rho(\boldsymbol{x})\, d\Omega},
\end{equation}
where $d\Omega=d(\cos\theta)\,d\phi$ is the solid-angle element. We treat this surface-based expression as a density-field analog of the local shape tensor. The angular integral is evaluated with the Fibonacci-lattice directions $\hat{\boldsymbol{n}}_i$ defined by Equation~\eqref{eq:fibonacci}: for each trial ellipsoid, these directions define surface sampling points, and the corresponding spherical Voronoi-cell areas provide the quadrature weights for $d\Omega$. Equation~\eqref{eq:angular_shape_tensor} is then computed as a weighted sum over directions. The smoothed density field suppresses particle shot noise and yields more stable axis ratios and orientations, especially in sparsely sampled regions.

\subsection{Ray-based superellipsoid fitting}
\label{sec:ray_based_fitting}

The iterative shape tensor provides a useful ellipsoidal baseline, but it cannot capture center offsets or higher-order departures from ellipsoidal symmetry.\footnote{A future extension of the density-field shape-tensor method could expand the angular density field in spherical harmonics, analogous to higher-order Fourier terms in 2D isophotal analysis \citep{jedrzejewskiCCDSurfacePhotometry1987}.} We therefore adopt a more flexible description by fitting generalized ellipsoids, or superellipsoids, to iso-density points extracted from the same smoothed density field. The procedure has three steps: we sample the density field along a nearly uniform set of rays, construct smooth monotonic radial density profiles along those rays and invert them to obtain iso-density points, and fit a superellipsoid to those points at each target density level.

\subsubsection{Radial density profiles from ray sampling}

We sample the density field along rays defined by the Fibonacci-lattice directions introduced in Section~\ref{sec:angular_sampling}. Denoting the galaxy center by $\boldsymbol{x}_c$ and the unit vector of ray $i$ by $\hat{\boldsymbol{n}}_i$, the position at radius $r$ along that ray is
\begin{equation}
\boldsymbol{x}(r,\hat{\boldsymbol{n}}_i)=\boldsymbol{x}_c+r\,\hat{\boldsymbol{n}}_i .
\end{equation}
Along each ray, the density is evaluated at $n=500$ logarithmically spaced radii between $r_{\rm inner}$ and $r_{\rm outer}$. We define $r_{\rm inner}$ as the smallest radius at which the density drops to $0.9$ of the central value and $r_{\rm outer}$ as the radius where $\rho = 100\,M_\odot\,\mathrm{kpc}^{-3}$. This gives a discrete radial profile $\{(r_\ell,\rho_\ell)\}_{\ell=1}^{n}$ for each ray.

Particle noise and substructure can introduce local upturns, so the raw profiles are not always strictly monotonic. To obtain a unique inverse relation $r(\rho)$, we regularize each ray profile into a smooth monotonic non-increasing function. We first define lower- and upper-envelope node sets,
\[
\mathcal{L}=\{\ell:\rho_\ell=\min_{k\le \ell}\rho_k\},\qquad
\mathcal{U}=\{\ell:\rho_\ell=\max_{k\ge \ell}\rho_k\}.
\]
In words, $\mathcal{L}$ consists of indices where the density attains a new record low moving outward from the center, while $\mathcal{U}$ consists of indices where the density attains a new record high moving inward from the outer edge. We interpolate these two node sets with PCHIP \citep{fritschMethodConstructingLocal1984,molerNumericalComputingMatlab2004}, which preserves monotonic trends and avoids the overshoot that can occur with standard cubic splines. The resulting interpolated envelopes are denoted by $\rho_{\rm lower}(r)$ and $\rho_{\rm upper}(r)$.

We then combine the two envelopes locally. In each radial interval $I_m$ bounded by shared envelope nodes, let $N_m$ be the number of raw density samples in interval $I_m$ and $N_{{\rm above},m}$ the number lying above the midpoint between $\rho_{\rm upper}(r)$ and $\rho_{\rm lower}(r)$. We set
\[
w_m=\frac{1+N_{{\rm above},m}}{2+N_m},
\]
and define
\[
\tilde{\rho}(r)
= w_m\,\rho_{\rm upper}(r)
+ \left(1-w_m\right)\rho_{\rm lower}(r),
\qquad r\in I_m .
\]
Because $\rho_{\rm upper}(r)$ and $\rho_{\rm lower}(r)$ are both monotonic decreasing and $w_m$ is constant within each interval, $\tilde{\rho}(r)$ is also monotonic decreasing on that interval. At interval boundaries, the construction uses shared envelope nodes, so the combined profile remains continuous across adjacent intervals. A final PCHIP interpolation of $\tilde{\rho}(r)$ gives the smooth monotonic profile used for inversion.

Panel (c) of Figure~\ref{fig:gal3d_method} illustrates this construction for one representative ray: the black points show the raw density samples, the green curves show the two monotonic envelopes, and the red curve shows the final weighted monotonic profile. The resulting $\rho(r)$ has a unique inverse $r(\rho)$, providing one iso-density point per ray at each target density level.

\subsubsection{Superellipsoid parameterization}
\label{sec:superellipsoid_model}

We model each iso-density surface with a generalized ellipsoid, or superellipsoid,
\begin{equation}
\label{eq:superellipsoid}
f(x,y,z) \;=\; \Bigl[\Bigl(\frac{x}{a}\Bigr)^2\Bigr]^{S_a}
+\Bigl[\Bigl(\frac{y}{b}\Bigr)^2\Bigr]^{S_b}
+\Bigl[\Bigl(\frac{z}{c}\Bigr)^2\Bigr]^{S_c}.
\end{equation}
Here, $a \ge b \ge c > 0$ are the semi-axes, and $S_a$, $S_b$, and $S_c$ are the superellipsoid shape indices. For convenience, we define the ellipticities $\varepsilon_{ab}\equiv 1-b/a$, $\varepsilon_{bc}\equiv 1-c/b$, and $\varepsilon_{ac}\equiv 1-c/a$, with $\varepsilon_{ac} = 1 - (1-\varepsilon_{ab})(1-\varepsilon_{bc})$. The standard ellipsoid is recovered when $S_a = S_b = S_c = 1$, whereas $S < 1$ produces more pointed or diamond-like shapes and $S > 1$ produces boxier shapes. For numerical stability, we restrict the indices to $0.2 < S_a, S_b, S_c < 2$. Figure~\ref{fig:superellipsoid_shape} illustrates how the ellipticities control flattening and elongation, while the shape indices describe departures from a pure ellipsoid. Together, these parameters provide a flexible description of bulges, disks, bars, and other complex galactic structures.

Each fitted iso-density surface is allowed to have its own center and orientation in the simulation coordinate system. We denote the center of each iso-density surface by $\Delta\boldsymbol{x}_c=(\Delta x_c, \Delta y_c, \Delta z_c)$ representing a small offset from the galaxy center, and parameterize the orientation by three Euler angles $(\alpha,\beta,\gamma)$. Let $R_{zyx}(\gamma,\beta,\alpha)$ denote the corresponding rotation matrix. For a set of points stacked by rows into $X\in\mathbb{R}^{n\times 3}$ in the simulation coordinate system, the coordinates in the principal-axis coordinate system of the fitted superellipsoid are
\begin{equation}
\label{eq:batch-transform}
X' \;=\; \bigl(X - \mathbf{1}_n \Delta\boldsymbol{x}_c\bigr)\, R_{zyx}(\gamma,\beta,\alpha)^{\!\top},
\end{equation}
where $\mathbf{1}_n$ is the $n\times 1$ column vector of ones. In this coordinate system, the model surface is the isosurface $f=1$ defined by Equation~\eqref{eq:superellipsoid}.

Each iso-density surface is described by 12 free parameters: the semi-axes $(a, b, c)$, the superellipsoid indices $(S_a, S_b, S_c)$, the center offsets $(\Delta x_c,\Delta y_c,\Delta z_c)$, and the Euler angles $(\alpha,\beta,\gamma)$. At fixed semi-major axis $a$, the intrinsic shape is controlled mainly by the ellipticities ($\varepsilon_{ab}$, $\varepsilon_{bc}$) and the superellipsoid indices ($S_a$, $S_b$, $S_c$). The center is left free to accommodate mild asymmetry (e.g., lopsidedness or tidal interactions); the fitted offsets are typically $\lesssim 0.5\,\mathrm{kpc}$, and the fitted orientations usually vary smoothly between adjacent density levels.

\subsubsection{Objective function and optimization}

For a target density $\rho_s$, the inverse ray profiles provide one 3D point $\boldsymbol{x}_i$ along each ray such that $\rho(\boldsymbol{x}_i)=\rho_s$. After transforming these points from the simulation coordinate system into the principal-axis coordinate system of the fitted superellipsoid using Equation~\eqref{eq:batch-transform}, we denote the transformed positions by $\boldsymbol{x}_i'$ and their Euclidean norms by $r_i'=\| \boldsymbol{x}_i' \|$.

We determine the best-fitting model by minimizing an approximately area-weighted mismatch between the extracted iso-density points and the superellipsoid surface. For numerical stability, we compare each point with the radius of the model surface along the same direction rather than penalizing the implicit-function residual directly. We define the surface-radius ratio
\begin{equation}
D(\boldsymbol{x}') \;=\; \frac{\| \boldsymbol{x}'\|}{\| \boldsymbol{x}'_e\|},
\end{equation}
where $\boldsymbol{x}_e'$ is the intersection of the ray through $\boldsymbol{x}'$ with the model surface, namely the point on the same half-line from the origin that satisfies $f(\boldsymbol{x}_e')=1$. We obtain $\boldsymbol{x}_e'$ by solving a one-dimensional root-finding problem for the radial coordinate along the ray. Panel (d) of Figure~\ref{fig:gal3d_method} illustrates this construction: the red point marks the extracted iso-density point $\boldsymbol{x}'$, and the green point marks $\boldsymbol{x}_e'$, where the same radial direction intersects the trial superellipsoid. By construction, $D=1$ on the model surface, so the fitting residual is $D-1$; points interior to and exterior to the surface have $D<1$ and $D>1$, respectively.

Our fiducial objective function is then
\begin{equation}
\chi^2_D \;=\; \sum_{i=1}^{N} \bigl(r_i'\bigr)^{2}\, \bigl[D(\boldsymbol{x}'_i)-1\bigr]^{2}.
\end{equation}
Because the ray directions are nearly uniformly distributed on the sphere, the factor $(r_i')^2$ provides an approximate equal-area weighting over the fitted surface. The quantity $D-1$ acts as a radial distance-like residual, so minimizing $\chi^2_D$ is analogous to fitting the surface by minimizing offsets between the iso-density points and the model. Panel (e) of Figure~\ref{fig:gal3d_method} shows one representative fitted surface, for which the point-to-model residuals are distributed approximately symmetrically around zero. Appendix~\ref{sec:numerical_caveats} compares this objective with a direct implicit-function objective and shows that the fiducial surface-radius-ratio form yields substantially more stable recovery of the superellipsoid parameters.

We solve the bounded nonlinear least-squares problem with \texttt{lmfit} \citep{newville2025limfit}, using the Trust Region Reflective (TRF) algorithm \citep{branchSubspaceInteriorConjugate1999}. The formal parameter uncertainties are estimated from the covariance matrix of the converged fit. Repeating the fit over a sequence of density levels yields the full radial 3D shape profile, as shown by the ellipticity and shape index profiles in panel (f) of Figure~\ref{fig:gal3d_method}.

\subsection{Model projection and method scope}
\label{sec:model_projection}

To validate the fitted three-dimensional models, we project each ellipsoid or superellipsoid along a chosen line of sight to construct a model surface-density map. For each image pixel, the corresponding line of sight intersects the sequence of fitted iso-density surfaces, defining a piecewise density profile whose integral gives the projected model surface density. This enables a direct comparison with the projected stellar density field reconstructed from the simulation.

Figure~\ref{fig:ellipsoid_vs_superellipsoid} compares projected stellar surface-density maps from ellipsoid and superellipsoid models for a representative simulated galaxy. Both models recover the large-scale extent and overall orientation of the stellar distribution in face-on and edge-on projections, but the superellipsoid model better captures the non-ellipsoidal structure and therefore yields smaller residuals. This illustrates the value of the additional shape parameters for describing intrinsic morphological variation. Additional examples in Section~\ref{sec:examples} further show that the fitted shape indices often depart substantially from unity, indicating structures that cannot be fully described by standard ellipsoids. For particle number $N_{\rm part}\gtrsim 3\times10^5$, the superellipsoid fit is also faster than the iterative shape-tensor method (see Appendix~\ref{sec:runtime_comparison} for a quantitative benchmark).

Nevertheless, a single superellipsoid cannot represent every structure exactly. For example, strongly X-shaped box/peanut bulges are not fully described by this model, although their presence can still be reflected by large values of $S_a$ and $S_c$, corresponding to enhanced boxiness in edge-on projection (see Section~\ref{sec:example_indices}). More generally, if a single superellipsoid is insufficient for a given morphology, the \texttt{Gal3D} framework can be extended by defining alternative shape functions as plug-in models within the same workflow. This paper focuses on overall radial shape profiles rather than multicomponent decomposition, but the same framework can also be applied to individual components identified through, for example, kinematic decomposition \citep{duIdentifyingKinematicStructures2019,duKinematicDecompositionIllustrisTNG2020}.

\section{Simulations and Galaxy Sample}
\label{sec:data}

\begin{figure*}[!ht]
\centering
	{\includegraphics[width=0.246\textwidth]{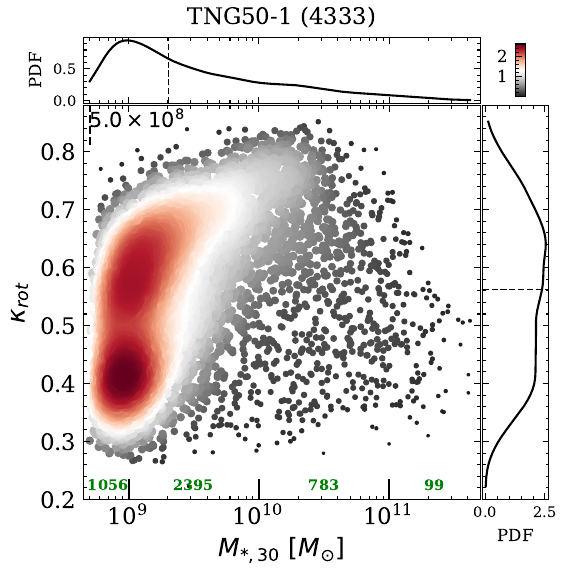}}
 	{\includegraphics[width=0.246\textwidth]{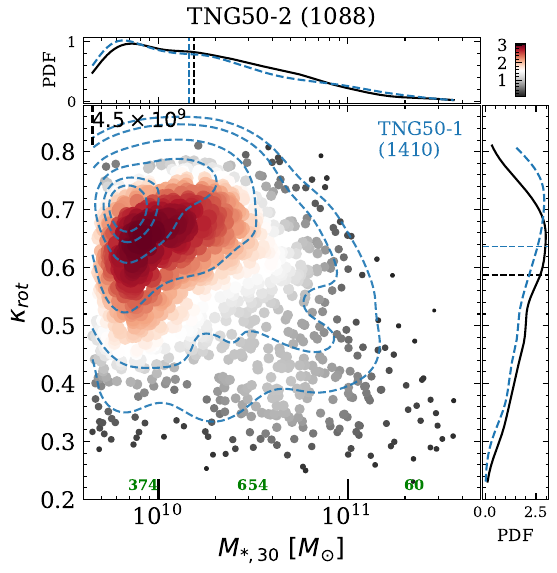}} 
    {\includegraphics[width=0.246\textwidth]{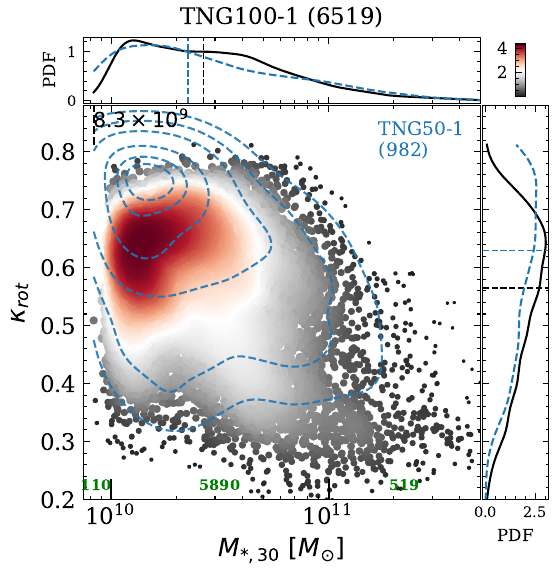}}
    {\includegraphics[width=0.246\textwidth]{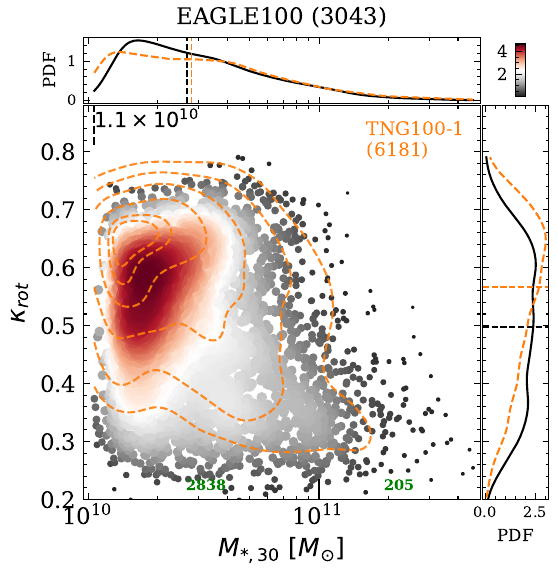}}
    \caption{Distribution of the selected $z=0$ galaxy samples in the $\kappa_{\rm rot}$--$M_{*,30}$ plane. From left to right, the panels show TNG50-1, TNG50-2, TNG100-1, and EAGLE100. Points represent individual galaxies and are coloured by local number density, with the colour bar shown at the upper right of each panel. The x-axis, $M_{*,30}$, is the stellar mass within a 30 kpc spherical aperture, and the y-axis is $\kappa_{\rm rot}$, the fraction of stellar kinetic energy in ordered azimuthal rotation within the same aperture. The solid black curves along the top and right axes show the marginal distributions of $M_{*,30}$ and $\kappa_{\rm rot}$ for the main sample; when comparison samples are overlaid, their distributions are shown as dashed coloured curves. Dashed lines in the corresponding colours mark the medians of each distribution. Vertical dashed lines mark the adopted lower stellar-mass limits, and the green numbers indicate galaxy counts in the corresponding mass bins. Panel titles list the simulation name and the total sample size. Dashed contours show the distribution of a comparison sample from another simulation after restricting it to the same stellar-mass range as the panel sample ; the contour levels enclose 10\%, 25\%, 50\%, 75\%, 90\%, and 95\% of the comparison sample (from inner to outer); the coloured label identifies the comparison sample and gives its size.}
    \label{fig:galaxy_samples}
\end{figure*}

We use cosmological hydrodynamical simulations from the IllustrisTNG and EAGLE projects to study the radial profiles of the intrinsic 3D shapes of galaxies. Here we summarize the simulations and describe how we construct the redshift $z=0$ galaxy samples analyzed in this work.

\subsection{IllustrisTNG simulation}
\label{sec:illustris}
The IllustrisTNG project \citep{marinacciFirstResultsIllustrisTNG2018,naimanFirstResultsIllustrisTNG2018,nelsonFirstResultsIllustrisTNG2018,nelsonIllustrisTNGSimulationsPublic2019,pillepichFirstResultsIllustrisTNG2018,pillepichFirstResultsTNG502019,springelFirstResultsIllustrisTNG2018} is a suite of magnetohydrodynamical simulations of large cosmological volumes carried out with the moving-mesh code AREPO \citep{springelPurSiMuove2010,pakmorMagnetohydrodynamicsUnstructuredMoving2011,pakmorImprovingConvergenceProperties2016}. The subgrid physics model in IllustrisTNG builds on that of its predecessor, Illustris \citep{vogelsbergerModelCosmologicalSimulations2013,vogelsbergerPropertiesGalaxiesReproduced2014,vogelsbergerIntroducingIllustrisProject2014,genelIntroducingIllustrisProject2014,nelsonIllustrisSimulationPublic2015,sijackiIllustrisSimulationEvolving2015}, with substantial revisions to the implementations of supernova feedback, active galactic nucleus (AGN) feedback, and chemical enrichment. 

We analyze three runs from the IllustrisTNG suite: TNG50-1, TNG100-1, and TNG50-2. TNG50-1 is the highest-resolution run in the TNG50 series, following a cosmological volume of $(51.7 \,\mathrm{cMpc})^{3}$ with a baryonic mass resolution of $8.4 \times 10^{4} \,M_{\odot}$ and a dark matter mass resolution of $4.5 \times 10^{5} \,M_{\odot}$. TNG100-1 covers a larger volume of $(110.7 \,\mathrm{cMpc})^{3}$ with a baryonic mass resolution of $1.4 \times 10^{6} \,M_{\odot}$ and a dark matter mass resolution of $7.5 \times 10^{6} \,M_{\odot}$. TNG50-2 shares the same initial conditions as TNG50-1 but at lower resolution, with a baryonic mass resolution of $6.8 \times 10^{5} \,M_{\odot}$ and a dark matter mass resolution of $3.7 \times 10^{6} \,M_{\odot}$. The gravitational softening lengths for stellar particles at $z=0$ are 288 pc, 740 pc, and 576 pc for TNG50-1, TNG100-1, and TNG50-2, respectively.

\subsection{EAGLE simulation}
\label{sec:eagle}
The EAGLE project is a suite of cosmological simulations performed with a modified version of the smoothed-particle hydrodynamics (SPH) code GADGET-3, based on GADGET-2 \citep{springelCosmologicalSimulationCode2005}. The subgrid physics models implemented in EAGLE include radiative cooling, star formation, stellar mass loss and metal enrichment, energy feedback from star formation, gas accretion onto and mergers of supermassive black holes (BHs), and AGN feedback \citep{schayeEAGLEProjectSimulating2015,crainEAGLESimulationsGalaxy2015}.

We use one EAGLE reference run, Ref-L0100N1504, hereafter EAGLE100. EAGLE100 evolves a volume of $(100 \,\mathrm{cMpc})^{3}$ with a baryonic mass resolution of $1.81 \times 10^{6} \,M_{\odot}$ and a dark matter mass resolution of $9.70 \times 10^{6} \,M_{\odot}$. The gravitational softening length for stellar particles at $z=0$ is 700 pc.

\subsection{Galaxy samples}

We select galaxy samples from these simulations at redshift $z=0$. To mitigate resolution effects and ensure robust 3D shape measurements, we require each galaxy to contain at least $10^{4}$ stellar particles. Because stellar particle masses differ across runs, this criterion corresponds to different lower stellar-mass limits in different simulations. We also impose a common upper limit of $M_{*,30}<5\times10^{11}\,M_\odot$ for all simulations, which removes only a handful of the most massive galaxies.

For each galaxy, we first (i) recenter the system using the iterative shrinking-sphere method \citep{powerInnerStructureLCDM2003}; (ii) compute the 3D stellar half-mass radius, $r_{\rm e}$, defined as the radius enclosing half of the total stellar mass within $30\,\mathrm{kpc}$; (iii) shift velocities so that the mass-weighted mean stellar velocity within $0.5\,r_{\rm e}$ is zero; and (iv) rotate the coordinate system so that the $z$-axis aligns with the total stellar angular momentum within $2\,r_{\rm e}$.

We adopt $M_{*,30}$ as our stellar mass measure, defined as the total stellar mass within a 30 kpc spherical aperture. We quantify rotational support with $\kappa_{\rm rot}$, defined as the fraction of stellar kinetic energy in ordered azimuthal rotation within the same aperture \citep{salesFeedbackStructureSimulated2010}:
\begin{equation}
    \kappa_{\rm rot} \;=\; \frac{\sum_{i} m_i\, v_{\phi,i}^{2}}{\sum_{i} m_i\, \| \boldsymbol{v}_i\|^{2}},
\end{equation}
where $m_i$, $v_{\phi,i}$, and $\boldsymbol{v}_i$ correspond, respectively, to the mass, azimuthal velocity in the cylindrical coordinate system, and 3D velocity vector of each stellar particle.

Figure~\ref{fig:galaxy_samples} shows the distribution of the selected galaxy samples in the $\kappa_{\rm rot}$--$M_{*,30}$ plane. The final samples contain 4333, 1088, and 6519 galaxies for TNG50-1, TNG50-2, and TNG100-1, respectively, and 3043 galaxies for EAGLE100.\footnote{EAGLE100 has a volume and resolution similar to those of TNG100-1 but yields a smaller galaxy sample, likely because differences in the adopted subgrid physics lead to different galaxy stellar mass functions in the two simulations \citep{schayeEAGLEProjectSimulating2015,pillepichFirstResultsIllustrisTNG2018}; see also Figure 4 of \citet{crainHydrodynamicalSimulationsGalaxy2023}.} 
The marginal distributions and dashed comparison contours in Figure~\ref{fig:galaxy_samples} show that offsets in $\kappa_{\rm rot}$ persist at fixed $M_{*,30}$ after matching the stellar-mass range. Comparing TNG50-1 with TNG50-2 indicates modestly stronger rotational support, with a median $\kappa_{\rm rot}$ higher by $\approx 0.05$ at higher TNG resolution. The orange contours in the EAGLE100 panel show that TNG galaxies also have systematically higher $\kappa_{\rm rot}$ than their EAGLE counterparts by approximately 0.06 in the median (dashed lines in Figure~\ref{fig:galaxy_samples}). Across all runs, however, the overall dependence of $\kappa_{\rm rot}$ on stellar mass remains qualitatively similar.

\section{Galaxy examples: diverse structures revealed by \GalThreeD}
\label{sec:examples}

\begin{figure*}[ht!]
\centering
{\includegraphics[width=\textwidth]{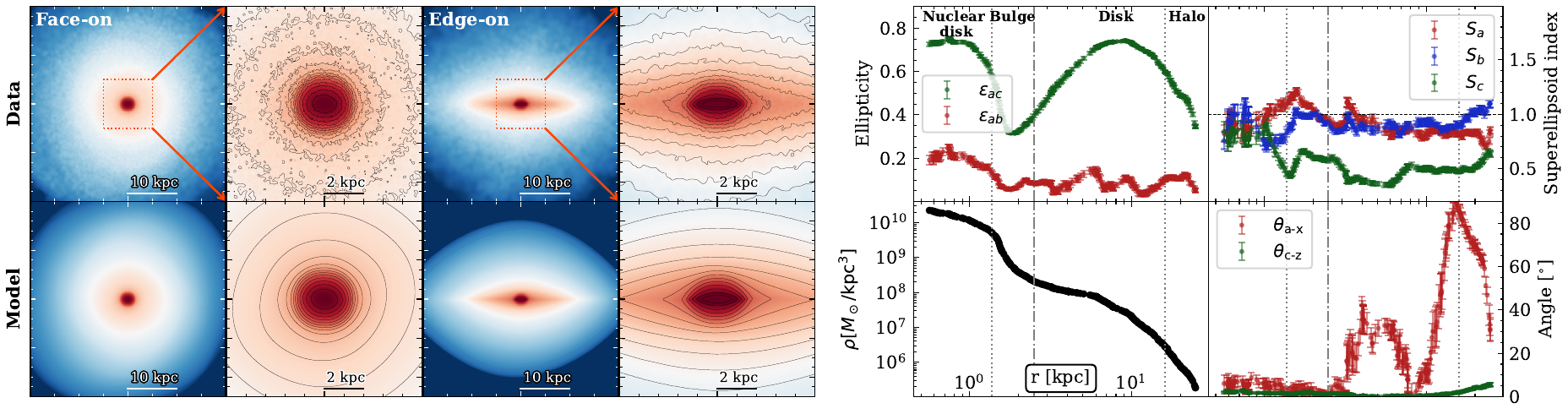}}
{\includegraphics[width=\textwidth]{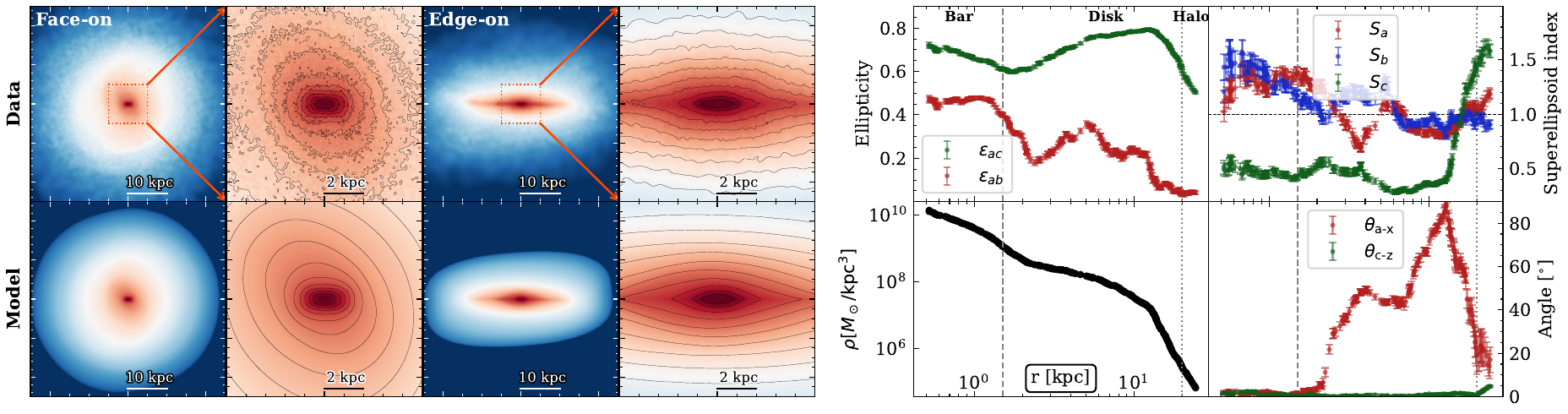}}
{\includegraphics[width=\textwidth]{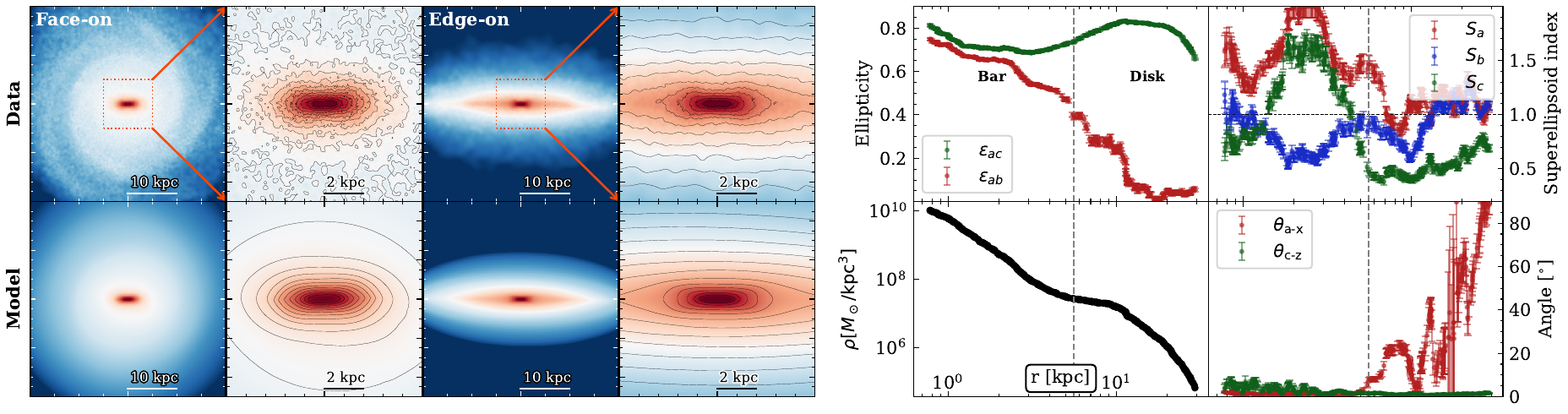}}
\caption{Radial 3D shape profiles for three representative disk galaxies from TNG50-1. From top to bottom, the panel groups show an unbarred galaxy (SubfindID 264885), a barred galaxy (SubfindID 485056) with a face-on boxy but vertically thin bar, and a barred galaxy (SubfindID 63869) with a prominent box/peanut-shaped bulge. In each group, the left panels show face-on and edge-on projections of the stellar density field. For each orientation, a wide-field view is shown together with a zoomed-in isodensity-contour view of the central region enclosed by the dashed orange box. Rows labeled ``Data'' show projected stellar density maps reconstructed with adaptive kernel density estimation, while rows labeled ``Model'' show the corresponding best-fit superellipsoidal models. The right panels show radial profiles as functions of radial distance $r$: ellipticities $\varepsilon_{ac}\equiv 1-c/a$ and $\varepsilon_{ab}\equiv 1-b/a$ (upper left), superellipsoid indices $S_a$, $S_b$, and $S_c$ (upper right), stellar density $\rho$ (lower left), and orientation angles $\theta_{a\text{-}x}$ and $\theta_{c\text{-}z}$ (lower right), defined as the angles between the major axis and the $x$-axis and between the minor axis and the $z$-axis, respectively. In the radial profiles, vertical dash-dotted, dashed, and dotted lines mark the bulge ($\varepsilon_{ac}=0.4$), bar ($\varepsilon_{ab}=0.4$), and disk ($\varepsilon_{ac}=0.6$) outer edges, respectively. Together, these examples illustrate how the superellipsoid parameters distinguish nuclear disks, bulges, main disks, bars, and box/peanut bulges, as well as higher-order structural differences among these components.}
\label{fig:disk_and_barred_galaxies}
\end{figure*}

\begin{figure*}[ht!]
\centering
{\includegraphics[width=\textwidth]{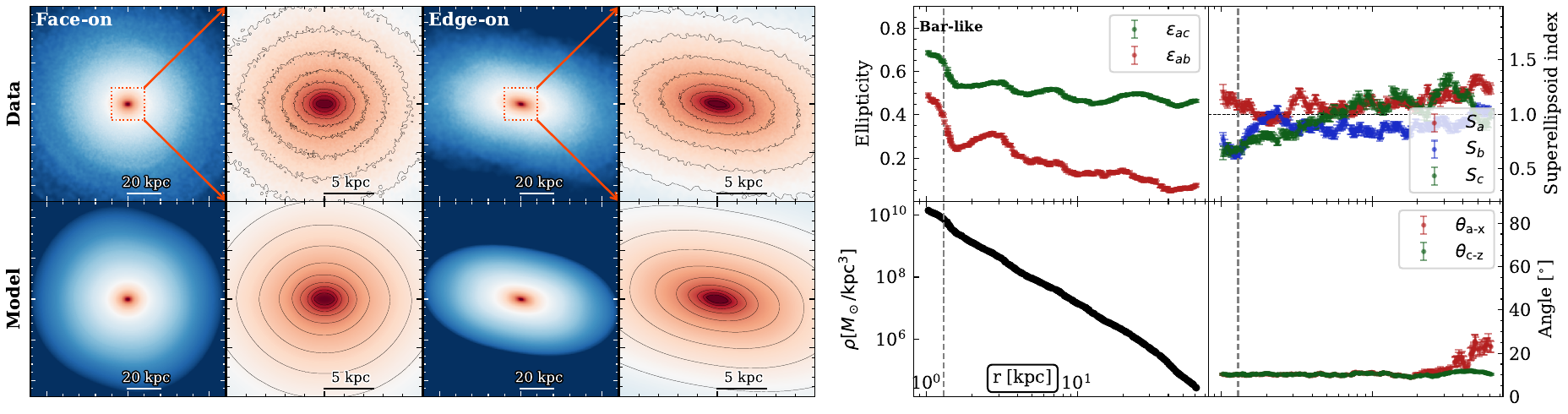}}
{\includegraphics[width=\textwidth]{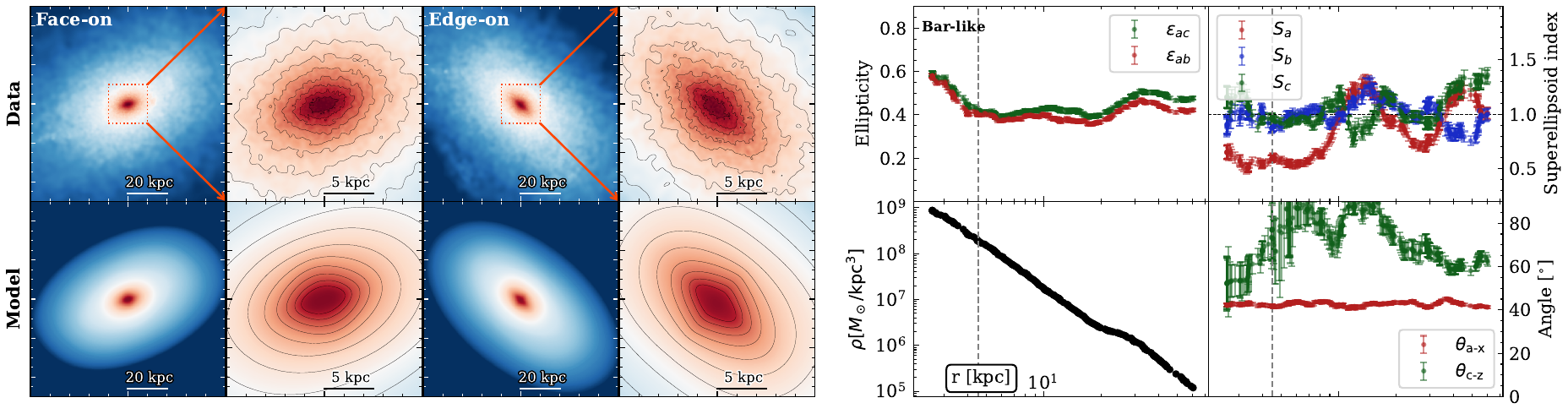}}
{\includegraphics[width=\textwidth]{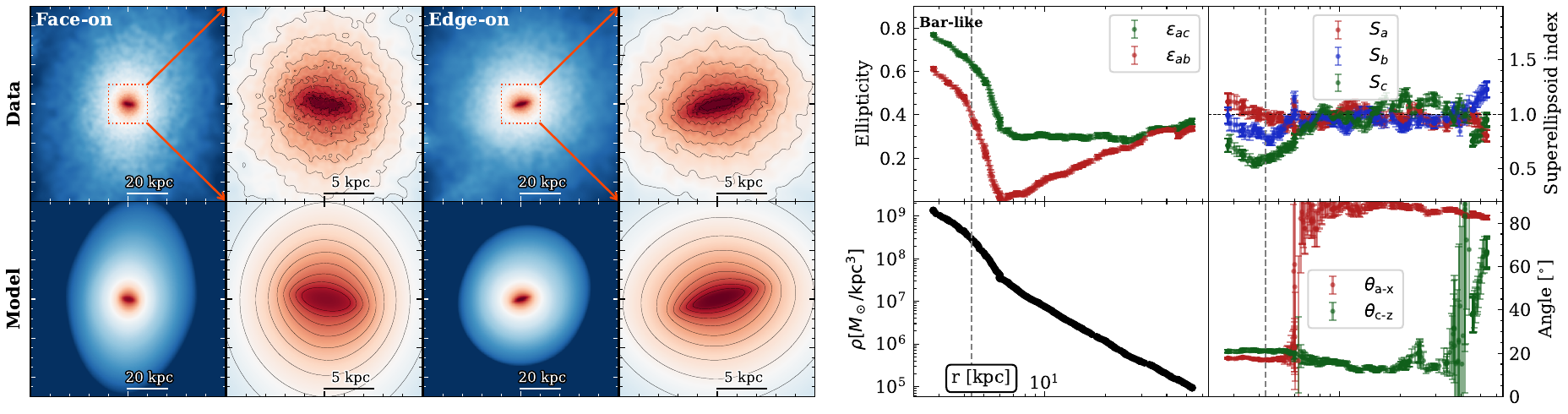}}
\caption{Similar to Figure~\ref{fig:disk_and_barred_galaxies}, but for three elliptical galaxies. The top galaxy (SubfindID 242789) is drawn from TNG50-1, while the middle and bottom galaxies are the central galaxies of HaloID 24 and HaloID 35, drawn from EAGLE100. These examples illustrate oblate, prolate, and radially varying spheroidal structures, respectively. Their main differences are encoded primarily in the ellipticity and orientation profiles, although in some radial ranges the shape indices also depart substantially from the standard ellipsoidal case. Vertical dashed lines in the radial profiles mark the outer edge of 
bar-like structures ($\varepsilon_{ab}=0.4$).}
\label{fig:elliptical_galaxy}
\end{figure*}

Radial superellipsoid profiles provide a 3D description of galaxy structure. The examples in Figures~\ref{fig:disk_and_barred_galaxies} and~\ref{fig:elliptical_galaxy} show how the fitted parameters map onto familiar stellar components, including nuclear disks, bulges, main disks, bars, box/peanut bulges, and spheroidal envelopes. In this framework, the ellipticities $\varepsilon_{ac}$ and $\varepsilon_{ab}$ describe the first-order intrinsic geometry, the orientation angles $\theta_{a\text{-}x}$ and $\theta_{c\text{-}z}$ trace the radial variation of the principal axes, and the shape indices $S_a$, $S_b$, and $S_c$ quantify higher-order departures from a standard ellipsoid. These two orientation angles are derived from the fitted Euler angles: $\theta_{a\text{-}x}$ is defined as the angle between the fitted major axis and the galaxy $x$-axis, while $\theta_{c\text{-}z}$ is defined as the angle between the fitted minor axis and the galaxy $z$-axis. As described in Section~\ref{sec:data}, the $z$-axis is aligned with the stellar angular momentum within $2\,r_{\rm e}$, and the $x$-axis is an arbitrary direction in the perpendicular plane. Figure~\ref{fig:superellipsoid_shape} illustrates the idealized behavior of the ellipticities and shape indices, and the examples show how these parameters appear in simulated galaxies.

\subsection{Disks, bars, and spheroids from ellipticity profiles}

Galactic structures are generally triaxial and can be characterized by principal semi-axes, $a \ge b \ge c$. The ellipticity $\varepsilon_{ac}$ measures flattening: for $\varepsilon_{ab}\sim 0$, a larger $\varepsilon_{ac}$ corresponds to a flatter, more disky geometry. In the disk galaxy examples shown in Figure~\ref{fig:disk_and_barred_galaxies}, disk-dominated regions maintain high flattening, with $\varepsilon_{ac}\gtrsim 0.6$, whereas pressure-supported components such as bulges, stellar halos, or the elliptical galaxies in Figure~\ref{fig:elliptical_galaxy} typically have $\varepsilon_{ac}\lesssim 0.5$. The disk regions in all three examples extend to $\gtrsim 15$ kpc, and their approximate outer boundaries are marked by vertical dotted lines in the radial-profile panels. Notably, as seen in the first example (top panels), a nuclear disk is embedded within a more spheroidal central component, likely a bulge. Compared with the two examples without nuclear disks, the nuclear disk significantly raises the central density, as shown in the $\rho$ profile panel.

Bars are identified primarily through enhanced elongation, quantified by $\varepsilon_{ab}$. Compared with the unbarred galaxy in the top panels of Figure~\ref{fig:disk_and_barred_galaxies}, the two barred galaxies show pronounced central increases in $\varepsilon_{ab}$. The approximate edge of each bar is marked by the threshold $\varepsilon_{ab}=0.4$ (vertical dashed lines), a simple criterion adopted as a visual guide. The middle-panel disk galaxy in Figure~\ref{fig:disk_and_barred_galaxies} exhibits double-barred morphology, evidenced by dual peaks in $\varepsilon_{ab}$ with different position angles $\theta_{a\text{-}x}$. Notably, the same diagnostic also identifies bar-like structures in elliptical galaxies. All three elliptical examples in Figure~\ref{fig:elliptical_galaxy} exhibit elevated $\varepsilon_{ab}$ in the inner few kiloparsecs. Such bar-like structures in elliptical galaxies have been reported in recent works \citep[e.g.,][]{naabATLAS3DProjectXXV2014,schulzeKinematicsSimulatedGalaxies2018,pulsoniStellarHalosETGs2020,lokasBarlikeGalaxiesIllustrisTNG2021,duRevisitingExcessBarlike2026}, and \texttt{Gal3D} is able to characterize them accurately in three dimensions.

Although the elliptical examples in Figure~\ref{fig:elliptical_galaxy} have comparable central elongations, their outer structures differ substantially. The first example (top panels) is predominantly oblate, with moderate flattening ($\varepsilon_{ac}\sim0.5$) and weak elongation ($\varepsilon_{ab}\lesssim0.2$) over a broad radial range. The second (middle panels) is largely prolate, with $\varepsilon_{ab}\approx\varepsilon_{ac}\approx0.5$ across much of the profile. The third (bottom panels) exhibits stronger radial variation: it is highly elongated inside $r\lesssim 7$ kpc, becomes rounder and more oblate near $r\sim 8$ kpc, and returns to a more prolate or triaxial configuration at larger radii, where $\varepsilon_{ab}\approx\varepsilon_{ac}\approx 0.3$.

\subsection{Warps and twists from orientation profiles}

The orientation angles provide an additional way to diagnose radial changes in galaxy structure. Disk warps, which are common in the outskirts of nearby edge-on galaxies and often have an $S$-shaped morphology \citep[e.g.,][]{sanchez-saavedraFrequencyWarpedSpiral1990,annWarpedDisksSpiral2006}, can be traced by $\theta_{c\text{-}z}$. This angle is derived from the fitted Euler angles and measures the misalignment between the minor axis of the fitted surface and the galaxy $z$-axis, and therefore traces changes in the disk plane. In the first and second disk examples in Figure~\ref{fig:disk_and_barred_galaxies}, $\theta_{c\text{-}z}$ remains nearly constant across the main disk but changes by about $5^{\circ}$ at larger radii, indicating a weak outer warp. Such warps may be driven by misaligned gas accretion, tidal interactions, or internal dynamical processes \citep[e.g.,][]{debattistaWarpedGalaxiesMisaligned1999,shenGalacticWarpsInduced2006,roskarMisalignedAngularMomentum2010,semczukTidallyInducedWarps2020,hanTiltedDarkHalos2023,jonssonTangledWarpMilky2024}. The superellipsoid framework thus provides a quantitative measure of intrinsic disk warping in three dimensions.

Radial twists of the major axis are traced by $\theta_{a\text{-}x}$, the angle between the fitted major axis and the galaxy $x$-axis. Such twists are common in double-barred galaxies \citep{erwinDoublebarredGalaxiesCatalog2004, bittnerGalaxiesGalaxiesTIMER2021}, and are also observed in some elliptical galaxies \citep{nietoOriginInnerIsophotal1992}. In the middle disk example of Figure~\ref{fig:disk_and_barred_galaxies}, the two $\varepsilon_{ab}$ peaks are accompanied by a $\sim50^{\circ}$ change in $\theta_{a\text{-}x}$, indicating misaligned inner and outer bars. In elliptical galaxies, different prolate or bar-like components can likewise coexist at different orientations. For example, the bottom elliptical example of Figure~\ref{fig:elliptical_galaxy} shows a twist in $\theta_{a\text{-}x}$ near $r\sim6$--$7$ kpc, indicating that the inner elongated structure is oriented differently from the outer body. Unlike projected isophotal twists, these measurements quantify intrinsic 3D twists and can therefore help distinguish projection effects from genuine structural misalignments.

\subsection{Boxiness and diskiness from shape indices}
\label{sec:example_indices}

The shape indices $S_a$, $S_b$, and $S_c$ describe higher-order deviations from ellipsoidal geometry and therefore capture structural information that is not contained in the axis ratios alone. Because galactic structures are supported by different stellar orbit families \citep{binneyGalacticDynamicsSecond2008}, their shapes can deviate from simple ellipsoids in ways that encode boxiness or diskiness.

For example, thinner disk structures generally have cuspy isodensity contours in edge-on views and are expected to have $S_c < 1$. In the disk galaxy examples shown in Figure~\ref{fig:disk_and_barred_galaxies}, the main disks have $S_c\sim 0.5$, consistent with this expectation. In comparison, spheroidal structures are typically characterized by $S_c \sim 1$, as seen in the elliptical examples in Figure~\ref{fig:elliptical_galaxy}.

Bars are complex 3D structures composed of different stellar orbit families \citep{contopoulosOrbitsBarredGalaxies1989,athanassoulaMorphologyBarOrbits1992,sellwoodDynamicsBarredGalaxies1993,valluriUnifiedFrameworkOrbital2016}. The two barred galaxies in Figure~\ref{fig:disk_and_barred_galaxies} both show central $\varepsilon_{ab}$ enhancements, but their shape index profiles separate their higher-order structure. In the middle example, the inner bar is boxy in face-on projection but remains vertically disky in edge-on projection. This morphology is reflected by $S_a>1$ and $S_b>1$, together with $S_c<1$ across the bar region. The bottom galaxy, previously identified as hosting an X-shaped box/peanut bulge by \citet{andersonInterplayAccretionGalaxy2023}, shows a different signature: $S_a>1$, $S_c>1$, and $S_b<1$. This indicates a bar that is boxy along the major axis and vertically thickened, but more tapered along the intermediate axis. The increase of $S_c$ from $\sim0.5$ near $r\sim1$ kpc to $\sim1.5$ by $r\sim2$ kpc marks the emergence of vertical boxiness, consistent with a box/peanut-shaped structure produced by bar buckling or related vertical heating processes \citep[e.g.,][]{combesFormationPropertiesPersisting1981,rahaDynamicalInstabilityBars1991,shenOurMilkyWay2010,erwinCaughtActDirect2016}.

The superellipsoid indices also remain informative outside barred regions. In the outskirts of the unbarred galaxy and the middle barred galaxy, the superellipsoid index along the minor axis, $S_c$, follows different trends: it remains $S_c\sim 0.5$ outside the main disk in the former, but rises sharply (from $S_c \lesssim 0.5$ to $\gtrsim 1.6$) beyond $r\sim15$ kpc in the latter. Thus, stellar envelopes with similar first-order flattening can differ substantially in higher-order morphology. The superellipsoid indices provide an efficient discriminator between boxy and disky isodensity structures that are degenerate in axis-ratio profiles.

\begin{figure*}[ht!]
\plotone{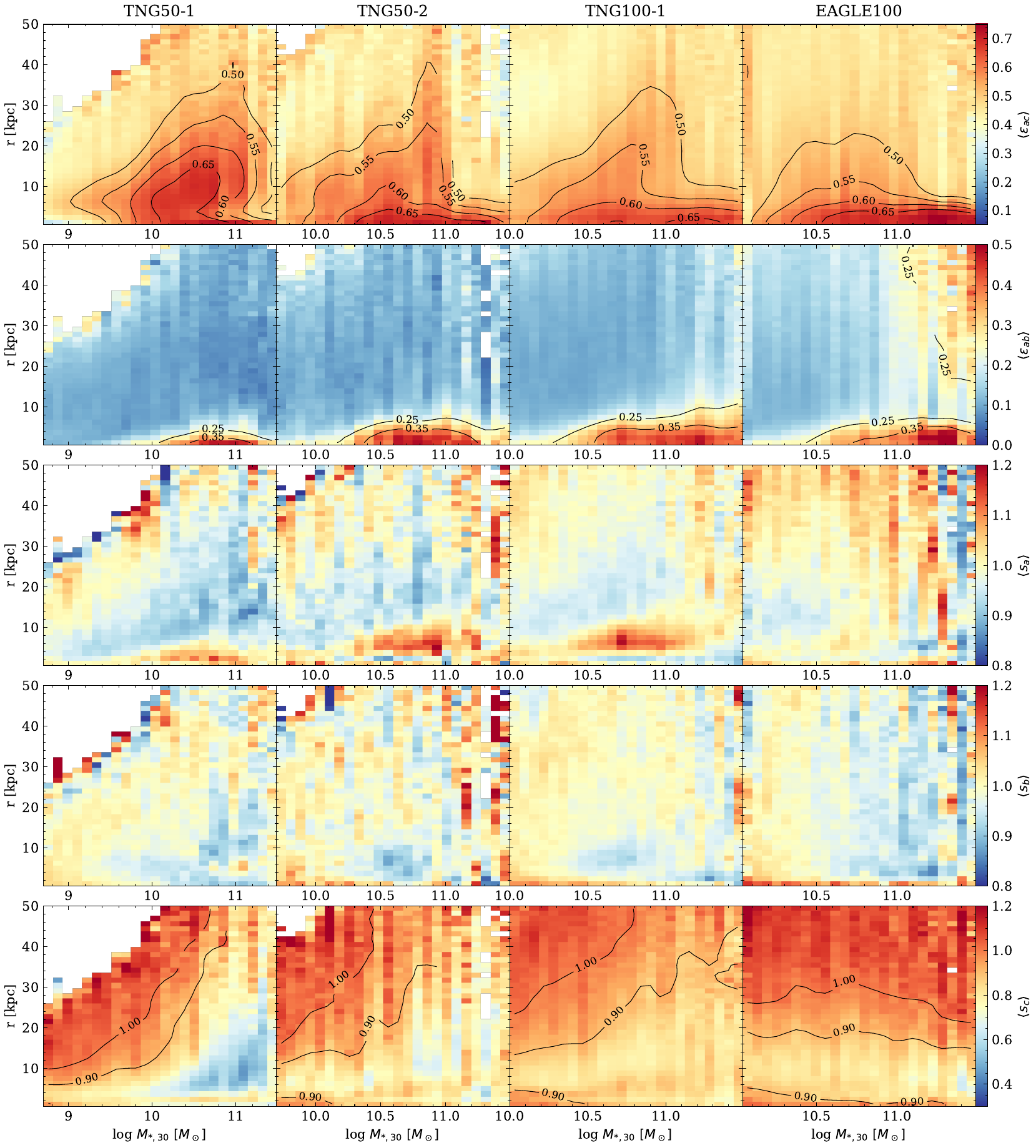}
    \caption{Two-dimensional maps of mean radial 3D shape profiles as functions of stellar mass $\log M_{*,30}$ (x-axis) and radius $r$ (y-axis) for the four simulation samples. Each column corresponds to one simulation, as labeled at the top. The five rows show, from top to bottom, the mean values of $\langle\varepsilon_{ac}\rangle$, $\langle\varepsilon_{ab}\rangle$, $\langle S_a\rangle$, $\langle S_b\rangle$, and $\langle S_c\rangle$. Blue (red) colors indicate lower (higher) values of each parameter. Contours mark constant levels as labeled.
    \label{fig:radial_shape_profile}}
\end{figure*}

\section{Comparison of Radial 3D Galaxy Structure across Cosmological Simulations}
\label{sec:statistics}
%  Systematic differences in radial 3D galaxy structure across simulations

This section compares the mean radial 3D shape profiles across the four simulation samples. For each galaxy, we average each shape parameter in radial bins, then group galaxies by stellar mass ($M_{*,30}$) and compute the mean profile at fixed radius within each mass bin. The resulting maps, shown in Figure~\ref{fig:radial_shape_profile}, therefore represent galaxy-averaged profiles in the $(M_{*,30}, r)$ plane; bins containing fewer than five galaxies are excluded. Although this averaging smooths over the full diversity of galaxy sizes and morphologies, it is intended to highlight systematic differences among the simulation suites. We use physical radius rather than $r/r_e$ to avoid biases from simulation-dependent differences in $r_e$, and we have verified that adopting $r/r_e$ does not change the qualitative trends discussed below. We therefore present the profiles in physical radius, which allows a more direct comparison of absolute structural scales across simulations.

Figure~\ref{fig:radial_shape_profile} reveals coherent trends in the stellar-mass-binned mean profiles, particularly signatures of flattened disk structure, bars, and elongated or triaxial structure in massive spheroids. The $\langle S_b\rangle$ map is included for completeness but shows no comparably clear global trend, so we do not discuss it in detail. We first discuss the flattened regions identified primarily by $\langle\varepsilon_{ac}\rangle$ and further characterized by $\langle S_c\rangle$, then discuss bars and box/peanut-shaped structures traced by $\langle\varepsilon_{ab}\rangle$ and $\langle S_a\rangle$, and finally the elongated or triaxial structures in the most massive galaxies.

\subsection{Disk structures: trends with resolution and simulation model}

The radial extent of flattened, disk structure increases with stellar mass up to $\log (M_{*,30}/M_\odot)\sim 11$ in all four simulations, but is strongly suppressed in galaxies above this mass. In Figure~\ref{fig:radial_shape_profile}, the top row shows the outward increase of the high-$\langle\varepsilon_{ac}\rangle$ ($>0.5$) region, accompanied by a corresponding locus of low $\langle S_c\rangle$ ($<1$) in the bottom row. Because high $\langle\varepsilon_{ac}\rangle$ and low $\langle S_c\rangle$ jointly indicate flattened and vertically thin structure, these maps show that disks become more radially extended over most of the sampled mass range, but less prominent in the most massive galaxies. The decline above $\sim 10^{11}\,M_\odot$ is plausibly related to the increasing importance of mergers, as suggested by the large ex-situ stellar fractions reported in this mass regime \citep{rodriguez-gomezRoleMergersHalo2017,tacchellaMorphologyStarFormation2019,davisonEAGLEsViewEx2020,angeloudiConstraintsSituEx2024}.

Across the TNG suite, higher resolution yields a thinner and more radially extended disk structure. At fixed stellar mass and radius, TNG50-1 systematically exhibits higher $\langle\varepsilon_{ac}\rangle$ and lower $\langle S_c\rangle$ than TNG50-2, with TNG100-1 showing broadly similar behavior over the overlapping mass range. This consistent ordering shows that, within TNG, resolution improvements promote greater flattening, reduced vertical thickness, and stronger vertically disky morphology.

At fixed stellar mass around $\log (M_{*,30}/M_\odot)\sim 10.5$--$11$, EAGLE100 exhibits less radially extended disks than the TNG runs. In Figure~\ref{fig:radial_shape_profile}, the TNG simulations show a sustained outward increase of the high-$\langle\varepsilon_{ac}\rangle$ region up to $\sim 10^{11}\,M_\odot$, whereas EAGLE100 shows a saturation near $\sim 10^{10.5}\,M_\odot$. Consistently, EAGLE100 also exhibits a mild reduction in the radial extent of low-$\langle S_c\rangle$ regions over the same mass range. Because EAGLE and TNG differ in both hydrodynamic solvers and subgrid models \citep{schayeEAGLEProjectSimulating2015,pillepichFirstResultsIllustrisTNG2018}, and simulated galaxy properties are sensitive to these choices \citep[e.g.,][]{scannapiecoAquilaComparisonProject2012,valentiniEffectGalacticOutflows2017,roca-fabregaAGORAHighresolutionGalaxy2021,grothCosmologicalSimulationCode2023}, these differences may contribute to the less radially extended and less vertically disky flattened structures in EAGLE100.

\subsection{Divergent bar and box/peanut bulge signatures in TNG and EAGLE}

For galaxies below the high-mass spheroid regime, $\log(M_{*,30}/M_\odot)\lesssim11.1$, most systems have disk structure. This is evident from the high-$\langle\varepsilon_{ac}\rangle$ (flattened disk) regions in Figure~\ref{fig:radial_shape_profile} and from Figure~\ref{fig:galaxy_samples}, where most galaxies in this mass range have $\kappa_{\rm rot} > 0.5$, a threshold commonly adopted to select rotation-supported disk galaxies in simulations \citep[e.g.,][]{zhaoBarredGalaxiesIllustrisTNG2020,luIllustrisTNGInsightsFactors2025}. In this mass range, the central high-$\langle\varepsilon_{ab}\rangle$ region in the second row of Figure~\ref{fig:radial_shape_profile} is therefore dominated mainly by stellar bars in disk galaxies. This bar signal becomes increasingly prominent above $\log (M_{*,30}/M_\odot)\sim10.5$ in all four simulations, consistent with previous studies showing that the bar fraction increases with stellar mass in both EAGLE and IllustrisTNG \citep{algorryBarredGalaxiesEAGLE2017,rosas-guevaraBuildupStronglyBarred2020,zhaoBarredGalaxiesIllustrisTNG2020,luIllustrisTNGInsightsFactors2025}. At lower masses, TNG100-1 and EAGLE100 show similarly weak mean $\langle\varepsilon_{ab}\rangle$ signals, consistent with the low bar fractions reported in this regime, possibly because bars are less well resolved at lower stellar masses \citep{algorryBarredGalaxiesEAGLE2017,zhaoBarredGalaxiesIllustrisTNG2020}.

At $\log (M_{*,30}/M_\odot)\sim10.5$--$11.1$, the bar-related $\langle\varepsilon_{ab}\rangle$ signal is weaker in EAGLE100 than in the TNG runs. This agrees with previous studies showing that EAGLE has a lower bar fraction than IllustrisTNG and that a substantial fraction of EAGLE bars are relatively weak \citep{algorryBarredGalaxiesEAGLE2017,zhaoBarredGalaxiesIllustrisTNG2020}. 

Within the TNG suite, the high-$\langle\varepsilon_{ab}\rangle$ region extends farthest in TNG100-1, is intermediate in TNG50-2, and is most compact in TNG50-1. The compact radial extent in TNG50-1 is consistent with the shorter bars reported in TNG50-1 galaxies by \citet{luIllustrisTNGInsightsFactors2025}. Gravitational softening can influence bar length and strength \citep{bauerCanStellarDiscs2019}, but the resolution dependence is likely complex because it may also reflect resolution-dependent subgrid behavior.

Box/peanut-shaped bulges appear to be more common in the TNG samples than in EAGLE100. In Figure~\ref{fig:radial_shape_profile}, the outer part of the bar-dominated region in the TNG runs, identified by high $\langle\varepsilon_{ab}\rangle$, is accompanied by elevated $\langle S_a\rangle$ and a weaker but co-spatial increase in $\langle S_c\rangle$. This pattern matches the bottom barred examples in Figure~\ref{fig:disk_and_barred_galaxies}, where the emergence of a box/peanut-shaped structure is marked by increases in both $S_a$ and $S_c$. The $\langle S_c\rangle$ enhancement is weaker than the $\langle S_a\rangle$ enhancement in the population maps, likely because the vertical boxiness signal is diluted by unbarred galaxies and bars without strong vertical thickening, which typically have $S_c\sim0.5$. By contrast, EAGLE100 shows weaker or no enhancements in either $\langle S_a\rangle$ or $\langle S_c\rangle$ near the outer part of the bar region, suggesting less prominent box/peanut-shaped morphology. The TNG--EAGLE difference is therefore visible both in $\langle\varepsilon_{ab}\rangle$ and in the higher-order structure of the barred region.

\subsection{Massive elliptical galaxies: central bar-like structure and outer triaxiality}
% Massive galaxies: central bar-like structure and outer triaxiality
% Central bar-like structure and outer triaxiality in massive elliptical galaxies

Massive galaxies with $\log(M_{*,30}/M_\odot)>11.1$ are generally elliptical systems, as shown by the examples in Figure~\ref{fig:elliptical_galaxy}. In this high-mass regime, the flattened disk (high-$\langle\varepsilon_{ac}\rangle$) region in Figure~\ref{fig:radial_shape_profile} shrinks dramatically, suggesting a strong influence of mergers. Figure~\ref{fig:galaxy_samples} also shows that these high-mass bins are dominated by slowly rotating systems with $\kappa_{\rm rot}<0.5$. Despite this, both the TNG and EAGLE100 simulations retain a central feature reminiscent of an elongated bar-like structure, characterized by elevated $\langle\varepsilon_{ab}\rangle$ and $\langle\varepsilon_{ac}\rangle$. This interpretation is further supported by Figure~\ref{fig:elliptical_galaxy}, where all three galaxies show centrally elevated $\varepsilon_{ab}$ without an extended surrounding disk, a morphology similar to previously reported bar-like structures \citep[e.g.,][]{naabATLAS3DProjectXXV2014,schulzeKinematicsSimulatedGalaxies2018,pulsoniStellarHalosETGs2020,lokasBarlikeGalaxiesIllustrisTNG2021,duRevisitingExcessBarlike2026}. The origin of these elongated or triaxial structures is beyond the scope of this paper and will be investigated in future work.

The outer regions of massive galaxies are more triaxial in EAGLE100 than in the TNG simulations. In the EAGLE100 high-mass region ($\log(M_{*,30}/M_\odot)>11.1$) of Figure~\ref{fig:radial_shape_profile}, the second row shows that $\langle\varepsilon_{ab}\rangle$ is high in the center ($r\lesssim10$ kpc, $\langle\varepsilon_{ab}\rangle\gtrsim0.4$), decreases at intermediate radii ($r\sim10$--$20$ kpc, $\langle\varepsilon_{ab}\rangle\lesssim0.2$), and rises again in the outskirts ($r\gtrsim20$ kpc). The outer rise reaches $\langle\varepsilon_{ab}\rangle \gtrsim 0.3$ in EAGLE100, whereas the TNG runs typically remain below $\sim0.2$, indicating that the outer stellar bodies of massive EAGLE ellipticals are, on average, more prolate or triaxial than their TNG counterparts. A galaxy with a similar radial behavior is shown by the bottom elliptical example in Figure~\ref{fig:elliptical_galaxy}. The inner and outer elongation signals are spatially separated in the population average, which may indicate different physical origins. Because the envelopes or stellar halos of elliptical galaxies are largely generated by mergers, especially major ones \citep[e.g.,][]{moodySimulatingMultipleMerger2014,liOriginPropertiesMassive2018,duEvolutionaryPathwaysDisk2021,lagosDiverseNatureFormation2022}, this difference likely reflects differences in their assembly histories or in the properties of their progenitor galaxies in the two simulation suites.

\section{Summary}
\label{sec:summary}

We have developed \texttt{Gal3D}, an open-source Python framework for measuring radial 3D shape profiles from simulated galaxy particle data. The framework reconstructs a continuous density field using adaptive kernel density estimation and characterizes radial structure with two complementary components: an iterative density-field shape tensor and a ray-based superellipsoid fit to iso-density surfaces. The superellipsoid formalism recovers axis ratios, orientations, center offsets, and higher-order shape indices that capture non-ellipsoidal morphology. The fitted models can also be projected for direct comparison with simulated surface-density maps. Applied to representative galaxies, \texttt{Gal3D} measures the radial shape profiles that distinguish nuclear disks, bulges, main disks, bars, box/peanut bulges, stellar halos, and early-type envelopes, while also quantifying higher-order features such as boxiness and diskiness.

The main results from comparing TNG50-1, TNG50-2, TNG100-1, and EAGLE100 are summarized in two stellar-mass regimes:

At stellar masses ($10.0\lesssim\log(M_{*,30}/M_\odot)\lesssim11.0$):
\begin{itemize}
\setlength{\itemsep}{0pt}
\setlength{\parskip}{0pt}
\setlength{\parsep}{0pt}
\item The radial extent of flattened disk regions increases with stellar mass up to $\sim10^{11}\,M_\odot$. Over this mass range, EAGLE galaxies have less radially extended and less vertically thinner disk structures than TNG galaxies and show a saturation of disk extent at $\log(M_{*,30}/M_\odot)\sim10.5$, while the TNG runs continue to increase up to $\sim10^{11}\,M_\odot$. Higher TNG resolution is associated with thinner disk structures.
\item The bar-related $\langle\varepsilon_{ab}\rangle$ signal strengthens above $\log(M_{*,30}/M_\odot)\sim10.5$ in all four simulations, but remains systematically weaker in EAGLE100 than in the TNG runs. Within the TNG suite, higher resolution is associated with a less radially extended bar-related region.
\item Box/peanut-shaped bulges are more prominent in TNG than in EAGLE100.
\end{itemize}

Across the transition interval $11.0\lesssim\log(M_{*,30}/M_\odot)\lesssim11.1$, the radial extent of flattened disk regions declines sharply, from tens of kiloparsecs to less than $\sim10$ kpc.

At the highest stellar masses ($\log(M_{*,30}/M_\odot)\gtrsim11.1$):
\begin{itemize}
\setlength{\itemsep}{0pt}
\setlength{\parskip}{0pt}
\setlength{\parsep}{0pt}
\item Prolate or triaxial stellar structures remain common in the inner regions of massive galaxies in both simulations.
\item The outer bodies of massive EAGLE galaxies are, on average, more prolate or triaxial than their TNG counterparts.
\end{itemize}

Overall, \texttt{Gal3D} provides a flexible framework for quantifying intrinsic radial 3D structure in cosmological simulations. It is well suited to future studies of stellar and dark matter components, to detailed investigations of the evolution of specific galaxy types and structural components, and to linking intrinsic 3D structure with projected morphology and observables.

The galaxy shape-profile data used in this work are publicly available at OSF at \dataset[doi: 10.17605/OSF.IO/K9X3E]{https://doi.org/10.17605/OSF.IO/K9X3E}. The \texttt{Gal3D} source code is publicly available on GitHub\footnote{\url{https://github.com/GalaxySimAnalytics/gal3d}} and an archival copy in Zenodo \citep{lu_2026_21471285}.

%% Please use the acknowledgment and contribution environments. This will 
%% be anonomyized when the "anonymous" style option is used. 
\begin{acknowledgments}
We thank the anonymous referee for the helpful comments that improved the quality of this work. We also thank V. P. Debattista for his valuable suggestions on the manuscript. This work is supported by the National Key R\&D Program of China (No. XY-2025-1459), the National Natural Science Foundation of China under grant No. 12573010, and the Science Fund for Creative Research Groups of the National Natural Science Foundation of China (No. 12221003). We thank the IllustrisTNG and EAGLE collaborations for making their simulation data publicly available. This work made use of the public IllustrisTNG and EAGLE data releases. This work also made use of computational support from the Computing Center in Xi’an, China.
\end{acknowledgments}

\begin{contribution}
%%This section gives authors the space to recognize author contributions. The text inside this environment is NOT counted towards the total word quanta. At a minimum, manuscripts are expected to include this text:

S.L. led the methodology development, analysis, visualization, and manuscript preparation. M.D. supervised the project and contributed to the interpretation of the results and the revision of the manuscript.

%% But authors are expected to provide more specific details, e.g. 
%%
%%SC was responsible for writing and submitting the manuscript.
%%WWM came up with the initial research concept and edited the manuscript.
%%OTS obtained the funding and edited the manuscript.
%%EBF provided the formal analysis and validation. He also edited the manuscript.
%%GEH Supervised the undergraduates, wrote the software and administers the project github and Zenodo repositories.
%%
%% Authors can use the Contributor Role Taxonomy (CRediT) at
%% for ideas on how write a good statement tailored to their needs.

\end{contribution}

%\facilities{HST(STIS), Swift(XRT and UVOT), AAVSO, CTIO:1.3m, CTIO:1.5m, CXO}

\software{Matplotlib \citep{Hunter:2007},  
          SciPy \citep{2020SciPy-NMeth}, 
          NumPy \citep{harris2020array},
          Cython \citep{behnelCythonBestBoth2011},
          lmfit \citep{newville2025limfit},
          Pynbody \citep{pynbody, andrew_pontzen_2026_pynbody}
          }


\appendix

\section{Numerical Sensitivity Tests for Superellipsoid Fitting}
\label{sec:numerical_caveats}

\begin{figure*}
\centering
{\includegraphics[width=0.9\textwidth]{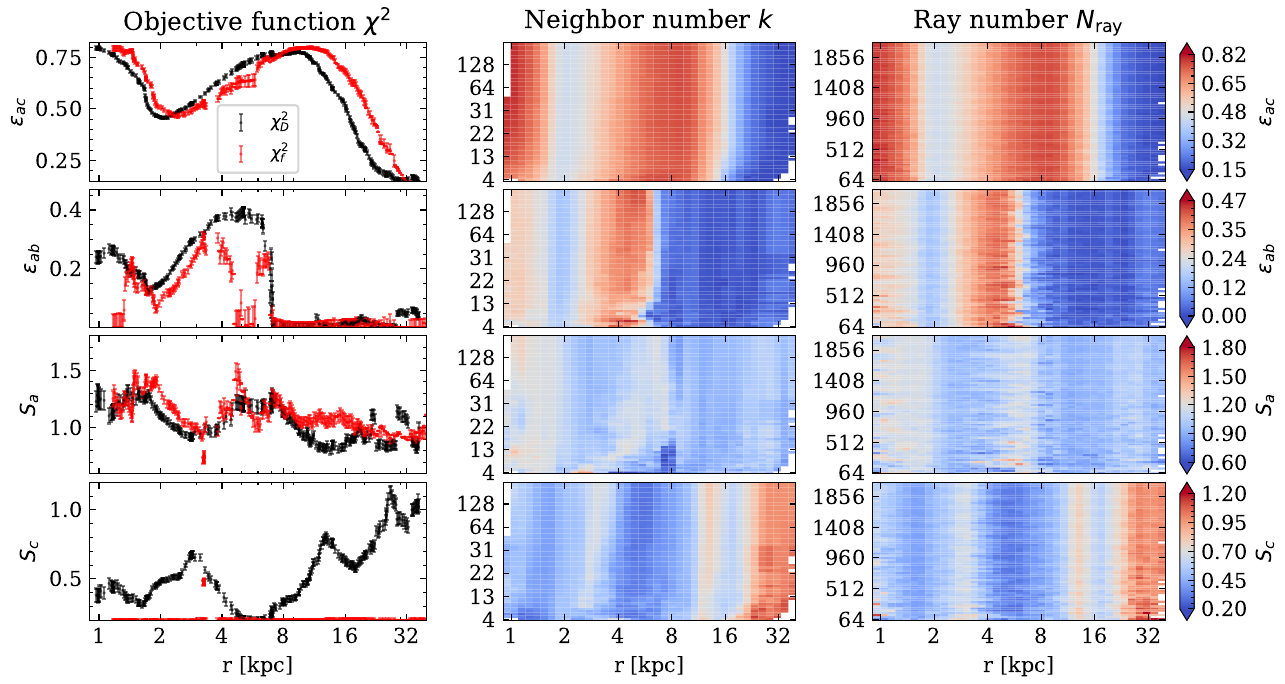}}
	\caption{Numerical tests of the superellipsoid fitting procedure for a representative TNG50-1 galaxy. The four rows show, from top to bottom, $\varepsilon_{ac}$, $\varepsilon_{ab}$, $S_a$, and $S_c$. The left column compares the radial profiles obtained with two objective functions: the fiducial surface-radius-ratio objective $\chi^2_D$ (black points with error bars) and the direct implicit-function objective $\chi^2_f$ (red points with error bars). The middle column shows the corresponding recovered parameter values as a function of radius and the number of neighbors, $k$, used in the adaptive kernel density estimation, and the right column shows the results as a function of radius and the number of rays, $N_{\rm ray}$. In the middle and right columns, color encodes the fitted parameter value according to the color bars shown at the right of each row. The figure illustrates that the surface-radius-ratio objective yields substantially more stable fits than the direct implicit-function objective, while the inferred profiles are largely insensitive to reasonable variations in $k$ and $N_{\rm ray}$.}
	\label{fig:numerical_caveats}
\end{figure*}

\begin{figure*}
\centering
\includegraphics[width=0.9\textwidth]{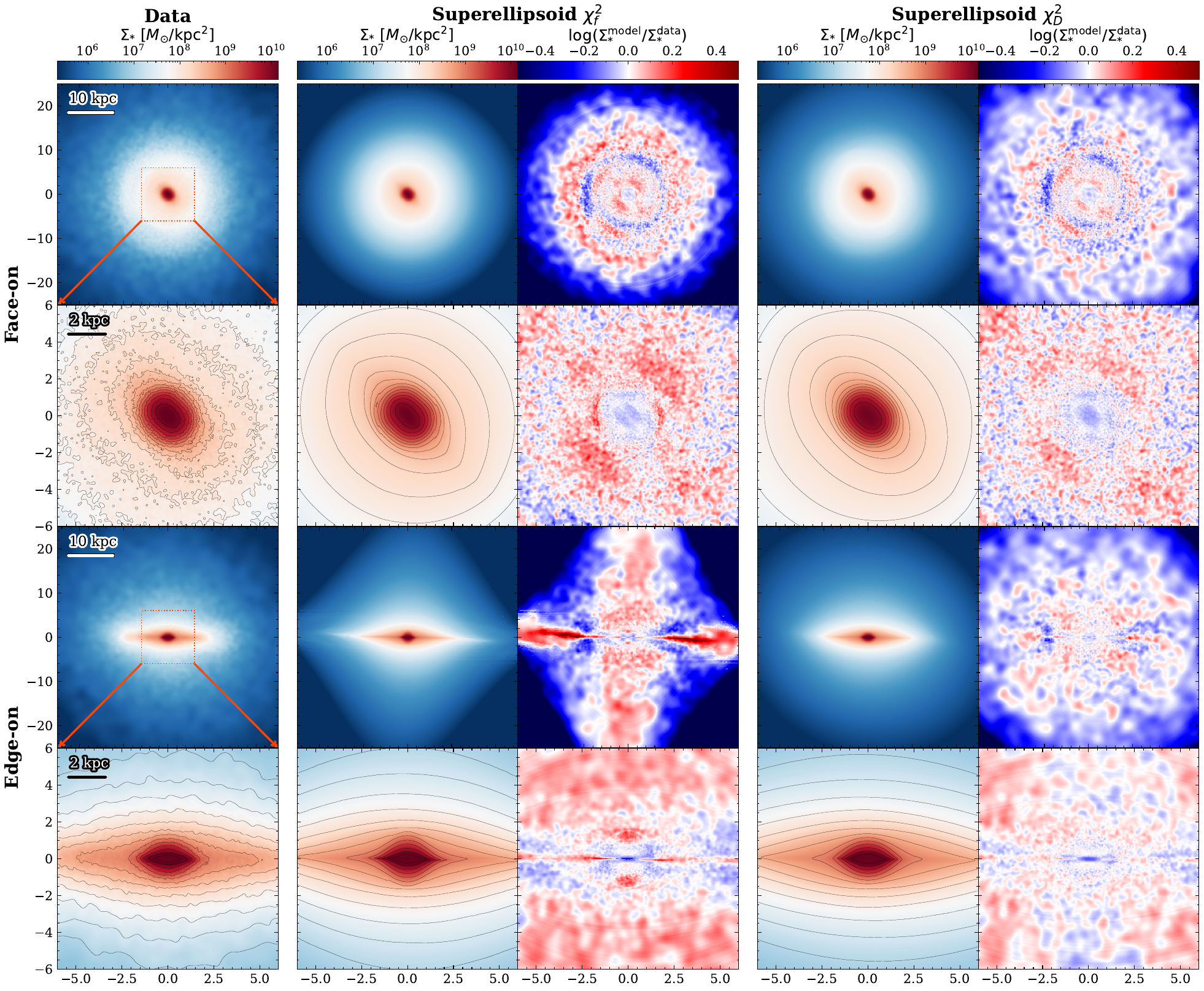}
\caption{Projected stellar surface-density maps of the $\chi^2_D$ and $\chi^2_f$ superellipsoid models for the test galaxy of
Figure~\ref{fig:numerical_caveats}, in the same format as
Figure~\ref{fig:ellipsoid_vs_superellipsoid}. The $\chi^2_f$ model appears
artificially pointed and disky in the edge-on view, reflecting the failure
to fit $S_c$.}
\label{fig:chi2_comparison}
\end{figure*}

In this appendix, we test the sensitivity of the superellipsoid fits to three numerical choices: the fitting objective function, the number of neighbors ($k$), and the number of rays ($N_{\rm ray}$). Figure~\ref{fig:numerical_caveats} summarizes the results for a representative TNG50-1 galaxy. The tests show that the choice of objective function matters for the reliability of the fit, whereas reasonable variations in $k$ and $N_{\rm ray}$ have little effect on the recovered radial shape profiles.

For comparison with the fiducial surface-radius-ratio objective, we define the direct implicit-function objective as
\begin{equation}
\chi^2_f \;=\; \sum_{i=1}^{N} \bigl(r_i'\bigr)^{2}\, \bigl[f(\boldsymbol{x}'_i)-1\bigr]^{2},
\end{equation}
which penalizes the residual in the superellipsoid function itself rather than the surface-radius ratio.
The left column of Figure~\ref{fig:numerical_caveats} compares the two objective functions. The fiducial surface-radius-ratio objective $\chi^2_D$ yields stable and physically sensible profiles across all fitted parameters. The direct implicit-function objective $\chi^2_f$ performs substantially worse, especially for $S_c$. In that case, most fitted values collapse to the lower bound $S_c=0.2$, indicating poor conditioning in the optimization. This behavior is expected because the direct implicit-function objective depends explicitly on the nonlinear superellipsoid indices, making the fit less stable when the surface departs from the ellipsoidal limit. We therefore adopt the surface-radius-ratio objective as the fiducial choice throughout the paper.

Figure~\ref{fig:chi2_comparison} provides a direct morphological comparison, showing the projected stellar surface-density maps of the $\chi^2_D$ and $\chi^2_f$ superellipsoid models. The $\chi^2_D$ model reproduces the galaxy morphology well in both face-on and edge-on projections. The $\chi^2_f$ model, however, appears artificially pointed and disky when viewed edge-on, reflecting the failure to fit $S_c$ (see left column of Figure~\ref{fig:numerical_caveats}).

The middle and right columns of Figure~\ref{fig:numerical_caveats} examine the dependence of the recovered profiles on the number of neighbors used in the adaptive kernel density estimation and on the number of rays. The profiles remain consistent across the tested values, indicating numerical robustness. Noticeable degradation occurs only for a very small number of neighbors, $k \lesssim 16$, where insufficient smoothing increases profile fluctuations, especially in the shape indices. For $k \gtrsim 16$, the profiles are mutually consistent. The test with varying numbers of rays shows similarly weak sensitivity: even $N_{\rm ray}=64$ reproduces the main trends obtained with larger ray samples. We therefore adopt $k=32$ and $N_{\rm ray}=1024$ as fiducial values because they give stable results while remaining computationally practical.

\section{Superellipsoid Morphologies for Independent Variations in $S_a$ and $S_b$}
\label{sec:superellipsoid_sa_sb}

\begin{figure*}
\centering
{\includegraphics[width=0.5\textwidth]{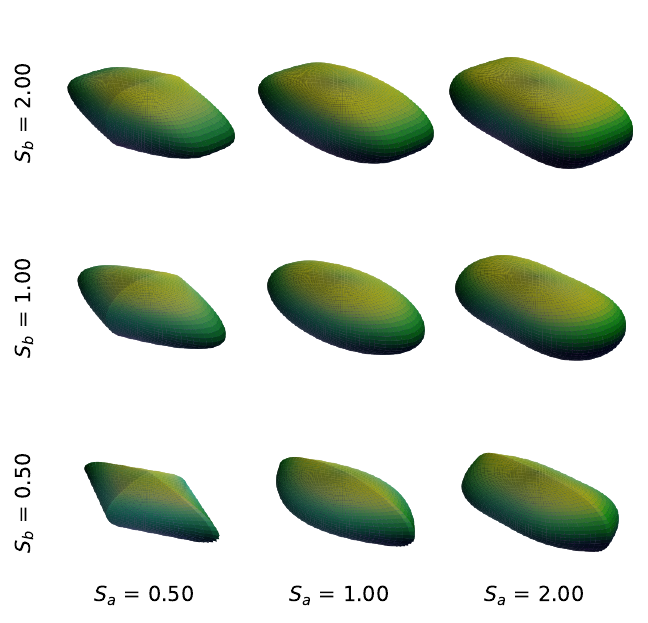}}
    \caption{Superellipsoids with fixed axis ratios $a:b:c=10:5:4$ and $S_c=1$, varying the shape indices $S_a$ and $S_b$ independently. This figure complements Figure~\ref{fig:superellipsoid_shape} in the main text, which shows the restricted case $S_a=S_b$ together with variations in $S_c$.}
    \label{fig:superellipsoid_shape_sa_sb_examples}
\end{figure*}

This appendix provides a complementary visualization of how the superellipsoid shape indices control morphology when the two indices, $S_a$ and $S_b$, vary independently. Figure~\ref{fig:superellipsoid_shape} illustrates the case in which $S_a=S_b$ and $S_c$ varies. Figure~\ref{fig:superellipsoid_shape_sa_sb_examples} instead fixes $S_c=1$ and varies $S_a$ and $S_b$ separately. The figure shows that decreasing either $S_a$ or $S_b$ below unity produces a more pointed, diamond-like morphology along the corresponding axis, whereas increasing either index above unity makes the surface broader and boxier in that direction.

\section{Computational Cost Comparison}
\label{sec:runtime_comparison}

\begin{figure*}
\centering
\includegraphics[width=0.6\textwidth]{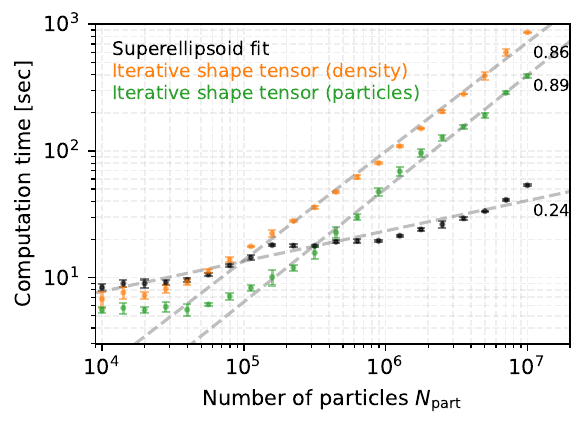}
\caption{Computation time as a function of particle number $N_{\rm part}$ for the
superellipsoid fit and two implementations of the iterative shape-tensor
method, measured on a single CPU core. Error bars show the mean and standard
deviation of five repeated measurements at each particle count. Dashed lines
are power-law fits to $N_{\rm part}>10^5$, with the best-fit slope
annotated. The superellipsoid fit runtime scales as $\propto N_{\rm part}^{0.24}$,
while both iterative implementations scale nearly linearly ($\propto
N_{\rm part}^{0.86-0.89}$). At $N_{\rm part}\lesssim
3\times10^5$, all three methods have comparable runtimes; at
$N_{\rm part}\sim10^7$, the superellipsoid fit is $\sim7$ times faster.}
\label{fig:runtime_comparison}
\end{figure*}

Figure~\ref{fig:runtime_comparison} compares the computation time of the
superellipsoid fit with two implementations of the iterative shape-tensor
method: the density-field version described in
Section~\ref{sec:shape_tensor} and a conventional particle-based
implementation. All measurements were performed on a single CPU core. For
the superellipsoid fit, the reported time is the total wall-clock time including data preprocessing (adaptive kernel density estimation and iso-density surface extraction) and the superellipsoid fitting. For the density-field iterative method (see Section~\ref{sec:shape_tensor}), the total time includes KD-tree construction and the iterative shape-tensor computation. For the
particle-based iterative method, no preprocessing is required and the
reported time is the fitting time alone. In each case, 200 nested
ellipsoidal (or superellipsoidal) surfaces were fitted to sample the
radial shape profile.

The superellipsoid fit runtime scales as $\propto N_{\rm part}^{0.24}$, while both
iterative shape-tensor implementations scale nearly linearly ($\propto
N_{\rm part}^{0.86-0.89}$). At particle counts below
$3\times10^5$, all three methods have comparable runtimes. At the largest
particle count tested ($\sim10^7$), the superellipsoid fit is approximately seven times faster than the conventional particle-based implementation.

%% For this sample we use BibTeX plus aasjournalv7.bst to generate the

\bibliography{sample701}{}

@ARTICLE{erwinPeanutsAngle2013,
       author = {{Erwin}, Peter and {Debattista}, Victor P.},
        title = "{Peanuts at an angle: detecting and measuring the three-dimensional structure of bars in moderately inclined galaxies}",
      journal = {\mnras},
         year = 2013,
        month = jun,
       volume = {431},
       number = {4},
        pages = {3060-3086},
          doi = {10.1093/mnras/stt385},
archivePrefix = {arXiv},
       eprint = {1301.0638},
 primaryClass = {astro-ph.CO},
       adsurl = {https://ui.adsabs.harvard.edu/abs/2013MNRAS.431.3060E}
}

@software{lu_2026_21471285,
  author       = {Lu, Shuai},
  title        = {Gal3D: Superellipsoid-Based 3D Galaxy Morphology
                   Modeling
                  },
  month        = jul,
  year         = 2026,
  publisher    = {Zenodo},
  version      = {v1.0.0},
  doi          = {10.5281/zenodo.21471285},
  url          = {https://doi.org/10.5281/zenodo.21471285},
}

@article{pulsoniStellarHalosETGs2020,
  title = {The Stellar Halos of ETGs in the IllustrisTNG Simulations: The Photometric and Kinematic Diversity of Galaxies at Large Radii},
  shorttitle = {The Stellar Halos of ETGs in the IllustrisTNG Simulations},
  author = {Pulsoni, C. and Gerhard, O. and Arnaboldi, M. and Pillepich, A. and Nelson, D. and Hernquist, L. and Springel, V.},
  year = 2020,
  month = sep,
  journal = {Astronomy and Astrophysics},
  volume = {641},
  pages = {A60},
  publisher = {EDP},
  issn = {0004-6361},
  doi = {10.1051/0004-6361/202038253},
  urldate = {2026-07-17}
}

@article{naabATLAS3DProjectXXV2014,
  title = {The ATLAS3D Project - XXV. Two-Dimensional Kinematic Analysis of Simulated Galaxies and the Cosmological Origin of Fast and Slow Rotators},
  author = {Naab, Thorsten and Oser, L. and Emsellem, E. and Cappellari, Michele and Krajnovi{\'c}, D. and McDermid, R. M. and Alatalo, K. and Bayet, E. and Blitz, L. and Bois, M. and Bournaud, F. and Bureau, M. and Crocker, A. and Davies, R. L. and Davis, T. A. and {de Zeeuw}, P. T. and Duc, P.-A. and Hirschmann, M. and Johansson, P. H. and Khochfar, S. and Kuntschner, H. and Morganti, R. and Oosterloo, T. and Sarzi, M. and Scott, N. and Serra, P. and {van de Ven}, G. and Weijmans, A. and Young, L. M.},
  year = 2014,
  month = nov,
  journal = {Monthly Notices of the Royal Astronomical Society},
  volume = {444},
  pages = {3357--3387},
  publisher = {OUP},
  issn = {0035-8711},
  doi = {10.1093/mnras/stt1919},
  urldate = {2026-07-17}
}

@article{schulzeKinematicsSimulatedGalaxies2018,
  title = {Kinematics of Simulated Galaxies - I. Connecting Dynamical and Morphological Properties of Early-Type Galaxies at Different Redshifts},
  author = {Schulze, Felix and Remus, Rhea-Silvia and Dolag, Klaus and Burkert, Andreas and Emsellem, Eric and {van de Ven}, Glenn},
  year = 2018,
  month = nov,
  journal = {Monthly Notices of the Royal Astronomical Society},
  volume = {480},
  pages = {4636--4658},
  publisher = {OUP},
  issn = {0035-8711},
  doi = {10.1093/mnras/sty2090},
  urldate = {2026-07-17}
}

@article{debattistaCausesHaloShape2008,
  title = {The Causes of Halo Shape Changes Induced by Cooling Baryons: Disks versus Substructures},
  shorttitle = {The Causes of Halo Shape Changes Induced by Cooling Baryons},
  author = {Debattista, Victor P. and Moore, Ben and Quinn, Thomas and Kazantzidis, Stelios and Maas, Ryan and Mayer, Lucio and Read, Justin and Stadel, Joachim},
  year = 2008,
  month = jul,
  journal = {The Astrophysical Journal},
  volume = {681},
  pages = {1076--1088},
  publisher = {IOP},
  issn = {0004-637X},
  doi = {10.1086/587977},
  urldate = {2026-03-02}
}

@article{jonssonTangledWarpMilky2024,
  title = {The Tangled Warp of the Milky Way},
  author = {J{\'o}nsson, Viktor Hrannar and McMillan, Paul J.},
  year = 2024,
  month = aug,
  journal = {Astronomy and Astrophysics},
  volume = {688},
  pages = {A38},
  publisher = {EDP},
  issn = {0004-6361},
  doi = {10.1051/0004-6361/202449744},
  urldate = {2026-05-16}
}

@article{mitsudaIsophoteShapesEarlytype2017,
  title = {Isophote Shapes of Early-Type Galaxies in Massive Clusters at z {$\sim$} 1 and 0},
  author = {Mitsuda, Kazuma and Doi, Mamoru and Morokuma, Tomoki and Suzuki, Nao and Yasuda, Naoki and Perlmutter, Saul and Aldering, Greg and Meyers, Joshua},
  year = 2017,
  month = jan,
  journal = {The Astrophysical Journal},
  volume = {834},
  pages = {109},
  publisher = {IOP},
  issn = {0004-637X},
  doi = {10.3847/1538-4357/834/2/109},
  urldate = {2026-03-10}
}

@article{naabPropertiesEarlyTypeDry2006,
  title = {Properties of Early-Type, Dry Galaxy Mergers and the Origin of Massive Elliptical Galaxies},
  author = {Naab, Thorsten and Khochfar, Sadegh and Burkert, Andreas},
  year = 2006,
  month = jan,
  journal = {The Astrophysical Journal},
  volume = {636},
  number = {2},
  pages = {L81},
  publisher = {IOP Publishing},
  issn = {0004-637X},
  doi = {10.1086/500205},
  urldate = {2026-07-07},
  langid = {english}
}

@article{monteiro-oliveiraNoEvidenceDichotomy2025,
  title = {No Evidence of a Dichotomy in the Elliptical Galaxy Population},
  author = {{Monteiro-Oliveira}, Rog{\'e}rio and Lin, Yen-Ting and Chen, Wei-Huai and Chuang, Chen-Yu and {Abdurro'uf} and Wu, Po-Feng},
  year = 2025,
  month = jul,
  journal = {The Astrophysical Journal},
  volume = {988},
  pages = {138},
  publisher = {IOP},
  issn = {0004-637X},
  doi = {10.3847/1538-4357/ade0ba},
  urldate = {2026-03-10}
}

@article{kleinShapeFIREboxGalaxies2026,
  title = {The Shape of FIREbox Galaxies and a Potential Tension with Low-Mass Disks},
  author = {Klein, Courtney and Wang, Jenny D. and Xia, Luke Y. and Bullock, James S. and Moreno, Jorge and Feldmann, Robert and Mercado, Francisco J. and {Faucher-Gigu{\`e}re}, Claude-Andr{\'e} and Stern, Jonathan and Sanchez, N. Nicole and Hussein, Abdelaziz and Park, Michelle S.},
  year = 2026,
  month = feb,
  journal = {The Astrophysical Journal},
  volume = {998},
  number = {1},
  pages = {125},
  publisher = {The American Astronomical Society},
  issn = {0004-637X},
  doi = {10.3847/1538-4357/ae31f4},
  urldate = {2026-03-10},
  langid = {english}
}

@article{yongGalaxy3DShape2024,
  title = {Galaxy 3D Shape Recovery Using Mixture Density Network},
  author = {Yong, Suk Yee and Harborne, K. E. and Foster, Caroline and Bassett, Robert and Poole, Gregory B. and Cavanagh, Mitchell},
  year = 2024,
  month = may,
  journal = {Publications of the Astronomical Society of Australia},
  volume = {41},
  pages = {e033},
  issn = {1323-3580},
  doi = {10.1017/pasa.2024.32},
  urldate = {2026-03-10}
}

@article{giocoliAIDATNGProject3D2026,
  title = {The AIDA-TNG Project: 3D Halo Shapes},
  shorttitle = {The AIDA-TNG Project},
  author = {Giocoli, C. and Despali, G. and Moscardini, L. and Meneghetti, M. and Sheth, R. K. and Pillepich, A. and Vogelsberger, M.},
  year = 2026,
  month = feb,
  journal = {Astronomy and Astrophysics},
  volume = {706},
  pages = {A340},
  publisher = {EDP},
  issn = {0004-6361},
  doi = {10.1051/0004-6361/202558400},
  urldate = {2026-05-17}
}

@article{chuaShapeDarkMatter2019,
  title = {Shape of Dark Matter Haloes in the Illustris Simulation: Effects of Baryons},
  shorttitle = {Shape of Dark Matter Haloes in the Illustris Simulation},
  author = {Chua, Kun Ting Eddie and Pillepich, Annalisa and Vogelsberger, Mark and Hernquist, Lars},
  year = 2019,
  month = mar,
  journal = {Monthly Notices of the Royal Astronomical Society},
  volume = {484},
  pages = {476--493},
  publisher = {OUP},
  issn = {0035-8711},
  doi = {10.1093/mnras/sty3531},
  urldate = {2026-03-02}
}

@inproceedings{feldmannCosmologicalSimulationsGalaxies2026,
  title = {Cosmological Simulations of Galaxies},
  booktitle = {Encyclopedia of Astrophysics, Volume 4},
  author = {Feldmann, Robert and Bieri, Rebekka},
  year = 2026,
  month = jan,
  volume = {4},
  pages = {576--599},
  address = {eprint: arXiv:2507.08925},
  doi = {10.1016/B978-0-443-21439-4.00111-5},
  urldate = {2026-05-17}
}

@misc{teyssierNumericalCosmology2025,
  title = {Numerical Cosmology},
  author = {Teyssier, Romain},
  year = 2025,
  month = oct,
  number = {arXiv:2510.13129},
  eprint = {2510.13129},
  primaryclass = {astro-ph},
  publisher = {arXiv},
  doi = {10.48550/arXiv.2510.13129},
  urldate = {2025-12-04},
  archiveprefix = {arXiv}
}

@misc{valentiniHydrodynamicMethodsSubresolution2025,
  title = {Hydrodynamic Methods and Sub-Resolution Models for Cosmological Simulations},
  author = {Valentini, Milena and Dolag, Klaus},
  year = 2025,
  month = feb,
  number = {arXiv:2502.06954},
  eprint = {2502.06954},
  primaryclass = {astro-ph},
  publisher = {arXiv},
  doi = {10.48550/arXiv.2502.06954},
  urldate = {2025-12-04},
  archiveprefix = {arXiv}
}

@article{vogelsbergerCosmologicalSimulationsGalaxy2020,
  title = {Cosmological Simulations of Galaxy Formation},
  author = {Vogelsberger, Mark and Marinacci, Federico and Torrey, Paul and Puchwein, Ewald},
  year = 2020,
  month = jan,
  journal = {Nature Reviews Physics},
  volume = {2},
  pages = {42--66},
  doi = {10.1038/s42254-019-0127-2},
  urldate = {2026-05-17}
}

@article{sotillo-ramosDiscFlaringTNG502023,
  title = {Disc Flaring with TNG50: Diversity across Milky Way and M31 Analogues},
  shorttitle = {Disc Flaring with TNG50},
  author = {{Sotillo-Ramos}, Diego and Donnari, Martina and Pillepich, Annalisa and Frankel, Neige and Nelson, Dylan and Springel, Volker and Hernquist, Lars},
  year = 2023,
  month = aug,
  journal = {Monthly Notices of the Royal Astronomical Society},
  volume = {523},
  pages = {3915--3938},
  publisher = {OUP},
  issn = {0035-8711},
  doi = {10.1093/mnras/stad1485},
  urldate = {2026-05-17}
}

@article{semenovFormationGalacticDisks2024,
  title = {Formation of Galactic Disks. I. Why Did the Milky Way's Disk Form Unusually Early?},
  author = {Semenov, Vadim A. and Conroy, Charlie and Chandra, Vedant and Hernquist, Lars and Nelson, Dylan},
  year = 2024,
  month = feb,
  journal = {The Astrophysical Journal},
  volume = {962},
  pages = {84},
  publisher = {IOP},
  issn = {0004-637X},
  doi = {10.3847/1538-4357/ad150a},
  urldate = {2026-05-17}
}

@article{gargiuloHighLowSersic2022,
  title = {High and Low S\'ersic Index Bulges in Milky Way- and M31-like Galaxies: Origin and Connection to the Bar with TNG50},
  shorttitle = {High and Low S\'ersic Index Bulges in Milky Way- and M31-like Galaxies},
  author = {Gargiulo, Ignacio D. and Monachesi, Antonela and G{\'o}mez, Facundo A. and Nelson, Dylan and Pillepich, Annalisa and Pakmor, R{\"u}diger and Grand, R. J. J. and Fragkoudi, Francesca and Hernquist, Lars and Lovell, Mark and Marinacci, Federico},
  year = 2022,
  month = may,
  journal = {Monthly Notices of the Royal Astronomical Society},
  volume = {512},
  pages = {2537--2555},
  publisher = {OUP},
  issn = {0035-8711},
  doi = {10.1093/mnras/stac629},
  urldate = {2026-05-17}
}

@article{schultheisNuclearStellarDiscs2025,
  title = {Nuclear Stellar Discs},
  author = {Schultheis, Mathias and Sormani, Mattia C. and Gadotti, Dimitri A.},
  year = 2025,
  month = nov,
  journal = {Astronomy and Astrophysics Review},
  volume = {33},
  pages = {7},
  publisher = {Springer},
  issn = {0935-4956},
  doi = {10.1007/s00159-025-00163-6},
  urldate = {2026-05-16}
}

@article{athanassoulaNatureBulgesGeneral2005,
  title = {On the Nature of Bulges in General and of Box/Peanut Bulges in Particular: Input from N-Body Simulations},
  shorttitle = {On the Nature of Bulges in General and of Box/Peanut Bulges in Particular},
  author = {Athanassoula, E.},
  year = 2005,
  month = apr,
  journal = {Monthly Notices of the Royal Astronomical Society},
  volume = {358},
  pages = {1477--1488},
  publisher = {OUP},
  issn = {0035-8711},
  doi = {10.1111/j.1365-2966.2005.08872.x},
  urldate = {2026-05-16}
}

@article{duFORMINGDOUBLEBARREDGALAXIES2015,
  title = {FORMING DOUBLE-BARRED GALAXIES FROM DYNAMICALLY COOL INNER DISKS},
  author = {Du, Min and Shen, Juntai and Debattista, Victor P.},
  year = 2015,
  month = may,
  journal = {The Astrophysical Journal},
  volume = {804},
  number = {2},
  pages = {139},
  publisher = {The American Astronomical Society},
  issn = {0004-637X},
  doi = {10.1088/0004-637X/804/2/139},
  urldate = {2024-07-21},
  langid = {english}
}

@article{erwinDependenceBarFrequency2018,
  title = {The Dependence of Bar Frequency on Galaxy Mass, Colour, and Gas Content -- and Angular Resolution -- in the Local Universe},
  author = {Erwin, Peter},
  year = 2018,
  month = mar,
  journal = {Monthly Notices of the Royal Astronomical Society},
  volume = {474},
  number = {4},
  pages = {5372--5392},
  issn = {0035-8711},
  doi = {10.1093/mnras/stx3117},
  urldate = {2024-07-27}
}

@article{garcia-ruizNeutralHydrogenOptical2002,
  title = {Neutral Hydrogen and Optical Observations of Edge-on Galaxies: Hunting for Warps},
  shorttitle = {Neutral Hydrogen and Optical Observations of Edge-on Galaxies},
  author = {{Garc{\'i}a-Ruiz}, I. and Sancisi, R. and Kuijken, K.},
  year = 2002,
  month = nov,
  journal = {Astronomy and Astrophysics},
  volume = {394},
  pages = {769--789},
  publisher = {EDP},
  issn = {0004-6361},
  doi = {10.1051/0004-6361:20020976},
  urldate = {2026-05-16}
}

@article{binneyWarps1992,
  title = {Warps.},
  author = {Binney, James},
  year = 1992,
  month = jan,
  journal = {Annual Review of Astronomy and Astrophysics},
  volume = {30},
  pages = {51--74},
  issn = {0066-4146},
  doi = {10.1146/annurev.aa.30.090192.000411},
  urldate = {2026-05-16}
}

@article{dehnenTwistedPrecessingCepheid2023,
  title = {A Twisted and Precessing Cepheid Warp in the Outer Milky Way Disc},
  author = {Dehnen, Walter and Semczuk, Marcin and Sch{\"o}nrich, Ralph},
  year = 2023,
  month = jul,
  journal = {Monthly Notices of the Royal Astronomical Society},
  volume = {523},
  pages = {1556--1564},
  publisher = {OUP},
  issn = {0035-8711},
  doi = {10.1093/mnras/stad1502},
  urldate = {2026-05-16}
}

@article{debattistaWarpedGalaxiesMisaligned1999,
  title = {Warped Galaxies from Misaligned Angular Momenta},
  author = {Debattista, Victor P. and Sellwood, J. A.},
  year = 1999,
  month = mar,
  journal = {The Astrophysical Journal},
  volume = {513},
  pages = {L107-L110},
  publisher = {IOP},
  issn = {0004-637X},
  doi = {10.1086/311913},
  urldate = {2026-05-16}
}

@article{roskarMisalignedAngularMomentum2010,
  title = {Misaligned Angular Momentum in Hydrodynamic Cosmological Simulations: Warps, Outer Discs and Thick Discs},
  shorttitle = {Misaligned Angular Momentum in Hydrodynamic Cosmological Simulations},
  author = {Ro{\v s}kar, Rok and Debattista, Victor P. and Brooks, Alyson M. and Quinn, Thomas R. and Brook, Chris B. and Governato, Fabio and Dalcanton, Julianne J. and Wadsley, James},
  year = 2010,
  month = oct,
  journal = {Monthly Notices of the Royal Astronomical Society},
  volume = {408},
  pages = {783--796},
  publisher = {OUP},
  issn = {0035-8711},
  doi = {10.1111/j.1365-2966.2010.17178.x},
  urldate = {2026-05-16}
}

@article{shenGalacticWarpsInduced2006,
  title = {Galactic Warps Induced by Cosmic Infall},
  author = {Shen, Juntai and Sellwood, J. A.},
  year = 2006,
  month = jul,
  journal = {Monthly Notices of the Royal Astronomical Society},
  volume = {370},
  pages = {2--14},
  publisher = {OUP},
  issn = {0035-8711},
  doi = {10.1111/j.1365-2966.2006.10477.x},
  urldate = {2026-05-16}
}

@article{zhangDiversePhysicalOrigins2025,
  title = {The Diverse Physical Origins of Stars in the Dynamically Hot Bulge: CALIFA versus IllustrisTNG},
  shorttitle = {The Diverse Physical Origins of Stars in the Dynamically Hot Bulge},
  author = {Zhang, Le and Zhu, Ling and Pillepich, Annalisa and Du, Min and Jiang, Fangzhou and {Falc{\'o}n-Barroso}, Jes{\'u}s},
  year = 2025,
  month = jul,
  journal = {Astronomy and Astrophysics},
  volume = {699},
  pages = {A320},
  publisher = {EDP},
  issn = {0004-6361},
  doi = {10.1051/0004-6361/202451292},
  urldate = {2026-05-16}
}

@article{kormendySecularEvolutionFormation2004,
  title = {Secular Evolution and the Formation of Pseudobulges in Disk Galaxies},
  author = {Kormendy, John and Kennicutt, Jr., Robert C.},
  year = 2004,
  month = sep,
  journal = {Annual Review of Astronomy and Astrophysics},
  volume = {42},
  pages = {603--683},
  issn = {0066-4146},
  doi = {10.1146/annurev.astro.42.053102.134024},
  urldate = {2026-05-16}
}

@book{sandageCarnegieAtlasGalaxies1994,
  title = {The Carnegie Atlas of Galaxies},
  author = {Sandage, Allan and Bedke, John},
  year = 1994,
  month = jan,
  journal = {The Carnegie Atlas of Galaxies. Volumes I},
  volume = {638},
  urldate = {2026-05-16}
}

@article{rydenIntrinsicShapeSpiral2006,
  title = {The Intrinsic Shape of Spiral Galaxies in the 2MASS Large Galaxy Atlas},
  author = {Ryden, Barbara S.},
  year = 2006,
  month = apr,
  journal = {The Astrophysical Journal},
  volume = {641},
  pages = {773--784},
  publisher = {IOP},
  issn = {0004-637X},
  doi = {10.1086/500497},
  urldate = {2026-05-16}
}

@article{kruitGalaxyDisks2011,
  title = {Galaxy Disks},
  author = {van der Kruit, P. C. and Freeman, K. C.},
  year = 2011,
  month = sep,
  journal = {Annual Review of Astronomy and Astrophysics},
  volume = {49},
  number = {Volume 49, 2011},
  pages = {301--371},
  publisher = {Annual Reviews},
  issn = {0066-4146, 1545-4282},
  doi = {10.1146/annurev-astro-083109-153241},
  urldate = {2026-05-16},
  langid = {english}
}

@book{binneyGalacticDynamics1987,
  title = {Galactic Dynamics},
  author = {Binney, James and Tremaine, Scott},
  year = 1987,
  month = jan,
  journal = {Princeton},
  urldate = {2026-05-16}
}

@misc{sellwoodGALAXYPackageNbody2014,
  title = {GALAXY Package for N-Body Simulation},
  author = {Sellwood, J. A.},
  year = 2014,
  month = jun,
  number = {arXiv:1406.6606},
  eprint = {1406.6606},
  primaryclass = {astro-ph},
  publisher = {arXiv},
  doi = {10.48550/arXiv.1406.6606},
  urldate = {2024-07-09},
  archiveprefix = {arXiv}
}

@article{bassettProspectsRecoveringGalaxy2019,
  title = {Prospects for Recovering Galaxy Intrinsic Shapes from Projected Quantities},
  author = {Bassett, Robert and Foster, Caroline},
  year = 2019,
  month = aug,
  journal = {Monthly Notices of the Royal Astronomical Society},
  volume = {487},
  number = {2},
  pages = {2354--2371},
  issn = {0035-8711},
  doi = {10.1093/mnras/stz1440},
  urldate = {2026-03-10}
}

@article{bailinInternalExternalAlignment2005,
  title = {Internal and External Alignment of the Shapes and Angular Momenta of {$\Lambda$}CDM Halos},
  author = {Bailin, Jeremy and Steinmetz, Matthias},
  year = 2005,
  month = jul,
  journal = {The Astrophysical Journal},
  volume = {627},
  pages = {647--665},
  publisher = {IOP},
  issn = {0004-637X},
  doi = {10.1086/430397},
  urldate = {2025-10-07}
}

@article{folsomCosmologicalSimulationsStellar2025,
  title = {Cosmological Simulations of Stellar Halos with Gaia Sausage-Enceladus Analogs: Two Sausages, One Bun?},
  shorttitle = {Cosmological Simulations of Stellar Halos with Gaia Sausage-Enceladus Analogs},
  author = {Folsom, Dylan and Lisanti, Mariangela and Necib, Lina and Horta, Danny and Vogelsberger, Mark and Hernquist, Lars},
  year = 2025,
  month = apr,
  journal = {The Astrophysical Journal},
  volume = {983},
  pages = {119},
  publisher = {IOP},
  issn = {0004-637X},
  doi = {10.3847/1538-4357/adbe31},
  urldate = {2026-05-16}
}

@article{heMorphologyChallengesDecomposing2025,
  title = {Beyond Morphology: Challenges in Decomposing Massive Stellar Halos in Sombrero-like, Halo-Embedded Disk Galaxies},
  shorttitle = {Beyond Morphology},
  author = {He, Wu-Tao and Du, Min and Li, Zhao-Yu and Li, Yuan},
  year = 2025,
  month = jul,
  journal = {Astronomy and Astrophysics},
  volume = {699},
  pages = {A99},
  publisher = {EDP},
  issn = {0004-6361},
  doi = {10.1051/0004-6361/202554551},
  urldate = {2026-05-16}
}

@misc{chenDownbendingBreaksGalactic2026,
  title = {Down-Bending Breaks in Galactic Disks Are an Intrinsic Byproduct of Inside-out Growth},
  author = {Chen, Liufei and Du, Min and Lu, Shuai and Li, Jing and Ho, Luis C.},
  year = 2026,
  month = jan,
  publisher = {arXiv},
  doi = {10.48550/arXiv.2602.00626},
  urldate = {2026-05-16}
}

@article{hanTiltedDarkHalos2023,
  title = {Tilted Dark Halos Are Common and Long-Lived, and Can Warp Galactic Disks},
  author = {Han, Jiwon Jesse and Semenov, Vadim and Conroy, Charlie and Hernquist, Lars},
  year = 2023,
  month = nov,
  journal = {The Astrophysical Journal},
  volume = {957},
  pages = {L24},
  publisher = {IOP},
  issn = {0004-637X},
  doi = {10.3847/2041-8213/ad0641},
  urldate = {2026-05-16}
}

@article{semczukTidallyInducedWarps2020,
  title = {Tidally Induced Warps of Spiral Galaxies in IllustrisTNG},
  author = {Semczuk, Marcin and {\L}okas, Ewa L. and D'Onghia, Elena and Athanassoula, E. and Debattista, Victor P. and Hernquist, Lars},
  year = 2020,
  month = nov,
  journal = {Monthly Notices of the Royal Astronomical Society},
  volume = {498},
  pages = {3535--3548},
  publisher = {OUP},
  issn = {0035-8711},
  doi = {10.1093/mnras/staa2609},
  urldate = {2026-05-16}
}

@article{zeeWarpedDiskGalaxies2022,
  title = {Warped Disk Galaxies. I. Linking U-Type Warps in Groups/Clusters to Jellyfish Galaxies},
  author = {Zee, Woong-Bae G. and Yoon, Suk-Jin and Moon, Jun-Sung and An, Sung-Ho and Paudel, Sanjaya and Yun, Kiyun},
  year = 2022,
  month = aug,
  journal = {The Astrophysical Journal},
  volume = {935},
  pages = {48},
  publisher = {IOP},
  issn = {0004-637X},
  doi = {10.3847/1538-4357/ac7462},
  urldate = {2026-05-16}
}

@article{nelsonFirstResultsIllustrisTNG2018,
  title = {First Results from the IllustrisTNG Simulations: The Galaxy Colour Bimodality},
  shorttitle = {First Results from the IllustrisTNG Simulations},
  author = {Nelson, Dylan and Pillepich, Annalisa and Springel, Volker and Weinberger, Rainer and Hernquist, Lars and Pakmor, R{\"u}diger and Genel, Shy and Torrey, Paul and Vogelsberger, Mark and Kauffmann, Guinevere and Marinacci, Federico and Naiman, Jill},
  year = 2018,
  month = mar,
  journal = {Monthly Notices of the Royal Astronomical Society},
  volume = {475},
  pages = {624--647},
  publisher = {OUP},
  issn = {0035-8711},
  doi = {10.1093/mnras/stx3040},
  urldate = {2026-05-16}
}

@article{naabFormationBoxyDisky1999,
  title = {On the Formation of Boxy and Disky Elliptical Galaxies},
  author = {Naab, Thorsten and Burkert, Andreas and Hernquist, Lars},
  year = 1999,
  month = oct,
  journal = {The Astrophysical Journal},
  volume = {523},
  pages = {L133-L136},
  publisher = {IOP},
  issn = {0004-637X},
  doi = {10.1086/312275},
  urldate = {2026-03-10}
}

@article{zeeuwStructureDynamicsElliptical1991,
  title = {Structure and Dynamics of Elliptical Galaxies},
  author = {de Zeeuw, Tim and Franx, Marijn},
  year = 1991,
  month = sep,
  journal = {Annual Review of Astronomy and Astrophysics},
  volume = {29},
  number = {Volume 29, 1991},
  pages = {239--274},
  publisher = {Annual Reviews},
  issn = {0066-4146, 1545-4282},
  doi = {10.1146/annurev.aa.29.090191.001323},
  urldate = {2026-03-12},
  langid = {english}
}

@article{benacchioTriaxialityEllipticalGalaxies1980,
  title = {Triaxiality in Elliptical Galaxies},
  author = {Benacchio, L. and Galletta, G.},
  year = 1980,
  month = dec,
  journal = {Monthly Notices of the Royal Astronomical Society},
  volume = {193},
  pages = {885--894},
  publisher = {OUP},
  issn = {0035-8711},
  doi = {10.1093/mnras/193.4.885},
  urldate = {2026-05-16}
}

@article{favaroIntrinsicFlatteningGalaxy2024,
  title = {The Intrinsic Flattening of Galaxy Disks},
  author = {Favaro, Jeremy and Courteau, St{\'e}phane and Comer{\'o}n, S{\'e}bastien and Stone, Connor},
  year = 2024,
  month = dec,
  journal = {The Astrophysical Journal},
  volume = {978},
  number = {1},
  pages = {63},
  publisher = {The American Astronomical Society},
  issn = {0004-637X},
  doi = {10.3847/1538-4357/ad932e},
  urldate = {2026-02-28},
  langid = {english}
}

@article{mendez-abreuIntrinsicThreedimensionalShape2018,
  title = {The Intrinsic Three-Dimensional Shape of Galactic Bars},
  author = {{M{\'e}ndez-Abreu}, J. and Costantin, L. and Aguerri, J. A. L. and {de Lorenzo-C{\'a}ceres}, A. and Corsini, E. M.},
  year = 2018,
  month = sep,
  journal = {Monthly Notices of the Royal Astronomical Society},
  volume = {479},
  pages = {4172--4186},
  publisher = {OUP},
  issn = {0035-8711},
  doi = {10.1093/mnras/sty1694},
  urldate = {2026-02-28}
}

@article{costantinIntrinsicShapeBulges2018,
  title = {The Intrinsic Shape of Bulges in the CALIFA Survey},
  author = {Costantin, L. and {M{\'e}ndez-Abreu}, J. and Corsini, E. M. and {Eliche-Moral}, M. C. and Tapia, T. and Morelli, L. and Bont{\`a}, E. Dalla and Pizzella, A.},
  year = 2018,
  month = jan,
  journal = {Astronomy \& Astrophysics},
  volume = {609},
  pages = {A132},
  publisher = {EDP Sciences},
  issn = {0004-6361, 1432-0746},
  doi = {10.1051/0004-6361/201731823},
  urldate = {2026-02-28},
  copyright = {\copyright{} ESO, 2018},
  langid = {english}
}

@article{weijmansATLAS3DProject2014,
  title = {The ATLAS 3D Project -- XXIV. The Intrinsic Shape Distribution of Early-Type Galaxies},
  author = {Weijmans, Anne-Marie and {de Zeeuw}, P. T. and Emsellem, Eric and Krajnovi{\'c}, Davor and Lablanche, Pierre-Yves and Alatalo, Katherine and Blitz, Leo and Bois, Maxime and Bournaud, Fr{\'e}d{\'e}ric and Bureau, Martin and Cappellari, Michele and Crocker, Alison F. and Davies, Roger L. and Davis, Timothy A. and Duc, Pierre-Alain and Khochfar, Sadegh and Kuntschner, Harald and McDermid, Richard M. and Morganti, Raffaella and Naab, Thorsten and Oosterloo, Tom and Sarzi, Marc and Scott, Nicholas and Serra, Paolo and Verdoes Kleijn, Gijs and Young, Lisa M.},
  year = 2014,
  month = nov,
  journal = {Monthly Notices of the Royal Astronomical Society},
  volume = {444},
  number = {4},
  pages = {3340--3356},
  issn = {0035-8711},
  doi = {10.1093/mnras/stu1603},
  urldate = {2026-03-10}
}

@article{roychowdhuryIntrinsicShapesDwarf2013,
  title = {The Intrinsic Shapes of Dwarf Irregular Galaxies.},
  author = {Roychowdhury, S. and Chengalur, J. N. and Karachentsev, I. D. and Kaisina, E. I.},
  year = 2013,
  month = nov,
  journal = {Monthly Notices of the Royal Astronomical Society},
  volume = {436},
  pages = {L104-L108},
  publisher = {OUP},
  issn = {0035-8711},
  doi = {10.1093/mnrasl/slt123},
  urldate = {2026-03-10}
}

@article{rodriguezIntrinsicShapeGalaxies2013,
  title = {The Intrinsic Shape of Galaxies in SDSS/Galaxy Zoo},
  author = {Rodr{\'i}guez, Silvio and Padilla, Nelson D.},
  year = 2013,
  month = sep,
  journal = {Monthly Notices of the Royal Astronomical Society},
  volume = {434},
  pages = {2153--2166},
  publisher = {OUP},
  issn = {0035-8711},
  doi = {10.1093/mnras/stt1168},
  urldate = {2026-02-28}
}

@article{chakrabortyIntrinsicShapesVery2011,
  title = {Intrinsic Shapes of Very Flat Elliptical Galaxies},
  author = {Chakraborty, D. K. and Diwakar, A. K. and Pandey, S. K.},
  year = 2011,
  month = mar,
  journal = {Monthly Notices of the Royal Astronomical Society},
  volume = {412},
  number = {1},
  pages = {585--590},
  issn = {0035-8711},
  doi = {10.1111/j.1365-2966.2010.17930.x},
  urldate = {2026-03-10}
}

@article{padillaShapesGalaxiesSloan2008,
  title = {The Shapes of Galaxies in the Sloan Digital Sky Survey},
  author = {Padilla, Nelson D. and Strauss, Michael A.},
  year = 2008,
  month = aug,
  journal = {Monthly Notices of the Royal Astronomical Society},
  volume = {388},
  pages = {1321--1334},
  publisher = {OUP},
  issn = {0035-8711},
  doi = {10.1111/j.1365-2966.2008.13480.x},
  urldate = {2026-05-16}
}

@article{lambasTrueShapesGalaxies1992,
  title = {On the True Shapes of Galaxies.},
  author = {Lambas, D. G. and Maddox, S. J. and Loveday, J.},
  year = 1992,
  month = sep,
  journal = {Monthly Notices of the Royal Astronomical Society},
  volume = {258},
  pages = {404--414},
  publisher = {OUP},
  issn = {0035-8711},
  doi = {10.1093/mnras/258.2.404},
  urldate = {2026-05-16}
}

@article{bakIntrinsicShapeDistribution2000,
  title = {The Intrinsic Shape Distribution of a Sample of Elliptical Galaxies},
  author = {Bak, Jakob and Statler, Thomas S.},
  year = 2000,
  month = jul,
  journal = {The Astronomical Journal},
  volume = {120},
  pages = {110--122},
  publisher = {IOP},
  issn = {0004-6256},
  doi = {10.1086/301437},
  urldate = {2026-05-16}
}

@article{binneyApparentTrueEllipticities1981,
  title = {The Apparent and True Ellipticities of Galaxies of Different Hubble Types in the Second Reference Catalogue.},
  author = {Binney, J. and {de Vaucouleurs}, G.},
  year = 1981,
  month = feb,
  journal = {Monthly Notices of the Royal Astronomical Society},
  volume = {194},
  pages = {679--691},
  publisher = {OUP},
  issn = {0035-8711},
  doi = {10.1093/mnras/194.3.679},
  urldate = {2026-05-16}
}

@article{sandageIntrinsicFlatteningSpiral1970,
  title = {The Intrinsic Flattening of e, so, and Spiral Galaxies as Related to Galaxy Formation and Evolution},
  author = {Sandage, Allan and Freeman, Kenneth C. and Stokes, N. R.},
  year = 1970,
  month = jun,
  journal = {The Astrophysical Journal},
  volume = {160},
  pages = {831},
  publisher = {IOP},
  issn = {0004-637X},
  doi = {10.1086/150475},
  urldate = {2026-05-16}
}

@article{erwinIMFITFASTFLEXIBLE2015,
  title = {IMFIT: A FAST, FLEXIBLE NEW PROGRAM FOR ASTRONOMICAL IMAGE FITTING},
  shorttitle = {IMFIT},
  author = {Erwin, Peter},
  year = 2015,
  month = jan,
  journal = {The Astrophysical Journal},
  volume = {799},
  number = {2},
  pages = {226},
  publisher = {The American Astronomical Society},
  issn = {0004-637X},
  doi = {10.1088/0004-637X/799/2/226},
  urldate = {2024-12-29},
  langid = {english}
}

@article{pengDetailedDecompositionGalaxy2010,
  title = {Detailed Decomposition of Galaxy Images. II. Beyond Axisymmetric Models},
  author = {Peng, Chien Y. and Ho, Luis C. and Impey, Chris D. and Rix, Hans-Walter},
  year = 2010,
  month = jun,
  journal = {The Astronomical Journal},
  volume = {139},
  pages = {2097--2129},
  publisher = {IOP},
  issn = {0004-6256},
  doi = {10.1088/0004-6256/139/6/2097},
  urldate = {2026-05-16}
}

@article{pignatelliGASPHOTToolGalaxy2006,
  title = {GASPHOT: A Tool for Galaxy Automatic Surface PHOTometry},
  shorttitle = {GASPHOT},
  author = {Pignatelli, E. and Fasano, G. and Cassata, P.},
  year = 2006,
  month = jan,
  journal = {Astronomy and Astrophysics},
  volume = {446},
  pages = {373--388},
  publisher = {EDP},
  issn = {0004-6361},
  doi = {10.1051/0004-6361:20041704},
  urldate = {2026-05-16}
}

@article{desouzaBUDDANewTwodimensional2004,
  title = {BUDDA: A New Two-Dimensional Bulge/Disk Decomposition Code for Detailed Structural Analysis of Galaxies},
  shorttitle = {BUDDA},
  author = {{de Souza}, R. E. and Gadotti, D. A. and {dos Anjos}, S.},
  year = 2004,
  month = aug,
  journal = {The Astrophysical Journal Supplement Series},
  volume = {153},
  pages = {411--427},
  publisher = {IOP},
  issn = {0067-0049},
  doi = {10.1086/421554},
  urldate = {2026-05-16}
}

@article{pengDetailedStructuralDecomposition2002,
  title = {Detailed Structural Decomposition of Galaxy Images},
  author = {Peng, Chien Y. and Ho, Luis C. and Impey, Chris D. and Rix, Hans-Walter},
  year = 2002,
  month = jul,
  journal = {The Astronomical Journal},
  volume = {124},
  pages = {266--293},
  publisher = {IOP},
  issn = {0004-6256},
  doi = {10.1086/340952},
  urldate = {2026-05-16}
}

@inproceedings{simardGIM2DIRAFPackage1998,
  title = {GIM2D: An IRAF Package for the Quantitative Morphology Analysis of Distant Galaxies},
  shorttitle = {GIM2D},
  booktitle = {Astronomical Data Analysis Software and Systems VII},
  author = {Simard, L.},
  year = 1998,
  month = jan,
  volume = {145},
  pages = {108},
  urldate = {2026-05-16}
}

@article{ciamburEllipsesAccuratelyModelling2015,
  title = {Beyond Ellipse(s): Accurately Modelling the Isophotal Structure of Galaxies with ISOFIT and CMODEL},
  shorttitle = {Beyond Ellipse(s)},
  author = {Ciambur, B. C.},
  year = 2015,
  month = sep,
  journal = {The Astrophysical Journal},
  volume = {810},
  pages = {120},
  publisher = {IOP},
  issn = {0004-637X},
  doi = {10.1088/0004-637X/810/2/120},
  urldate = {2026-05-16}
}

@inproceedings{buskoErrorEstimationElliptical1996,
  title = {Error Estimation in Elliptical Isophote Fitting},
  booktitle = {Astronomical Data Analysis Software and Systems V},
  author = {Busko, I. C.},
  year = 1996,
  month = jan,
  volume = {101},
  pages = {139},
  urldate = {2026-05-16}
}

@misc{kormendySecularEvolutionDisk2013,
  title = {Secular Evolution in Disk Galaxies},
  author = {Kormendy, John},
  year = 2013,
  month = nov,
  number = {arXiv:1311.2609},
  eprint = {1311.2609},
  publisher = {arXiv},
  doi = {10.48550/arXiv.1311.2609},
  urldate = {2024-11-25},
  archiveprefix = {arXiv}
}

@article{freemanDisksSpiralS01970,
  title = {On the Disks of Spiral and S0 Galaxies},
  author = {Freeman, K. C.},
  year = 1970,
  month = jun,
  journal = {The Astrophysical Journal},
  volume = {160},
  pages = {811},
  publisher = {IOP},
  issn = {0004-637X},
  doi = {10.1086/150474},
  urldate = {2026-05-16}
}

@article{hubbleExtragalacticNebulae1926,
  title = {Extragalactic Nebulae.},
  author = {Hubble, E. P.},
  year = 1926,
  month = dec,
  journal = {The Astrophysical Journal},
  volume = {64},
  pages = {321--369},
  publisher = {IOP},
  issn = {0004-637X},
  doi = {10.1086/143018},
  urldate = {2026-05-16}
}

@article{vandesandeRelationCharacteristicStellar2018,
  title = {A Relation between the Characteristic Stellar Ages of Galaxies and Their Intrinsic Shapes},
  author = {{van de Sande}, Jesse and Scott, Nicholas and {Bland-Hawthorn}, Joss and Brough, Sarah and Bryant, Julia J. and Colless, Matthew and Cortese, Luca and Croom, Scott M. and {d'Eugenio}, Francesco and Foster, Caroline and Goodwin, Michael and Konstantopoulos, Iraklis S. and Lawrence, Jon S. and McDermid, Richard M. and Medling, Anne M. and Owers, Matt S. and Richards, Samuel N. and Sharp, Rob},
  year = 2018,
  month = apr,
  journal = {Nature Astronomy},
  volume = {2},
  pages = {483--488},
  issn = {2397-3366},
  doi = {10.1038/s41550-018-0436-x},
  urldate = {2026-05-16}
}

@article{lokasBarlikeGalaxiesIllustrisTNG2021,
  title = {Bar-like Galaxies in IllustrisTNG},
  author = {{\L}okas, Ewa L.},
  year = 2021,
  month = mar,
  journal = {Astronomy and Astrophysics},
  volume = {647},
  pages = {A143},
  publisher = {EDP},
  issn = {0004-6361},
  doi = {10.1051/0004-6361/202040056},
  urldate = {2026-05-14}
}

@article{duEvolutionaryPathwaysDisk2021,
  title = {The Evolutionary Pathways of Disk-, Bulge-, and Halo-Dominated Galaxies},
  author = {Du, Min and Ho, Luis C. and Debattista, Victor P. and Pillepich, Annalisa and Nelson, Dylan and Hernquist, Lars and Weinberger, Rainer},
  year = 2021,
  month = oct,
  journal = {The Astrophysical Journal},
  volume = {919},
  number = {2},
  eprint = {2101.12373},
  primaryclass = {astro-ph},
  pages = {135},
  issn = {0004-637X, 1538-4357},
  doi = {10.3847/1538-4357/ac0e98},
  urldate = {2024-07-27},
  archiveprefix = {arXiv}
}

@article{moodySimulatingMultipleMerger2014,
  title = {Simulating Multiple Merger Pathways to the Central Kinematics of Early-Type Galaxies},
  author = {Moody, Christopher E. and Romanowsky, Aaron J. and Cox, Thomas J. and Novak, G. S. and Primack, Joel R.},
  year = 2014,
  month = oct,
  journal = {Monthly Notices of the Royal Astronomical Society},
  volume = {444},
  pages = {1475--1485},
  publisher = {OUP},
  issn = {0035-8711},
  doi = {10.1093/mnras/stu1444},
  urldate = {2026-05-13}
}

@article{lagosDiverseNatureFormation2022,
  title = {The Diverse Nature and Formation Paths of Slow Rotator Galaxies in the EAGLE Simulations},
  author = {Lagos, Claudia del P. and Emsellem, Eric and {van de Sande}, Jesse and Harborne, Katherine E. and Cortese, Luca and Davison, Thomas and Foster, Caroline and Wright, Ruby J.},
  year = 2022,
  month = jan,
  journal = {Monthly Notices of the Royal Astronomical Society},
  volume = {509},
  pages = {4372--4391},
  publisher = {OUP},
  issn = {0035-8711},
  doi = {10.1093/mnras/stab3128},
  urldate = {2026-05-13}
}

@article{liOriginPropertiesMassive2018,
  title = {The Origin and Properties of Massive Prolate Galaxies in the Illustris Simulation},
  author = {Li, Hongyu and Mao, Shude and Emsellem, Eric and Xu, Dandan and Springel, Volker and Krajnovi{\'c}, Davor},
  year = 2018,
  month = jan,
  journal = {Monthly Notices of the Royal Astronomical Society},
  volume = {473},
  pages = {1489--1511},
  publisher = {OUP},
  issn = {0035-8711},
  doi = {10.1093/mnras/stx2374},
  urldate = {2026-05-13}
}

@article{scannapiecoAquilaComparisonProject2012,
  title = {The Aquila Comparison Project: The Effects of Feedback and Numerical Methods on Simulations of Galaxy Formation},
  shorttitle = {The Aquila Comparison Project},
  author = {Scannapieco, C. and Wadepuhl, M. and Parry, O. H. and Navarro, J. F. and Jenkins, A. and Springel, V. and Teyssier, R. and Carlson, E. and Couchman, H. M. P. and Crain, R. A. and Dalla Vecchia, C. and Frenk, C. S. and Kobayashi, C. and Monaco, P. and Murante, G. and Okamoto, T. and Quinn, T. and Schaye, J. and Stinson, G. S. and Theuns, T. and Wadsley, J. and White, S. D. M. and Woods, R.},
  year = 2012,
  month = jun,
  journal = {Monthly Notices of the Royal Astronomical Society},
  volume = {423},
  pages = {1726--1749},
  publisher = {OUP},
  issn = {0035-8711},
  doi = {10.1111/j.1365-2966.2012.20993.x},
  urldate = {2026-05-13}
}

@article{roca-fabregaAGORAHighresolutionGalaxy2021,
  title = {The AGORA High-Resolution Galaxy Simulations Comparison Project. III. Cosmological Zoom-in Simulation of a Milky Way-Mass Halo},
  author = {{Roca-F{\`a}brega}, Santi and Kim, Ji-Hoon and Hausammann, Loic and Nagamine, Kentaro and Lupi, Alessandro and Powell, Johnny W. and Shimizu, Ikkoh and Ceverino, Daniel and Primack, Joel R. and Quinn, Thomas R. and Revaz, Yves and Vel{\'a}zquez, H{\'e}ctor and Abel, Tom and Buehlmann, Michael and Dekel, Avishai and Dong, Bili and Hahn, Oliver and Hummels, Cameron and Kim, Ki-Won and Smith, Britton D. and Strawn, Clayton and Teyssier, Romain and Turk, Matthew J. and {AGORA Collaboration}},
  year = 2021,
  month = aug,
  journal = {The Astrophysical Journal},
  volume = {917},
  pages = {64},
  publisher = {IOP},
  issn = {0004-637X},
  doi = {10.3847/1538-4357/ac088a},
  urldate = {2026-05-13}
}

@article{grothCosmologicalSimulationCode2023,
  title = {The Cosmological Simulation Code OPENGADGET3 - Implementation of Meshless Finite Mass},
  author = {Groth, Frederick and Steinwandel, Ulrich P. and Valentini, Milena and Dolag, Klaus},
  year = 2023,
  month = nov,
  journal = {Monthly Notices of the Royal Astronomical Society},
  volume = {526},
  pages = {616--644},
  publisher = {OUP},
  issn = {0035-8711},
  doi = {10.1093/mnras/stad2717},
  urldate = {2026-05-13}
}

@article{valentiniEffectGalacticOutflows2017,
  title = {On the Effect of Galactic Outflows in Cosmological Simulations of Disc Galaxies},
  author = {Valentini, Milena and Murante, Giuseppe and Borgani, Stefano and Monaco, Pierluigi and Bressan, Alessandro and Beck, Alexander M.},
  year = 2017,
  month = sep,
  journal = {Monthly Notices of the Royal Astronomical Society},
  volume = {470},
  pages = {3167--3193},
  publisher = {OUP},
  issn = {0035-8711},
  doi = {10.1093/mnras/stx1352},
  urldate = {2026-05-13}
}

@article{bauerCanStellarDiscs2019,
  title = {Can Stellar Discs in a Cosmological Setting Avoid Forming Strong Bars?},
  author = {Bauer, Jacob S and Widrow, Lawrence M},
  year = 2019,
  month = jun,
  journal = {Monthly Notices of the Royal Astronomical Society},
  volume = {486},
  number = {1},
  pages = {523--537},
  issn = {0035-8711},
  doi = {10.1093/mnras/stz478},
  urldate = {2024-10-09}
}

@article{rosas-guevaraBuildupStronglyBarred2020,
  title = {The Buildup of Strongly Barred Galaxies in the TNG100 Simulation},
  author = {{Rosas-Guevara}, Yetli and Bonoli, Silvia and Dotti, Massimo and Zana, Tommaso and Nelson, Dylan and Pillepich, Annalisa and Ho, Luis C. and {Izquierdo-Villalba}, David and Hernquist, Lars and Pakmor, R{\"u}ediger},
  year = 2020,
  month = jan,
  journal = {Monthly Notices of the Royal Astronomical Society},
  volume = {491},
  pages = {2547--2564},
  publisher = {OUP},
  issn = {0035-8711},
  doi = {10.1093/mnras/stz3180},
  urldate = {2026-05-10}
}

@article{angeloudiConstraintsSituEx2024,
  title = {Constraints on the in Situ and Ex Situ Stellar Masses in Nearby Galaxies Obtained with Artificial Intelligence},
  author = {Angeloudi, Eirini and {Falc{\'o}n-Barroso}, Jes{\'u}s and {Huertas-Company}, Marc and Boecker, Alina and Sarmiento, Regina and Eisert, Lukas and Pillepich, Annalisa},
  year = 2024,
  month = oct,
  journal = {Nature Astronomy},
  volume = {8},
  pages = {1310--1320},
  issn = {2397-3366},
  doi = {10.1038/s41550-024-02327-3},
  urldate = {2026-05-10}
}

@article{davisonEAGLEsViewEx2020,
  title = {An EAGLE's View of Ex Situ Galaxy Growth},
  author = {Davison, Thomas A. and Norris, Mark A. and Pfeffer, Joel L. and Davies, Jonathan J. and Crain, Robert A.},
  year = 2020,
  month = sep,
  journal = {Monthly Notices of the Royal Astronomical Society},
  volume = {497},
  pages = {81--93},
  publisher = {OUP},
  issn = {0035-8711},
  doi = {10.1093/mnras/staa1816},
  urldate = {2026-05-10}
}

@article{tacchellaMorphologyStarFormation2019,
  title = {Morphology and Star Formation in IllustrisTNG: The Build-up of Spheroids and Discs},
  shorttitle = {Morphology and Star Formation in IllustrisTNG},
  author = {Tacchella, Sandro and Diemer, Benedikt and Hernquist, Lars and Genel, Shy and Marinacci, Federico and Nelson, Dylan and Pillepich, Annalisa and {Rodriguez-Gomez}, Vicente and Sales, Laura V. and Springel, Volker and Vogelsberger, Mark},
  year = 2019,
  month = aug,
  journal = {Monthly Notices of the Royal Astronomical Society},
  volume = {487},
  pages = {5416--5440},
  publisher = {OUP},
  issn = {0035-8711},
  doi = {10.1093/mnras/stz1657},
  urldate = {2026-05-10}
}

@article{rodriguez-gomezRoleMergersHalo2017,
  title = {The Role of Mergers and Halo Spin in Shaping Galaxy Morphology},
  author = {{Rodriguez-Gomez}, Vicente and Sales, Laura V. and Genel, Shy and Pillepich, Annalisa and Zjupa, Jolanta and Nelson, Dylan and Griffen, Brendan and Torrey, Paul and Snyder, Gregory F. and Vogelsberger, Mark and Springel, Volker and Ma, Chung-Pei and Hernquist, Lars},
  year = 2017,
  month = may,
  journal = {Monthly Notices of the Royal Astronomical Society},
  volume = {467},
  pages = {3083--3098},
  publisher = {OUP},
  issn = {0035-8711},
  doi = {10.1093/mnras/stx305},
  urldate = {2026-05-10}
}

@article{combesFormationPropertiesPersisting1981,
  title = {Formation and Properties of Persisting Stellar Bars.},
  author = {Combes, F. and Sanders, R. H.},
  year = 1981,
  month = mar,
  journal = {Astronomy and Astrophysics},
  volume = {96},
  pages = {164--173},
  publisher = {EDP},
  issn = {0004-6361},
  urldate = {2026-05-08}
}

@article{valluriUnifiedFrameworkOrbital2016,
  title = {A Unified Framework for the Orbital Structure of Bars and Triaxial Ellipsoids},
  author = {Valluri, Monica and Shen, Juntai and Abbott, Caleb and Debattista, Victor P.},
  year = 2016,
  month = feb,
  journal = {The Astrophysical Journal},
  volume = {818},
  pages = {141},
  publisher = {IOP},
  issn = {0004-637X},
  doi = {10.3847/0004-637X/818/2/141},
  urldate = {2026-05-08}
}

@article{contopoulosOrbitsBarredGalaxies1989,
  title = {Orbits in Barred Galaxies},
  author = {Contopoulos, G. and Grosbol, P.},
  year = 1989,
  month = nov,
  journal = {Astronomy and Astrophysics Review},
  volume = {1},
  pages = {261--289},
  publisher = {Springer},
  issn = {0935-4956},
  doi = {10.1007/BF00873080},
  urldate = {2026-05-08}
}

@article{rahaDynamicalInstabilityBars1991,
  title = {A Dynamical Instability of Bars in Disk Galaxies},
  author = {Raha, N. and Sellwood, J. A. and James, R. A. and Kahn, F. D.},
  year = 1991,
  month = aug,
  journal = {Nature},
  volume = {352},
  pages = {411--412},
  issn = {0028-0836},
  doi = {10.1038/352411a0},
  urldate = {2026-05-07}
}

@article{shenOurMilkyWay2010,
  title = {Our Milky Way as a Pure-Disk Galaxy---A Challenge for Galaxy Formation},
  author = {Shen, Juntai and Rich, R. Michael and Kormendy, John and Howard, Christian D. and De Propris, Roberto and Kunder, Andrea},
  year = 2010,
  month = sep,
  journal = {The Astrophysical Journal},
  volume = {720},
  pages = {L72-L76},
  publisher = {IOP},
  issn = {0004-637X},
  doi = {10.1088/2041-8205/720/1/L72},
  urldate = {2026-05-07}
}

@book{binneyGalacticDynamicsSecond2008,
  title = {Galactic Dynamics: Second Edition},
  shorttitle = {Galactic Dynamics},
  author = {Binney, James and Tremaine, Scott},
  year = 2008,
  month = jan,
  journal = {Galactic Dynamics: Second Edition},
  urldate = {2026-04-30}
}

@article{athanassoulaMorphologyBarOrbits1992,
  title = {Morphology of Bar Orbits.},
  author = {Athanassoula, E.},
  year = 1992,
  month = nov,
  journal = {Monthly Notices of the Royal Astronomical Society},
  volume = {259},
  pages = {328--344},
  publisher = {OUP},
  issn = {0035-8711},
  doi = {10.1093/mnras/259.2.328},
  urldate = {2026-04-30}
}

@article{jedrzejewskiCCDSurfacePhotometry1987,
  title = {CCD Surface Photometry of Elliptical Galaxies - I. Observations, Reduction and Results.},
  author = {Jedrzejewski, Robert I.},
  year = 1987,
  month = jun,
  journal = {Monthly Notices of the Royal Astronomical Society},
  volume = {226},
  pages = {747--768},
  publisher = {OUP},
  issn = {0035-8711},
  doi = {10.1093/mnras/226.4.747},
  urldate = {2026-04-30}
}

@article{dubinskiStructureColdDark1991,
  title = {The Structure of Cold Dark Matter Halos},
  author = {Dubinski, John and Carlberg, R. G.},
  year = 1991,
  month = sep,
  journal = {The Astrophysical Journal},
  volume = {378},
  pages = {496},
  publisher = {IOP},
  issn = {0004-637X},
  doi = {10.1086/170451},
  urldate = {2026-04-30}
}

@article{katzDissipationlessCollapseExpanding1991,
  title = {Dissipationless Collapse in an Expanding Universe},
  author = {Katz, Neal},
  year = 1991,
  month = feb,
  journal = {The Astrophysical Journal},
  volume = {368},
  pages = {325},
  publisher = {IOP},
  issn = {0004-637X},
  doi = {10.1086/169696},
  urldate = {2026-04-30}
}

@article{sellwoodDynamicsBarredGalaxies1993,
  title = {Dynamics of Barred Galaxies},
  author = {Sellwood, J. A. and Wilkinson, A.},
  year = 1993,
  month = feb,
  journal = {Rep. Prog. Phys.},
  volume = {56},
  number = {2},
  eprint = {astro-ph/0608665},
  pages = {173--256},
  issn = {0034-4885, 1361-6633},
  doi = {10.1088/0034-4885/56/2/001},
  urldate = {2024-07-04},
  archiveprefix = {arXiv}
}

@article{nietoOriginInnerIsophotal1992,
  title = {The Origin of Inner Isophotal Twits in Elliptical Galaxies.},
  author = {Nieto, J. -L. and Bender, R. and Poulain, P. and Surma, P.},
  year = 1992,
  month = apr,
  journal = {Astronomy and Astrophysics},
  volume = {257},
  pages = {97--117},
  publisher = {EDP},
  issn = {0004-6361},
  urldate = {2025-10-07}
}

@article{bittnerGalaxiesGalaxiesTIMER2021,
  title = {Galaxies within Galaxies in the TIMER Survey: Stellar Populations of Inner Bars Are Scaled Replicas of Main Bars},
  shorttitle = {Galaxies within Galaxies in the TIMER Survey},
  author = {Bittner, Adrian and {de Lorenzo-C{\'a}ceres}, Adriana and Gadotti, Dimitri A. and {S{\'a}nchez-Bl{\'a}zquez}, Patricia and Neumann, Justus and Coelho, Paula and {Falc{\'o}n-Barroso}, Jes{\'u}s and Fragkoudi, Francesca and Kim, Taehyun and {Mart{\'i}n-Navarro}, Ignacio and {M{\'e}ndez-Abreu}, Jairo and P{\'e}rez, Isabel and Querejeta, Miguel and {van de Ven}, Glenn},
  year = 2021,
  month = feb,
  journal = {Astronomy and Astrophysics},
  volume = {646},
  pages = {A42},
  publisher = {EDP},
  issn = {0004-6361},
  doi = {10.1051/0004-6361/202039505},
  urldate = {2026-04-30}
}

@article{erwinDoublebarredGalaxiesCatalog2004,
  title = {Double-Barred Galaxies - I. A Catalog of Barred Galaxies with Stellar Secondary Bars and Inner Disks},
  author = {Erwin, Peter},
  year = 2004,
  month = mar,
  journal = {A\&A},
  volume = {415},
  number = {3},
  pages = {941--957},
  publisher = {EDP Sciences},
  issn = {0004-6361, 1432-0746},
  doi = {10.1051/0004-6361:20034408},
  urldate = {2024-07-06},
  copyright = {{\copyright} ESO, 2004},
  langid = {english}
}

@article{annWarpedDisksSpiral2006,
  title = {Warped Disks in Spiral Galaxies},
  author = {Ann, H. B. and Park, J.-C.},
  year = 2006,
  month = jan,
  journal = {New Astronomy},
  volume = {11},
  pages = {293--305},
  publisher = {Elsevier},
  issn = {1384-1076},
  doi = {10.1016/j.newast.2005.08.006},
  urldate = {2026-04-30}
}

@article{sanchez-saavedraFrequencyWarpedSpiral1990,
  title = {Frequency of Warped Spiral Galaxies at Visible Wavelengths.},
  author = {{S{\'a}nchez-Saavedra}, M. L. and Battaner, E. and Florido, E.},
  year = 1990,
  month = oct,
  journal = {Monthly Notices of the Royal Astronomical Society},
  volume = {246},
  pages = {458},
  publisher = {OUP},
  issn = {0035-8711},
  urldate = {2026-04-30}
}

@article{duKinematicDecompositionIllustrisTNG2020,
  title = {Kinematic Decomposition of IllustrisTNG Disk Galaxies: Morphology and Relation with Morphological Structures},
  shorttitle = {Kinematic Decomposition of IllustrisTNG Disk Galaxies},
  author = {Du, Min and Ho, Luis C. and Debattista, Victor P. and Pillepich, Annalisa and Nelson, Dylan and Zhao, Dongyao and Hernquist, Lars},
  year = 2020,
  month = jun,
  journal = {The Astrophysical Journal},
  volume = {895},
  pages = {139},
  publisher = {IOP},
  issn = {0004-637X},
  doi = {10.3847/1538-4357/ab8fa8},
  urldate = {2026-04-23}
}

@article{duIdentifyingKinematicStructures2019,
  title = {Identifying Kinematic Structures in Simulated Galaxies Using Unsupervised Machine Learning},
  author = {Du, Min and Ho, Luis C. and Zhao, Dongyao and Shi, Jingjing and Debattista, Victor P. and Hernquist, Lars and Nelson, Dylan},
  year = 2019,
  month = oct,
  journal = {ApJ},
  volume = {884},
  number = {2},
  pages = {129},
  publisher = {The American Astronomical Society},
  issn = {0004-637X},
  doi = {10.3847/1538-4357/ab43cc},
  urldate = {2026-04-23},
  langid = {english}
}

@article{athanassoulaBoxyPeanutBulges2016,
  title = {Boxy/Peanut/X Bulges, Barlenses and the Thick Part of Galactic Bars: What Are They and How Did They Form?},
  shorttitle = {Boxy/Peanut/X Bulges, Barlenses and the Thick Part of Galactic Bars},
  author = {Athanassoula, E.},
  year = 2016,
  month = jan,
  volume = {418},
  pages = {391},
  address = {eprint: arXiv:1503.04804},
  doi = {10.1007/978-3-319-19378-6_14},
  urldate = {2024-06-24}
}

@article{crainHydrodynamicalSimulationsGalaxy2023,
  title = {Hydrodynamical Simulations of the Galaxy Population: Enduring Successes and Outstanding Challenges},
  shorttitle = {Hydrodynamical Simulations of the Galaxy Population},
  author = {Crain, Robert A. and van de Voort, Freeke},
  year = 2023,
  month = aug,
  journal = {Annual Review of Astronomy and Astrophysics},
  volume = {61},
  number = {Volume 61, 2023},
  pages = {473--515},
  publisher = {Annual Reviews},
  issn = {0066-4146, 1545-4282},
  doi = {10.1146/annurev-astro-041923-043618},
  urldate = {2026-04-14},
  langid = {english}
}

@article{sellwoodSecularEvolutionDisk2014,
  title = {Secular Evolution in Disk Galaxies},
  author = {Sellwood, J. A.},
  year = 2014,
  month = jan,
  journal = {Rev. Mod. Phys.},
  volume = {86},
  number = {1},
  pages = {1--46},
  publisher = {American Physical Society},
  doi = {10.1103/RevModPhys.86.1},
  urldate = {2024-07-12}
}

@misc{andersonInterplayAccretionGalaxy2023,
  title = {The Interplay between Accretion, Galaxy Downsizing and the Formation of Box/Peanut Bulges in TNG50},
  author = {Anderson, Stuart Robert and {Gough-Kelly}, Steven and Debattista, Victor P. and Du, Min and Erwin, Peter and Cuomo, Virginia and Caruana, Joseph and Hernquist, Lars and Vogelsberger, Mark},
  year = 2023,
  month = nov,
  number = {arXiv:2302.12788},
  eprint = {2302.12788},
  primaryclass = {astro-ph},
  publisher = {arXiv},
  doi = {10.48550/arXiv.2302.12788},
  urldate = {2024-08-22},
  archiveprefix = {arXiv}
}

@article{erwinCaughtActDirect2016,
  title = {Caught in the Act: Direct Detection of Galactic Bars in the Buckling Phase},
  shorttitle = {Caught in the Act},
  author = {Erwin, Peter and Debattista, Victor P.},
  year = 2016,
  month = jul,
  journal = {ApJL},
  volume = {825},
  number = {2},
  eprint = {1607.01290},
  primaryclass = {astro-ph},
  pages = {L30},
  issn = {2041-8205, 2041-8213},
  doi = {10.3847/2041-8205/825/2/L30},
  urldate = {2024-07-04},
  archiveprefix = {arXiv}
}

@misc{duRevisitingExcessBarlike2026,
  title = {Revisiting the Excess of Bar-like Structures in TNG50 Early-Type Galaxies: Consistency and Tension with Observations},
  shorttitle = {Revisiting the Excess of Bar-like Structures in TNG50 Early-Type Galaxies},
  author = {Du, Hangci and Wang, Yougang and Ge, Junqiang},
  year = 2026,
  month = mar,
  publisher = {arXiv},
  doi = {10.48550/arXiv.2603.21279},
  urldate = {2026-03-30}
}

@software{andrew_pontzen_2026_pynbody,
  author       = {Andrew Pontzen and
                  Rok Roškar and
                  Corentin Cadiou and
                  Greg Stinson and
                  Ben Keller and
                  Michele Mastropietro and
                  Shuai Lu and
                  A R Duffy and
                  Michael Tremmel and
                  Jorge Sarrato Alós and
                  Jo Bovy and
                  mkrets and
                  Martin P. Rey and
                  Jon Davies and
                  Thomas Quinn and
                  Eva Franck and
                  Rick Sarmento and
                  Rory Woods and
                  Isaac Alonso and
                  nroth0815 and
                  Jonathan Coles and
                  mtryan83 and
                  Alex Ji and
                  Claude and
                  Elaad Applebaum and
                  Tommaso Zana and
                  Tom Callingham and
                  Pawel Biernacki},
  title        = {pynbody/pynbody: Version 2.4.1},
  month        = jan,
  year         = 2026,
  publisher    = {Zenodo},
  version      = {v2.4.1},
  doi          = {10.5281/zenodo.18148085},
  url          = {https://doi.org/10.5281/zenodo.18148085},
}

@software{newville2025limfit,
  author       = {Newville, Matthew and
                  Otten, Renee and
                  Nelson, Andrew and
                  Stensitzki, Till and
                  Ingargiola, Antonino and
                  Allan, Daniel and
                  Fox, Austin and
                  Carter, Faustin and
                  Rawlik, Michal},
  title        = {LMFIT: Non-Linear Least-Squares Minimization and
                   Curve-Fitting for Python
                  },
  month        = jul,
  year         = 2025,
  publisher    = {Zenodo},
  version      = {1.3.4},
  doi          = {10.5281/zenodo.16175987},
  url          = {https://doi.org/10.5281/zenodo.16175987},
  swhid        = {swh:1:dir:76742b0e41b1d2bff5a3716dd2376531f3a21db8
                   ;origin=https://doi.org/10.5281/zenodo.598352;visi
                   t=swh:1:snp:89f98ce93be53a573d85de0145f9174c827b7e
                   92;anchor=swh:1:rel:9528e5133d2d78c4036335f023b212
                   487cf1b632;path=lmfit-lmfit-py-0566445
                  },
}

@article{behnelCythonBestBoth2011,
  title = {Cython: The Best of Both Worlds},
  shorttitle = {Cython},
  author = {Behnel, Stefan and Bradshaw, Robert and Citro, Craig and Dalcin, Lisandro and Seljebotn, Dag Sverre and Smith, Kurt},
  year = 2011,
  month = mar,
  journal = {Computing in Science and Engineering},
  volume = {13},
  pages = {31--39},
  publisher = {IEEE},
  doi = {10.1109/MCSE.2010.118},
  urldate = {2026-04-07}
}

@techreport{caroliRobustEfficientDelaunay2009,
  type = {Research Report},
  title = {Robust and Efficient Delaunay Triangulations of Points on or Close to a Sphere},
  author = {Caroli, Manuel and {Machado Manh{\~a}es de Castro}, Pedro and Loriot, Sebastien and Rouiller, Olivier and Teillaud, Monique and Wormser, Camille},
  year = 2009,
  number = {RR-7004},
  institution = {INRIA},
  urldate = {2026-04-07}
}

@article{springelGADGETCodeCollisionless2001,
  title = {GADGET: A Code for Collisionless and Gasdynamical Cosmological Simulations},
  shorttitle = {GADGET},
  author = {Springel, Volker and Yoshida, Naoki and White, Simon D. M.},
  year = 2001,
  month = apr,
  journal = {New Astronomy},
  volume = {6},
  number = {2},
  pages = {79--117},
  issn = {1384-1076},
  doi = {10.1016/S1384-1076(01)00042-2},
  urldate = {2026-04-07}
}

@article{zempDETERMININGSHAPEMATTER2011,
  title = {ON DETERMINING THE SHAPE OF MATTER DISTRIBUTIONS},
  author = {Zemp, Marcel and Gnedin, Oleg Y. and Gnedin, Nickolay Y. and Kravtsov, Andrey V.},
  year = 2011,
  month = nov,
  journal = {ApJS},
  volume = {197},
  number = {2},
  pages = {30},
  publisher = {The American Astronomical Society},
  issn = {0067-0049},
  doi = {10.1088/0067-0049/197/2/30},
  urldate = {2026-03-02},
  langid = {english}
}

@misc{pynbody,
  author = {{Pontzen}, A. and {Ro{\v s}kar}, R. and {Stinson}, G.~S. and {Woods},
     R. and {Reed}, D.~M. and {Coles}, J. and {Quinn}, T.~R.},
  title = "{pynbody: Astrophysics Simulation Analysis for Python}",
  note = {Astrophysics Source Code Library, ascl:1305.002},
  year = 2013
}

@Article{         harris2020array,
 title         = {Array programming with {NumPy}},
 author        = {Charles R. Harris and K. Jarrod Millman and St{\'{e}}fan J.
                 van der Walt and Ralf Gommers and Pauli Virtanen and David
                 Cournapeau and Eric Wieser and Julian Taylor and Sebastian
                 Berg and Nathaniel J. Smith and Robert Kern and Matti Picus
                 and Stephan Hoyer and Marten H. van Kerkwijk and Matthew
                 Brett and Allan Haldane and Jaime Fern{\'{a}}ndez del
                 R{\'{i}}o and Mark Wiebe and Pearu Peterson and Pierre
                 G{\'{e}}rard-Marchant and Kevin Sheppard and Tyler Reddy and
                 Warren Weckesser and Hameer Abbasi and Christoph Gohlke and
                 Travis E. Oliphant},
 year          = {2020},
 month         = sep,
 journal       = {Nature},
 volume        = {585},
 number        = {7825},
 pages         = {357--362},
 doi           = {10.1038/s41586-020-2649-2},
 publisher     = {Springer Science and Business Media {LLC}},
 url           = {https://doi.org/10.1038/s41586-020-2649-2}
}

@ARTICLE{2020SciPy-NMeth,
  author  = {Virtanen, Pauli and Gommers, Ralf and Oliphant, Travis E. and
            Haberland, Matt and Reddy, Tyler and Cournapeau, David and
            Burovski, Evgeni and Peterson, Pearu and Weckesser, Warren and
            Bright, Jonathan and {van der Walt}, St{\'e}fan J. and
            Brett, Matthew and Wilson, Joshua and Millman, K. Jarrod and
            Mayorov, Nikolay and Nelson, Andrew R. J. and Jones, Eric and
            Kern, Robert and Larson, Eric and Carey, C J and
            Polat, {\.I}lhan and Feng, Yu and Moore, Eric W. and
            {VanderPlas}, Jake and Laxalde, Denis and Perktold, Josef and
            Cimrman, Robert and Henriksen, Ian and Quintero, E. A. and
            Harris, Charles R. and Archibald, Anne M. and
            Ribeiro, Ant{\^o}nio H. and Pedregosa, Fabian and
            {van Mulbregt}, Paul and {SciPy 1.0 Contributors}},
  title   = {{{SciPy} 1.0: Fundamental Algorithms for Scientific
            Computing in Python}},
  journal = {Nature Methods},
  year    = {2020},
  volume  = {17},
  pages   = {261--272},
  adsurl  = {https://rdcu.be/b08Wh},
  doi     = {10.1038/s41592-019-0686-2},
}

@Article{Hunter:2007,
  Author    = {Hunter, J. D.},
  Title     = {Matplotlib: A 2D graphics environment},
  Journal   = {Computing in Science \& Engineering},
  Volume    = {9},
  Number    = {3},
  Pages     = {90--95},
  publisher = {IEEE COMPUTER SOC},
  doi       = {10.1109/MCSE.2007.55},
  year      = 2007
}

@article{algorryBarredGalaxiesEAGLE2017,
  author = {Algorry, David G. and Navarro, Julio F. and Abadi, Mario G. and Sales, Laura V. and Bower, Richard G. and Crain, Robert A. and Dalla Vecchia, Claudio and Frenk, Carlos S. and Schaller, Matthieu and Schaye, Joop and Theuns, Tom},
  year = {2017},
  month = jul,
  journal = {MNRAS},
  volume = {469},
  number = {1},
  pages = {1054--1064},
  issn = {0035-8711},
  doi = {10.1093/mnras/stx1008},
  note = {\url{https://doi.org/10.1093/mnras/stx1008}}
}

@article{zhaoBarredGalaxiesIllustrisTNG2020,
  author = {Zhao, Dongyao and Du, Min and Ho, Luis C. and Debattista, Victor P. and Shi, Jingjing},
  year = {2020},
  month = dec,
  journal = {ApJ},
  volume = {904},
  number = {2},
  pages = {170},
  publisher = {The American Astronomical Society},
  issn = {0004-637X},
  doi = {10.3847/1538-4357/abbe1b},
  langid = {english},
  note = {\url{https://dx.doi.org/10.3847/1538-4357/abbe1b}}
}

@article{luIllustrisTNGInsightsFactors2025,
  title = {IllustrisTNG Insights: Factors Affecting the Presence of Bars in Disk Galaxies},
  shorttitle = {IllustrisTNG Insights},
  author = {Lu, Shuai and Du, Min and Debattista, Victor P.},
  year = {2025},
  month = may,
  journal = {A\&A},
  volume = {697},
  pages = {A236},
  publisher = {EDP},
  issn = {0004-6361},
  doi = {10.1051/0004-6361/202453143},
  urldate = {2025-07-13}
}

@article{genelIntroducingIllustrisProject2014,
  shorttitle = {Introducing the Illustris Project},
  author = {Genel, Shy and Vogelsberger, Mark and Springel, Volker and Sijacki, Debora and Nelson, Dylan and Snyder, Greg and {Rodriguez-Gomez}, Vicente and Torrey, Paul and Hernquist, Lars},
  year = {2014},
  month = nov,
  journal = {MNRAS},
  volume = {445},
  number = {1},
  pages = {175--200},
  issn = {0035-8711},
  doi = {10.1093/mnras/stu1654},
  note = {\url{https://doi.org/10.1093/mnras/stu1654}}
}

@article{salesFeedbackStructureSimulated2010,
  author = {Sales, Laura V. and Navarro, Julio F. and Schaye, Joop and Vecchia, Claudio Dalla and Springel, Volker and Booth, C. M.},
  year = {2010},
  month = dec,
  journal = {MNRAS},
  volume = {409},
  number = {4},
  pages = {1541--1556},
  issn = {0035-8711},
  doi = {10.1111/j.1365-2966.2010.17391.x},
  note = {\url{https://doi.org/10.1111/j.1365-2966.2010.17391.x}}
}

@article{powerInnerStructureLCDM2003,
  title = {The Inner Structure of {$\Lambda$}CDM Haloes - I. A Numerical Convergence Study},
  author = {Power, C. and Navarro, J. F. and Jenkins, A. and Frenk, C. S. and White, S. D. M. and Springel, V. and Stadel, J. and Quinn, T.},
  year = 2003,
  month = jan,
  journal = {Monthly Notices of the Royal Astronomical Society},
  volume = {338},
  pages = {14--34},
  publisher = {OUP},
  issn = {0035-8711},
  doi = {10.1046/j.1365-8711.2003.05925.x},
  urldate = {2025-10-17}
}

@article{crainEAGLESimulationsGalaxy2015,
  title = {The EAGLE Simulations of Galaxy Formation: Calibration of Subgrid Physics and Model Variations},
  shorttitle = {The EAGLE Simulations of Galaxy Formation},
  author = {Crain, Robert A. and Schaye, Joop and Bower, Richard G. and Furlong, Michelle and Schaller, Matthieu and Theuns, Tom and Dalla Vecchia, Claudio and Frenk, Carlos S. and McCarthy, Ian G. and Helly, John C. and Jenkins, Adrian and {Rosas-Guevara}, Yetli M. and White, Simon D. M. and Trayford, James W.},
  year = 2015,
  month = jun,
  journal = {Monthly Notices of the Royal Astronomical Society},
  volume = {450},
  pages = {1937--1961},
  publisher = {OUP},
  issn = {0035-8711},
  doi = {10.1093/mnras/stv725},
  urldate = {2025-12-05}
}

@article{schayeEAGLEProjectSimulating2015,
  shorttitle = {The EAGLE Project},
  author = {Schaye, Joop and Crain, Robert A. and Bower, Richard G. and Furlong, Michelle and Schaller, Matthieu and Theuns, Tom and Dalla Vecchia, Claudio and Frenk, Carlos S. and McCarthy, I. G. and Helly, John C. and Jenkins, Adrian and {Rosas-Guevara}, Y. M. and White, Simon D. M. and Baes, Maarten and Booth, C. M. and Camps, Peter and Navarro, Julio F. and Qu, Yan and Rahmati, Alireza and Sawala, Till and Thomas, Peter A. and Trayford, James},
  year = {2015},
  month = jan,
  journal = {MNRAS},
  volume = {446},
  number = {1},
  pages = {521--554},
  issn = {0035-8711},
  doi = {10.1093/mnras/stu2058},
  note = {\url{https://doi.org/10.1093/mnras/stu2058}}
}

@article{springelCosmologicalSimulationCode2005,
  title = {The Cosmological Simulation Code GADGET-2},
  author = {Springel, Volker},
  year = 2005,
  month = dec,
  journal = {Monthly Notices of the Royal Astronomical Society},
  volume = {364},
  pages = {1105--1134},
  publisher = {OUP},
  issn = {0035-8711},
  doi = {10.1111/j.1365-2966.2005.09655.x},
  urldate = {2025-12-19}
}

@article{sijackiIllustrisSimulationEvolving2015,
  shorttitle = {The Illustris Simulation},
  author = {Sijacki, Debora and Vogelsberger, Mark and Genel, Shy and Springel, Volker and Torrey, Paul and Snyder, Gregory F. and Nelson, Dylan and Hernquist, Lars},
  year = {2015},
  month = sep,
  journal = {MNRAS},
  volume = {452},
  number = {1},
  pages = {575--596},
  issn = {0035-8711},
  doi = {10.1093/mnras/stv1340},
  note = {\url{https://doi.org/10.1093/mnras/stv1340}}
}

@article{nelsonIllustrisSimulationPublic2015,
  shorttitle = {The Illustris Simulation},
  author = {Nelson, D. and Pillepich, A. and Genel, S. and Vogelsberger, M. and Springel, V. and Torrey, P. and {Rodriguez-Gomez}, V. and Sijacki, D. and Snyder, G. F. and Griffen, B. and Marinacci, F. and Blecha, L. and Sales, L. and Xu, D. and Hernquist, L.},
  year = {2015},
  month = nov,
  journal = {Astronomy and Computing},
  volume = {13},
  pages = {12--37},
  issn = {2213-1337},
  doi = {10.1016/j.ascom.2015.09.003},
  langid = {english},
  note = {\url{https://www.sciencedirect.com/science/article/pii/S2213133715000864}}
}

@article{vogelsbergerIntroducingIllustrisProject2014,
  shorttitle = {Introducing the Illustris Project},
  author = {Vogelsberger, Mark and Genel, Shy and Springel, Volker and Torrey, Paul and Sijacki, Debora and Xu, Dandan and Snyder, Greg and Nelson, Dylan and Hernquist, Lars},
  year = {2014},
  month = oct,
  journal = {MNRAS},
  volume = {444},
  number = {2},
  pages = {1518--1547},
  issn = {0035-8711},
  doi = {10.1093/mnras/stu1536},
  note = {\url{https://doi.org/10.1093/mnras/stu1536}}
}

@article{vogelsbergerModelCosmologicalSimulations2013,
  author = {Vogelsberger, Mark and Genel, Shy and Sijacki, Debora and Torrey, Paul and Springel, Volker and Hernquist, Lars},
  year = {2013},
  month = dec,
  journal = {MNRAS},
  volume = {436},
  number = {4},
  pages = {3031--3067},
  issn = {0035-8711},
  doi = {10.1093/mnras/stt1789},
  note = {\url{https://doi.org/10.1093/mnras/stt1789}}
}

@article{vogelsbergerPropertiesGalaxiesReproduced2014,
  author = {Vogelsberger, M. and Genel, S. and Springel, V. and Torrey, P. and Sijacki, D. and Xu, D. and Snyder, G. and Bird, S. and Nelson, D. and Hernquist, L.},
  year = {2014},
  month = may,
  journal = {Nature},
  volume = {509},
  number = {7499},
  pages = {177--182},
  publisher = {Nature Publishing Group},
  issn = {1476-4687},
  doi = {10.1038/nature13316},
  copyright = {2014 Nature Publishing Group, a division of Macmillan Publishers Limited. All Rights Reserved.},
  langid = {english},
  note = {\url{https://www.nature.com/articles/nature13316}}
}

@article{pakmorImprovingConvergenceProperties2016,
  author = {Pakmor, R{\"u}diger and Springel, Volker and Bauer, Andreas and Mocz, Philip and Munoz, Diego J. and Ohlmann, Sebastian T. and Schaal, Kevin and Zhu, Chenchong},
  year = {2016},
  month = jan,
  journal = {MNRAS},
  volume = {455},
  pages = {1134--1143},
  issn = {0035-8711},
  doi = {10.1093/mnras/stv2380},
  note = {\url{https://ui.adsabs.harvard.edu/abs/2016MNRAS.455.1134P}}
}

@article{pakmorMagnetohydrodynamicsUnstructuredMoving2011,
  author = {Pakmor, Ruediger and Bauer, Andreas and Springel, Volker},
  year = {2011},
  month = dec,
  journal = {MNRAS},
  volume = {418},
  pages = {1392--1401},
  issn = {0035-8711},
  doi = {10.1111/j.1365-2966.2011.19591.x},
  note = {\url{https://ui.adsabs.harvard.edu/abs/2011MNRAS.418.1392P}}
}

@article{springelPurSiMuove2010,
  shorttitle = {E Pur Si Muove},
  author = {Springel, Volker},
  year = {2010},
  month = jan,
  journal = {MNRAS},
  volume = {401},
  pages = {791--851},
  issn = {0035-8711},
  doi = {10.1111/j.1365-2966.2009.15715.x},
  note = {\url{https://ui.adsabs.harvard.edu/abs/2010MNRAS.401..791S}}
}

@article{springelFirstResultsIllustrisTNG2018,
  shorttitle = {First Results from the IllustrisTNG Simulations},
  author = {Springel, Volker and Pakmor, R{\"u}diger and Pillepich, Annalisa and Weinberger, Rainer and Nelson, Dylan and Hernquist, Lars and Vogelsberger, Mark and Genel, Shy and Torrey, Paul and Marinacci, Federico and Naiman, Jill},
  year = {2018},
  month = mar,
  journal = {MNRAS},
  volume = {475},
  pages = {676--698},
  issn = {0035-8711},
  doi = {10.1093/mnras/stx3304},
  note = {\url{https://ui.adsabs.harvard.edu/abs/2018MNRAS.475..676S}}
}

@article{pillepichFirstResultsTNG502019,
  shorttitle = {First Results from the TNG50 Simulation},
  author = {Pillepich, Annalisa and Nelson, Dylan and Springel, Volker and Pakmor, R{\"u}diger and Torrey, Paul and Weinberger, Rainer and Vogelsberger, Mark and Marinacci, Federico and Genel, Shy and {van der Wel}, Arjen and Hernquist, Lars},
  year = {2019},
  month = dec,
  journal = {MNRAS},
  volume = {490},
  pages = {3196--3233},
  issn = {0035-8711},
  doi = {10.1093/mnras/stz2338},
  note = {\url{https://ui.adsabs.harvard.edu/abs/2019MNRAS.490.3196P}}
}

@article{pillepichFirstResultsIllustrisTNG2018,
  shorttitle = {First Results from the IllustrisTNG Simulations},
  author = {Pillepich, Annalisa and Nelson, Dylan and Hernquist, Lars and Springel, Volker and Pakmor, R{\"u}diger and Torrey, Paul and Weinberger, Rainer and Genel, Shy and Naiman, Jill P and Marinacci, Federico and Vogelsberger, Mark},
  year = {2018},
  month = mar,
  journal = {MNRAS},
  volume = {475},
  number = {1},
  pages = {648--675},
  issn = {0035-8711},
  doi = {10.1093/mnras/stx3112},
  note = {\url{https://doi.org/10.1093/mnras/stx3112}}
}

@article{nelsonIllustrisTNGSimulationsPublic2019,
  shorttitle = {The IllustrisTNG Simulations},
  author = {Nelson, Dylan and Springel, Volker and Pillepich, Annalisa and {Rodriguez-Gomez}, Vicente and Torrey, Paul and Genel, Shy and Vogelsberger, Mark and Pakmor, Ruediger and Marinacci, Federico and Weinberger, Rainer and Kelley, Luke and Lovell, Mark and Diemer, Benedikt and Hernquist, Lars},
  year = {2019},
  month = may,
  journal = {Computational Astrophysics and Cosmology},
  volume = {6},
  pages = {2},
  doi = {10.1186/s40668-019-0028-x},
  note = {\url{https://ui.adsabs.harvard.edu/abs/2019ComAC...6....2N}}
}

@article{naimanFirstResultsIllustrisTNG2018,
  shorttitle = {First Results from the IllustrisTNG Simulations},
  author = {Naiman, Jill P. and Pillepich, Annalisa and Springel, Volker and {Ramirez-Ruiz}, Enrico and Torrey, Paul and Vogelsberger, Mark and Pakmor, R{\"u}diger and Nelson, Dylan and Marinacci, Federico and Hernquist, Lars and Weinberger, Rainer and Genel, Shy},
  year = {2018},
  month = jun,
  journal = {MNRAS},
  volume = {477},
  pages = {1206--1224},
  issn = {0035-8711},
  doi = {10.1093/mnras/sty618},
  note = {\url{https://ui.adsabs.harvard.edu/abs/2018MNRAS.477.1206N}}
}

@article{marinacciFirstResultsIllustrisTNG2018,
  shorttitle = {First Results from the IllustrisTNG Simulations},
  author = {Marinacci, Federico and Vogelsberger, Mark and Pakmor, R{\"u}diger and Torrey, Paul and Springel, Volker and Hernquist, Lars and Nelson, Dylan and Weinberger, Rainer and Pillepich, Annalisa and Naiman, Jill and Genel, Shy},
  year = {2018},
  month = nov,
  journal = {MNRAS},
  volume = {480},
  pages = {5113--5139},
  issn = {0035-8711},
  doi = {10.1093/mnras/sty2206},
  note = {\url{https://ui.adsabs.harvard.edu/abs/2018MNRAS.480.5113M}}
}

@article{branchSubspaceInteriorConjugate1999,
  title = {A Subspace, Interior, and Conjugate Gradient Method for Large-Scale Bound-Constrained Minimization Problems},
  author = {Branch, Mary Ann and Coleman, Thomas F. and Li, Yuying},
  year = 1999,
  month = jan,
  journal = {SIAM J. Sci. Comput.},
  volume = {21},
  number = {1},
  pages = {1--23},
  publisher = {{Society for Industrial and Applied Mathematics}},
  issn = {1064-8275},
  doi = {10.1137/S1064827595289108},
  urldate = {2025-10-22}
}

@article{fritschMethodConstructingLocal1984,
  title = {A Method for Constructing Local Monotone Piecewise Cubic Interpolants},
  author = {Fritsch, F. N. and Butland, J.},
  year = 1984,
  month = jun,
  journal = {SIAM J. Sci. and Stat. Comput.},
  volume = {5},
  number = {2},
  pages = {300--304},
  publisher = {{Society for Industrial and Applied Mathematics}},
  issn = {0196-5204},
  doi = {10.1137/0905021},
  urldate = {2025-10-17}
}

@book{molerNumericalComputingMatlab2004,
  title = {Numerical Computing with Matlab},
  author = {Moler, Cleve B.},
  year = 2004,
  month = jan,
  series = {Other Titles in Applied Mathematics},
  publisher = {{Society for Industrial and Applied Mathematics}},
  doi = {10.1137/1.9780898717952},
  urldate = {2025-10-17},
  isbn = {978-0-89871-660-3}
}

@misc{maneewongvatanaAnalysisApproximateNearest1999,
  title = {Analysis of Approximate Nearest Neighbor Searching with Clustered Point Sets},
  author = {Maneewongvatana, Songrit and Mount, David M.},
  year = 1999,
  month = jan,
  publisher = {arXiv},
  doi = {10.48550/arXiv.cs/9901013},
  urldate = {2025-10-17}
}

@article{monaghanRefinedParticleMethod1985,
  title = {A Refined Particle Method for Astrophysical Problems},
  author = {Monaghan, J. J. and Lattanzio, J. C.},
  year = 1985,
  month = aug,
  journal = {Astronomy and Astrophysics},
  volume = {149},
  pages = {135--143},
  publisher = {EDP},
  issn = {0004-6361},
  urldate = {2025-10-17}
}
\bibliographystyle{aasjournalv7}


\end{CJK*}
\end{document}